\documentclass[journal]{IEEEtran}
\usepackage[T1]{fontenc}
\usepackage{textcomp}
\usepackage[latin9]{inputenc}
\usepackage{array}
\usepackage{booktabs}
\usepackage{multirow}
\usepackage{varwidth}
\usepackage{amsmath}
\usepackage{amssymb}
\usepackage{graphicx}
\usepackage{setspace}
\usepackage[bookmarks=true,bookmarksnumbered=true,bookmarksopen=true,bookmarksopenlevel=1,
 breaklinks=false,pdfborder={0 0 0},pdfborderstyle={},backref=false,colorlinks=false]
 {hyperref}
\hypersetup{pdftitle={Your Title},
 pdfauthor={Your Name},
 pdfpagelayout=OneColumn, pdfnewwindow=true, pdfstartview=XYZ, plainpages=false}

\makeatletter

\providecommand{\tabularnewline}{\\}
\newenvironment{cellvarwidth}[1][t]
    {\begin{varwidth}[#1]{\linewidth}}
    {\@finalstrut\@arstrutbox\end{varwidth}}

\let\oldforeign@language\foreign@language
\DeclareRobustCommand{\foreign@language}[1]{%
  \lowercase{\oldforeign@language{#1}}}
\newenvironment{lyxlist}[1]
	{\begin{list}{}
		{\settowidth{\labelwidth}{#1}
		 \setlength{\leftmargin}{\labelwidth}
		 \addtolength{\leftmargin}{\labelsep}
		 }}
	{\end{list}}

\usepackage[caption=false,font=footnotesize]{subfig}
\usepackage{cite}

\makeatother

\begin{document}
\title{A Hybrid Transmitting and Reflecting Beyond Diagonal Reconfigurable
Intelligent Surface with Independent Beam Control and Power Splitting}
\author{Zhaoyang~Ming,~\IEEEmembership{Student Member,~IEEE,} Shanpu~Shen,~\IEEEmembership{Senior Member,~IEEE,}
\\
Junhui~Rao,~\IEEEmembership{Member,~IEEE,} Zan~Li,~\IEEEmembership{Student Member,~IEEE,}
Jichen~Zhang,~\IEEEmembership{Student Member,~IEEE,} \\
Chi Yuk~Chiu,~\IEEEmembership{Senior Member,~IEEE,} and Ross~Murch,~\IEEEmembership{Fellow,~IEEE}\thanks{This work was supported by Hong Kong Research Grants Council for the
Area of Excellence Grant, AoE/E-601/22-R.}\thanks{Zhaoyang Ming, Junhui Rao, Zan Li, Jichen Zhang and Chi Yuk Chiu are
with the Department of Electronic and Computer Engineering, University
of Hong Kong Science and Technology, Hong Kong. (e-mail: \protect\href{mailto:zming@connect.ust.hk}{zming@connect.ust.hk},
\protect\href{mailto:jraoaa@connect.ust.hk}{jraoaa@connect.ust.hk}).}\thanks{Shanpu Shen is with the Department of Electrical Engineering and Electronics,
University of University of Liverpool, Liverpool, U.K. (e-mail: \protect\href{http://Shanpu.Shen@liverpool.ac.uk}{Shanpu.Shen@liverpool.ac.uk}).}\thanks{R. Murch is with the Department of Electronic and Computer Engineering
and Institute for Advanced Study (IAS) at the Hong Kong University
of Science and Technology, Hong Kong. (e-mail: \protect\href{http://eermurch@ust.hk}{eermurch@ust.hk}).}}
\markboth{}{Zhaoyang Ming \MakeLowercase{\emph{et al.}}: bababababa}
\maketitle
\begin{abstract}
A hybrid transmitting and reflecting beyond diagonal reconfigurable
intelligent surface (BD-RIS) design is proposed. Operating in the
same aperture, frequency band and polarization, the proposed BD-RIS
features independent beam steering control of its reflected and transmitted
waves. In addition it provides a hybrid mode with both reflected and
transmitted waves using tunable power splitting between beams. The
BD-RIS comprises two phase reconfigurable antenna arrays interconnected
by an array of tunable two-port power splitters. The two-port power
splitter in each BD-RIS cell is built upon a varactor in parallel
with a bias inductor to exert tunable impedance variations on transmission
lines. Provided with variable reverse DC voltages, the two-port power
splitter can control the power ratio of $\mathbf{S_{11}}$ over $\mathbf{S_{21}}$
from $\mathbf{-}$20 dB to 20 dB, thus allowing tunable power splitting.
Each antenna is 2-bit phase reconfigurable with 200 MHz bandwidth
at 2.4 GHz so that each cell of BD-RIS can also achieve independent
reflection and transmission phase control. To characterize and optimize
the electromagnetic response of the proposed BD-RIS design, a Thévenin
equivalent model and corresponding analytical method is provided.
A BD-RIS with 4$\mathbf{\times}$4 cells was also prototyped and tested.
Experiments show that in reflection and transmission mode, the fabricated
BD-RIS can realize beam steering in reflection and transmission space,
respectively. It is also verified that when operating in hybrid mode,
the BD-RIS enables independent beam steering of the reflected and
transmitted waves. This work helps fill the gap between realizing
practical hardware design and establishing an accurate physical model
for the hybrid transmitting and reflecting BD-RIS, enabling hybrid
transmitting and reflecting BD-RIS assisted wireless communications.
\end{abstract}


\section{Introduction}

\begin{singlespace}
\IEEEPARstart{T}{he} forthcoming sixth-generation (6G) mobile communication
system is envisioned to provide extremely high data rates, high spectral-
and energy-efficiency, extremely low latency, and ubiquitous network
coverage \cite{saad2019vision,wang2023road}. Reconfigurable intelligent
surfaces (RISs), are a promising technology for 6G, and have attracted
significant attention. RISs consist of a large number of reconfigurable
scattering elements and the electromagnetic response of each element
can be smartly adjusted to manipulate the wireless environment \cite{DiRenzo2020,wu2019towards,wu2021intelligent}.

On top of the conventional RISs, which are characterized by a diagonal
scattering matrix, the novel concept of beyond diagonal RIS (BD-RIS)
has been recently proposed \cite{li2022beyond,li2023reconfigurable}.
BD-RISs represent a family of RISs, where the adjacent elements can
be interconnected through tunable impedance components, and can be
characterized by a beyond diagonal scattering matrix, which provides
more degrees-of-freedom to manipulate the wireless environment and
thus brings an enhanced gain compared to conventional RISs \cite{li2022beyond,li2023reconfigurable,zhou2023optimizing,li2024beyond}.
\end{singlespace}

One of the important BD-RIS modes is when it is configured to simultaneously
transmit and reflect waves. When configured in this mode the BD-RIS
is denoted here as a hybrid transmitting and reflecting BD-RIS and
is also known in other literature as a simultaneously transmitting
and reflecting RIS (STAR-RIS) \cite{xu2021star} or intelligent omni-surface
(IOS) \cite{Zhang2020}. Compared with conventional RISs which only
support reflection from one side of the surface and provide half-space
coverage only, the hybrid transmitting and reflecting BD-RIS supports
simultaneous reflection and transmission from both sides of the surface
and enables full-space coverage \cite{xu2021star,mu2021simultaneously,Zhang2020,zhang2022intelligent,li2022beyond}.
The vast majority of the research performed on hybrid transmitting
and reflecting BD-RIS has been devoted to system optimization \cite{ahmed2023survey},
such as maximizing sum rate \cite{niu2021weighted,zhou2023optimizing},
and enhancing spectral and energy efficiency \cite{zhou2023optimizing,katwe2023improved}.
However, the limitation of these works \cite{xu2021star,mu2021simultaneously,Zhang2020,zhang2022intelligent,li2022beyond,katwe2023improved,zhou2023optimizing,niu2021weighted}
is that only numerical simulation is considered and there is no prototype
design to experimentally verify the hybrid transmitting and reflecting
BD-RIS.

Previously the implementation of bidirectional beam steering surfaces
has been provided using metasurfaces \cite{wang2021reconfigurable,bao2021programmable,hu2022intelligent,yin2024reconfigurable,duan2024non,yin2024highly,bao2021full,wang2018simultaneous,xu2016multifunctional,cai2017high,hou2020helicity,wu2022circular,zhang2018transmission,zhang2019multifunction,guo2019transmission,li2022transmission},
antennas \cite{lau2010planar,wang20191,yu2022reconfigurable,xiang2023simultaneous,zhao2024design,cao20231,li2021dual,yang2021multifunctional,liu2020wideband,feng2022reflect},
and frequency selective surfaces \cite{li2020fss,zhong2019fss}. To
achieve independent transmitting and reflecting beam steering using
a surface is however challenging. One approach has been to control
the surface's electromagnetic properties, including phases or amplitudes,
by using different frequency bands for the transmitted and reflected
waves \cite{zhong2019fss,xu2016multifunctional,bao2021full,wang2018simultaneous,li2021dual,yang2021multifunctional}.
However, this approach is not suitable for wireless communication
as spectrum efficiency is reduced due to the need for two frequency
bands. An alternative approach to achieve independent beam control
is to utilize polarization multiplexing to respectively control the
transmitting and reflecting beams at two orthogonal polarizations
\cite{bao2021programmable,cai2017high,hou2020helicity,wu2022circular,liu2020wideband,zhang2018transmission,feng2022reflect}.
However this is not conducive for extensions to dual-polarization
designs and has the issue of polarization conversion \cite{wang20191,li2022transmission,yu2022reconfigurable,xiang2023simultaneous}.
In addition to frequency and polarization multiplexing, exploiting
spatial multiplexing such as different propagation directions \cite{zhang2019multifunction,guo2019transmission,zhang2018transmission},
can be used to respectively control the reflected and transmitted
beams. However, using spatial multiplexing comes at the cost of structural
symmetry. That is when excited by incident waves with the same electromagnetic
properties from opposing directions, the designed surface may not
maintain the same functionality or operate normally as it does when
approached from the original direction. More importantly, many of
the abovementioned designs \cite{duan2024non,bao2021full,wang2018simultaneous,xu2016multifunctional,cai2017high,hou2020helicity,wu2022circular,zhang2018transmission,guo2019transmission,li2022transmission,li2021dual,yang2021multifunctional,liu2020wideband,feng2022reflect},
have their corresponding functions and properties entirely fixed once
fabricated, and cannot be changed or reconfigured, greatly limiting
their application in real-time wireless communication systems which
require adaptation.

In addition to the aforementioned limitations of frequency, polarization,
spatial multiplexing, and reconfigurability, there is also the issue
that independent and flexible mode switching between reflection, transmission,
and hybrid modes has not been fully implemented. Although some designs
have allowed for dynamic switching between reflection and transmission
modes \cite{wang2021reconfigurable,hu2022intelligent,yin2024highly,cao20231,yu2022reconfigurable}
and achieved bidirectional functionality by using aperture multiplexing
techniques \cite{hu2022intelligent,yu2022reconfigurable,yin2024highly},
the issue is that a single unit cannot directly realize mode switching
and arbitrary power splitting. Therefore, it remains a challenge to
design hybrid transmitting and reflecting BD-RIS to implement independent
beam control and mode switching with arbitrary power splitting between
beams.

\begin{figure}[t]
\begin{centering}
\includegraphics[width=1\columnwidth]{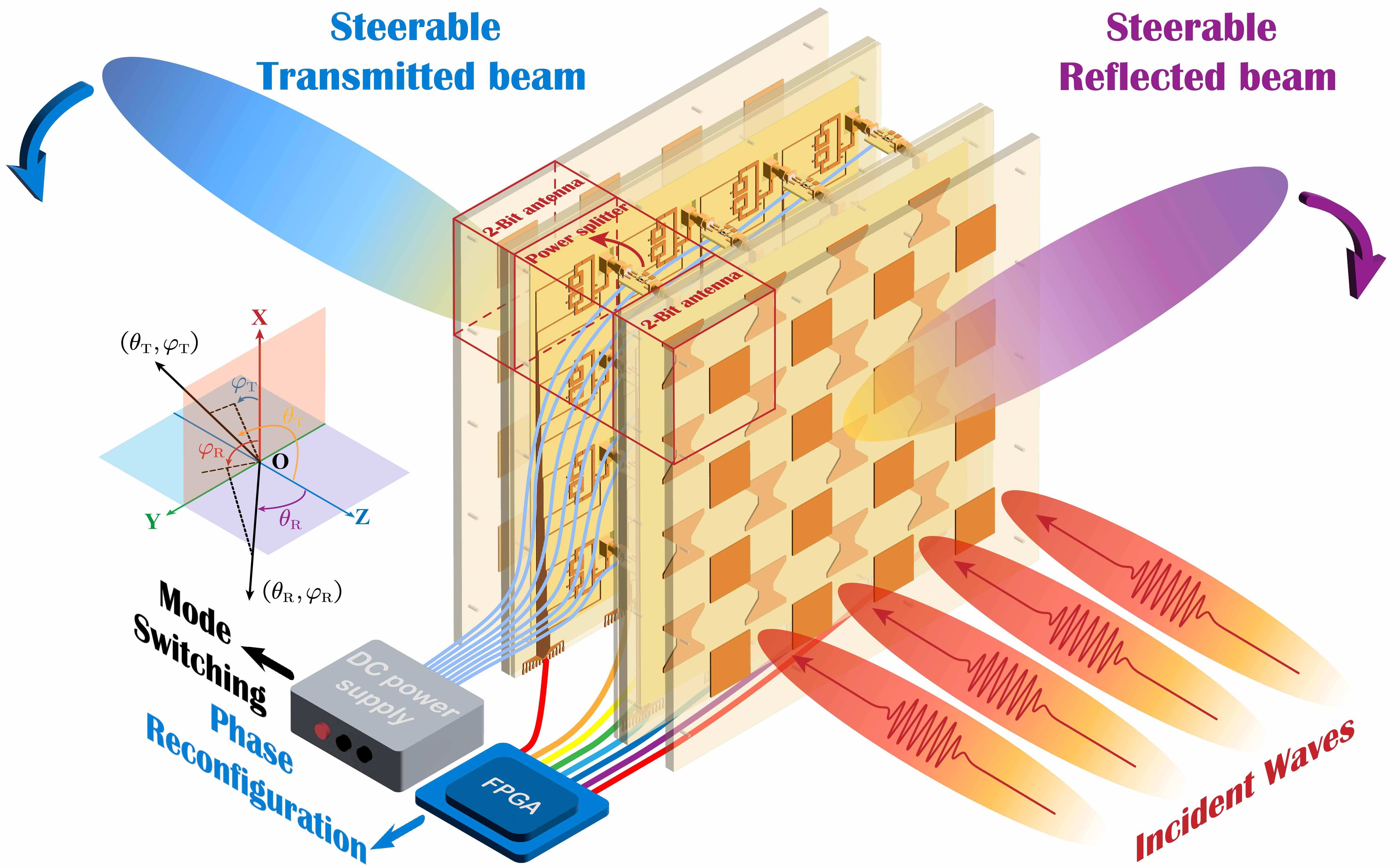}
\par\end{centering}
\centering{}\caption{\label{4=0000D74_BD-RIS} Illustration of the proposed hybrid transmitting
and reflecting BD-RIS with 4$\times$4 cells. In this example waves
incident from the right hand side can be all reflected, all transmitted
or simultaneously transmitted and reflected in a hybrid mode.}
\end{figure}

To overcome the aforementioned challenges, in this work we propose
a highly reconfigurable hybrid transmitting and reflecting BD-RIS
with independent beam control and power splitting with mode switching
in the same frequency band, polarization, and aperture with a symmetric
structure. An overview of the proposed BD-RIS with 4$\times$4 cells
is shown in Fig. \ref{4=0000D74_BD-RIS}, where two 4$\times$4 phase
reconfigurable antenna arrays are interconnected via 16 tunable two-port
power splitters. The phase states of all antennas and the operation
mode of the power splitters are controlled by field-programmable gate
arrays (FPGAs) and a DC power supply, respectively. By setting the
operation mode of the power splitters and configuring all the antennas
with the desired phases, we can realize dynamic mode switching with
independent beam steering of the reflected and transmitted beams.
This work fill the gap between realizing practical hardware design
and establishing an accurate physical model for the hybrid transmitting
and reflecting BD-RIS, helping enable hybrid transmitting and reflecting
BD-RIS assisted wireless communications. The contributions of this
work are summarized as follows.

\textit{1) Hardware Design Requirements: }We propose and justify the
hardware design requirements for each BD-RIS cell. Our proposed requirement
is that each cell should achieve at least eight reconfigurable states,
including two states for reflection mode, two for transmission mode
and at least four states for hybrid mode to enable independent beam
control and mode switching.

\textit{2) Hardware Design Methodology and Architecture:} Based on
the proposed design requirements, we propose a multiple-module-integrated
architecture for BD-RIS. This architecture is formed by interconnecting
two 2-bit phase reconfigurable antennas with a tunable two-port power
splitter, thus leading to a symmetrical structure and maintaining
the same polarization for all reflected and transmitted waves.

\textit{3) Physical Model, Analytical Method, and Optimization:} We
establish the corresponding Thévenin equivalent model for the proposed
architecture and derive an analytical method to efficiently calculate
the reflected and transmitted waves of the proposed BD-RIS under plane
wave excitation. The genetic algorithm (GA) is used to produce beamforming
designs.

\textit{4) Prototype and Experimental Verification:} We fabricate
a hybrid transmitting and reflecting BD-RIS prototype with 4$\times$4
cells and build up a testbed to measure the reflected and transmitted
scattered patterns to verify its multifunctionalities.

This paper is organized as follows. Section II introduces the model
for the hybrid transmitting and reflecting BD-RIS as proposed for
wireless communications. Section III elaborates the design specification
and the cell design methodology. Section IV illustrates the design
and implementation of multiple modules used in each cell of BD-RIS,
including a two-port power splitter and a 2-bit phase reconfigurable
antenna. Section V builds up the physical model and describes the
analytical and beamforming optimization methods for the proposed BD-RIS.
Section VI provides experimental verification using a prototype of
the proposed BD-RIS with 4$\times$4 cells, and discussions. Finally,
Section VII concludes this work.

\section{BD-RIS Background}

\begin{figure}[t]
\begin{centering}
\includegraphics[width=0.9\columnwidth]{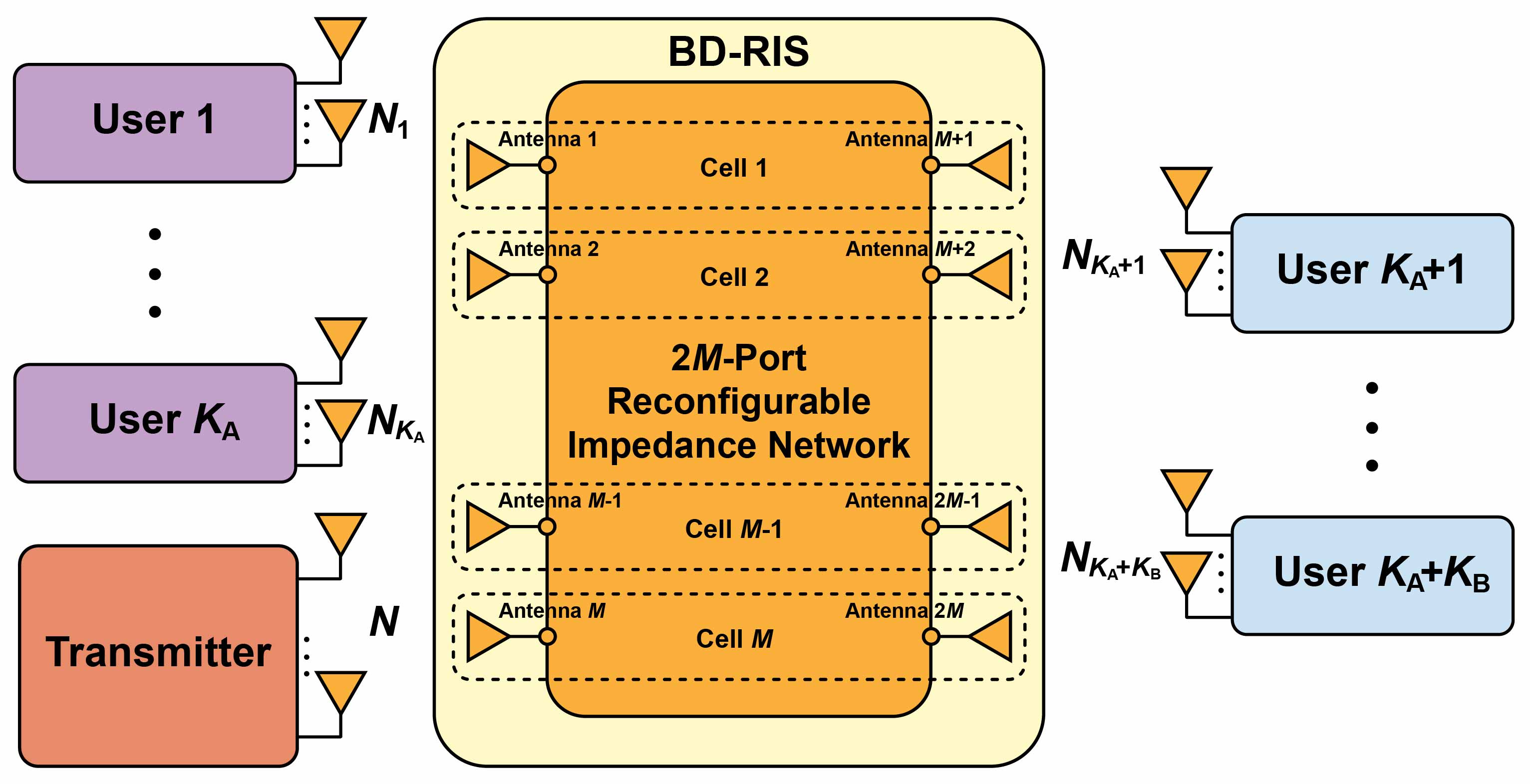}
\par\end{centering}
\centering{}\caption{\label{BD-RIS_communication_system}System diagram of an $M$-cell
BD-RIS-aided communication system.}
\end{figure}

We consider a hybrid transmitting and reflecting BD-RIS which consists
of $M$ cells as shown in Fig. \ref{BD-RIS_communication_system}.
The $m$th cell, for $m\in\mathcal{M}=\left\{ 1,\ldots,M\right\} $,
is constructed by antenna $m$ and antenna $m+M$ connected to a two-port
reconfigurable network, whose scattering matrix is denoted as
\begin{equation}
\mathbf{S}_{m}=\left[\begin{array}{cl}
\phi_{\mathrm{r,}m} & \phi_{\mathrm{t,}m}\\
\phi_{\mathrm{t,}m} & \phi_{\mathrm{r,}m+M}
\end{array}\right],\label{scattering_matrix_of_each_cell}
\end{equation}
where the subscript $\mathrm{r}$ refers to the reflection scattering
parameter and the subscript $\mathrm{t}$ refers to the transmission
scattering parameter between the $m$ and $m+M$ ports (and also by
symmetry the $m+M$ and $m$ ports). When the two-port reconfigurable
network is lossless, $\phi_{\mathrm{r},m}$ and $\phi_{\mathrm{t},m}$
should satisfy
\begin{equation}
\left|\phi_{\mathrm{r},m}\right|^{2}+\left|\phi_{\mathrm{t},m}\right|^{2}=1,\forall\,m\in\mathcal{M},\label{lossless_constraints_for_STAR-RIS-1}
\end{equation}
and this arises from energy conservation for reflection and transmission. 

The entire $M$-cell hybrid transmitting and reflecting BD-RIS contains
2$M$ antennas and a 2$M$-port reconfigurable network. Using the
notation in \eqref{scattering_matrix_of_each_cell}, the scattering
matrix $\mathbf{\Phi}\in\mathbb{C}^{2M\times2M}$ of the entire $2M$-port
reconfigurable network can be written as,
\begin{equation}
\left[\mathbf{\Phi}\right]_{i,j}=\begin{cases}
\:\phi_{\mathrm{r,}i} & ,\textrm{for }i=j,\\
\:\phi_{\mathrm{t,}i} & ,\textrm{for }j-i=M,\\
\:\phi_{\mathrm{t,}j} & ,\textrm{for }i-j=M,\\
\:0 & \textrm{, otherwise}.
\end{cases}\label{entry_of_scattering_matrix-1}
\end{equation}
This scattering matrix has both diagonal and off-diagonal elements,
and that is why it is referred to as a BD-RIS \cite{li2022beyond}.
Using \eqref{entry_of_scattering_matrix-1} the complete channel model
for the hybrid transmitting and reflecting BD-RIS system can also
be developed as described in \cite{li2022beyond}.

\begin{figure}[t]
\begin{centering}
\includegraphics[width=0.9\columnwidth]{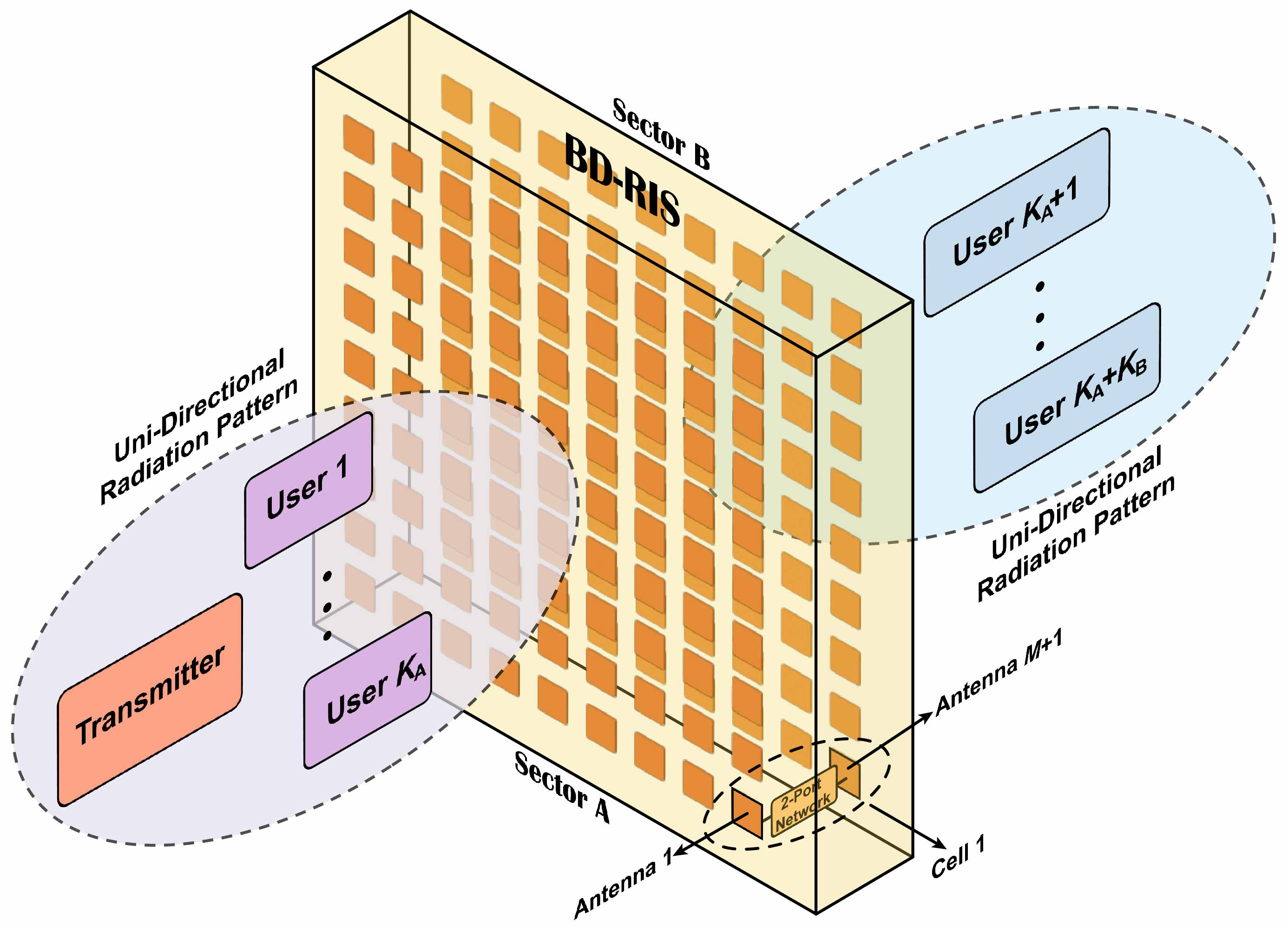}
\par\end{centering}
\centering{}\caption{\label{BD-RIS_communication_system-top_view} An $M$-cell BD-RIS
with 2$M$ back to back uni-directional antennas partitioning the
whole space into reflection (sector A) and transmission (sector B)
half spaces. }
\end{figure}

In the hybrid transmitting and reflecting BD-RIS, each antenna has
a uni-directional radiation pattern and the two antennas in each cell
are back to back, so that each antenna covers a half space (which
we refer to as a sector in the remainder of the paper) as shown in
Fig. \ref{BD-RIS_communication_system-top_view}. We denote the two
sectors as, sector A, and it contains antennas 1 to $M$ and sector
B which contains antennas $M+1$ to 2$M$. We assume that the transmitter
and $K_{\mathrm{A}}$ users, indexed by $\mathcal{K}_{\textrm{r}}=\left\{ 1,\ldots,K_{\mathrm{A}}\right\} $,
$0<K_{\mathrm{A}}<K$, are located in sector A (and therefore where
the BD-RIS reflected waves are present), and the remaining $K_{\mathrm{B}}=K-K_{\mathrm{A}}$
users, indexed by $\mathcal{K}_{\textrm{t}}=\left\{ K_{\mathrm{A}}+1,\ldots,K\right\} $,
are located in sector B (and where the BD-RIS transmitted waves are
present. Since our proposed BD-RIS structure is completely symmetric,
the reverse configuration (transmitter in sector B) is also automatically
satisfied and therefore we do not consider the return loss $\phi_{\mathrm{r,}i}$
for $i>M$ further.

Based on the relation \eqref{lossless_constraints_for_STAR-RIS-1},
we can adjust the phase $\theta_{\mathrm{r},m}=\arg\left(\phi_{\mathrm{r},m}\right)$
and $\theta_{\mathrm{t},m}=\arg\left(\phi_{\mathrm{t},m}\right)$,
$\forall\,m\in\mathcal{M}$ to control the beams for reflection and
transmission, respectively. On the other hand, we can adjust the magnitude
$\left|\phi_{\mathrm{r},m}\right|$ and $\left|\phi_{\mathrm{t},m}\right|$,
$\forall\,m\in\mathcal{M}$ to control the power allocation for reflection
and transmission under the constraint \eqref{lossless_constraints_for_STAR-RIS-1}.

\section{BD-RIS Cell Design Methodology}

In this section, we propose and specify the hardware design requirements
for each cell of the hybrid transmitting and reflecting BD-RIS and
propose a multiple-module-integrated architecture based on scattering
matrix analysis.

\subsection{Design Specification}

To obtain the optimal performance, an ideal hybrid transmitting and
reflecting BD-RIS should be able to continuously adjust the phase
and magnitude of $\phi_{\mathrm{r,\mathit{m}}}$ and $\phi_{\mathrm{t,\mathit{m}}}$,
$\forall\,m\in\mathcal{M}$. This is very challenging in practice
and as a compromise we discretize the magnitude and phase of $\phi_{\mathrm{r},m}$
and $\phi_{\mathrm{t},m}$ and use RF switches to implement the hybrid
transmitting and reflecting BD-RIS. Furthermore, to reduce circuit
complexity and insertion loss resulting from RF switches in practice,
we also restrict the configurations to the following three modes:

\textit{1) Reflection Mode}: Each cell purely reflects an incident
electromagnetic wave with at least 1-bit phase reconfigurability,
i.e. $\left|\phi_{\mathrm{r,\mathit{m}}}\right|\approx1$, $\left|\phi_{\mathrm{t,\mathit{m}}}\right|\approx0$,
$\theta_{\mathrm{r},m}\in\left\{ 0,\pi\right\} $, $\forall\,m\in\mathcal{M}$.

\textit{2) Transmission Mode}: Each cell purely transmits an incident
electromagnetic wave with at least 1-bit phase reconfigurability,
i.e. $\left|\phi_{\mathrm{r,\mathit{m}}}\right|\approx0$, $\left|\phi_{\mathrm{t,\mathit{m}}}\right|\approx1$,
$\theta_{\mathrm{t},m}\in\left\{ 0,\pi\right\} $, $\forall\,m\in\mathcal{M}$.

\textit{3) Hybrid Mode}: Each cell can both reflect and transmit an
incident electromagnetic wave with equal power. Therefore, half of
the incident electromagnetic wave is reflected and the other half
is transmitted both with at least 1-bit independent phase reconfigurability,
i.e. $\left|\phi_{\mathrm{r,\mathit{m}}}\right|\approx\mathrm{\frac{\mathrm{1}}{\sqrt{2}}}$,
$\left|\phi_{\mathrm{t,\mathit{m}}}\right|\approx\frac{1}{\sqrt{2}}$,
$\theta_{\mathrm{r},m}\in\left\{ 0,\pi\right\} $, $\theta_{\mathrm{t},m}\in\left\{ 0,\pi\right\} $,
$\forall\,m\in\mathcal{M}$.

To achieve these three modes and enable independent beam control,
each cell should achieve at least eight reconfigurable states. This
includes two states for the reflection mode, two for the transmission
mode and at least four states for the hybrid mode. 

\subsection{Cell Design}

\begin{figure}[t]
\begin{centering}
\textsf{\includegraphics[width=1\columnwidth]{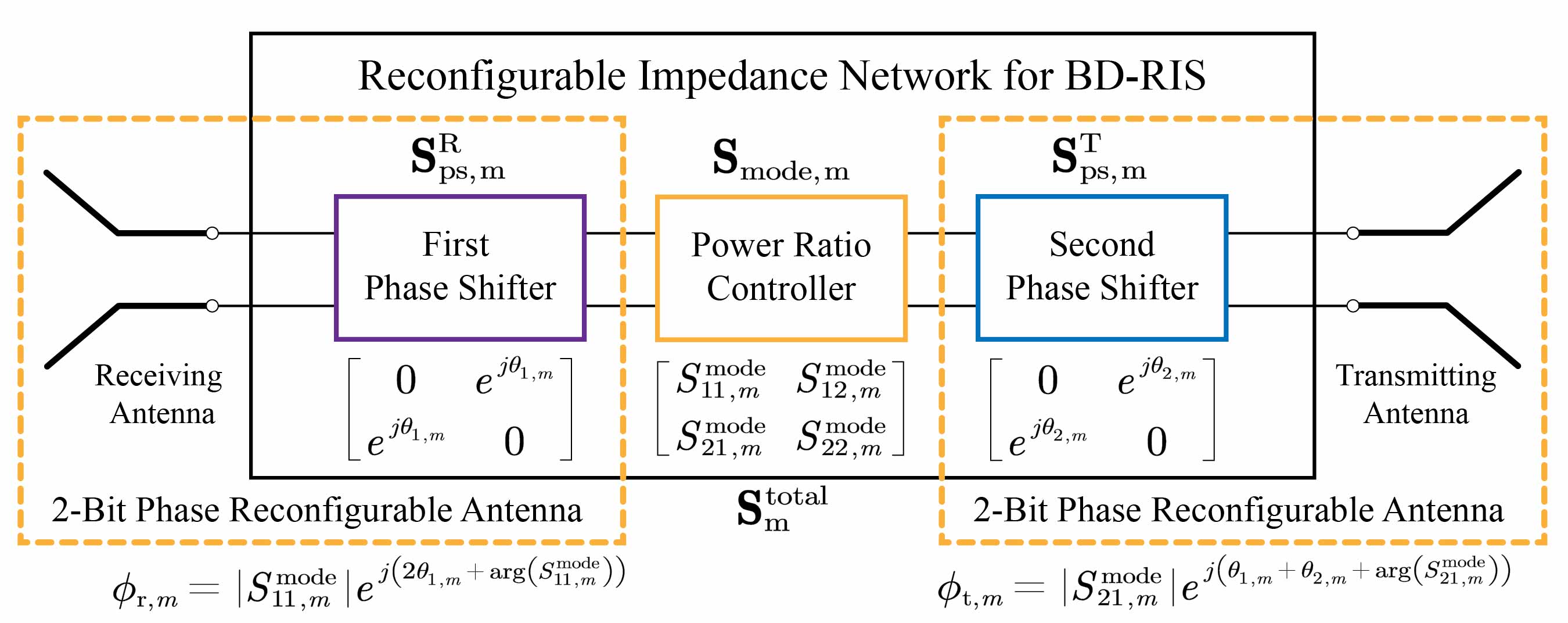}}
\par\end{centering}
\begin{centering}
(a)
\par\end{centering}
\begin{centering}
\textsf{\includegraphics[width=1\columnwidth]{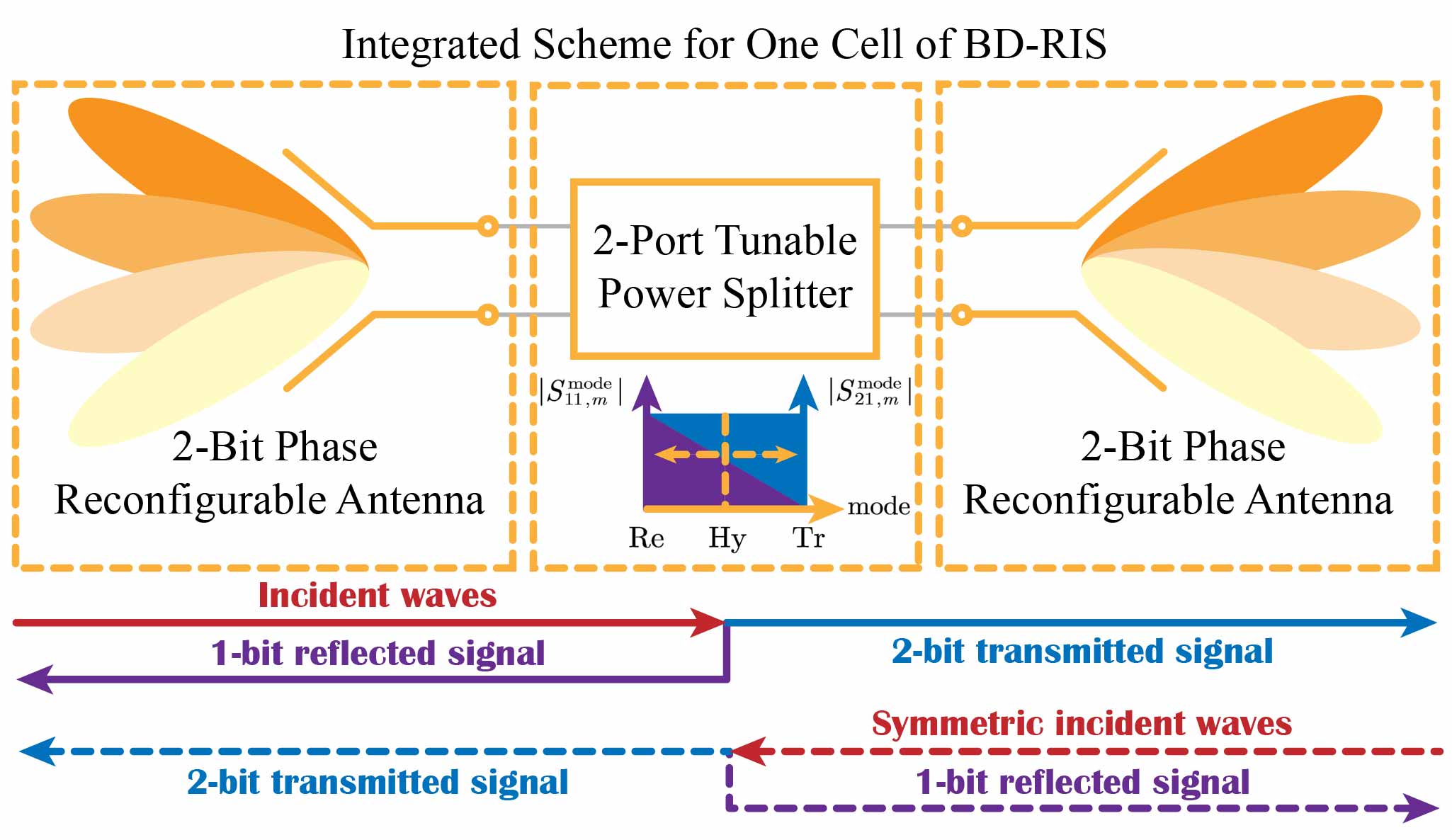}}
\par\end{centering}
\begin{centering}
(b)
\par\end{centering}
\caption{\label{proposed_scheme}The proposed architecture for a single cell
of the hybrid transmitting and reflecting BD-RIS. (a) A single cell
of BD-RIS containing reconfigurable impedance network. (b) Integrated
scheme for a single cell of BD-RIS. Re: Reflection, Hy: Hybrid, Tr:
Transmission.}
\end{figure}

To realize the design specification, we propose a two-port reconfigurable
impedance network for each cell of the hybrid transmitting and reflecting
BD-RIS as shown in Fig. \ref{proposed_scheme}(a). Each two-port reconfigurable
impedance network consists of three cascaded modules, including two
phase shifters responsible for phase reconfigurability of the reflected
and transmitted waves, and one power splitter responsible for dynamic
mode switching. For the $m$th cell, each of these modules is equivalent
to a two-port network and the scattering matrices for the first and
second phase shifter and the power splitter can be expressed respectively
as
\begin{equation}
\mathbf{S}_{\mathrm{ps,}m}^{\mathrm{R}}=\left[\begin{array}{cc}
0 & e^{j\theta_{1,m}}\\
e^{j\theta_{1,m}} & 0
\end{array}\right],\label{reflection phase shifter}
\end{equation}
\begin{equation}
\mathbf{S}_{\mathrm{ps,}m}^{\mathrm{T}}=\left[\begin{array}{cc}
0 & e^{j\theta_{2,m}}\\
e^{j\theta_{2,m}} & 0
\end{array}\right],\label{transmission phase shifter}
\end{equation}
\begin{equation}
\mathbf{S}_{\mathrm{mode,}m}=\left[\begin{array}{cc}
S_{11,m}^{\mathrm{mode}} & S_{12,m}^{\mathrm{mode}}\\
S_{21,m}^{\mathrm{mode}} & S_{22,m}^{\mathrm{mode}}
\end{array}\right],\label{power ratio controller}
\end{equation}
where $\theta_{1,m}$ and $\theta_{2,m}$ are the phase shifts introduced
by the two phase shifters and $S_{ij,m}^{\mathrm{mode}}$, $\forall\,i,j\in\left\{ 1,2\right\} $
are the S-parameters of the power splitter in the $m$th cell, $\forall\,m\in\mathcal{M}$.
The total scattering matrix $\mathbf{S}_{m}^{\mathrm{total}}$ of
the $m$th reconfigurable impedance network can be found as
\begin{equation}
\mathbf{S}_{m}^{\mathrm{total}}=\left[\begin{array}{cc}
S_{11,m}^{\mathrm{mode}}e^{j2\theta_{1,m}} & S_{12,m}^{\mathrm{mode}}e^{j\left(\theta_{1,m}+\theta_{2,m}\right)}\\
S_{21,m}^{\mathrm{mode}}e^{j\left(\theta_{1,m}+\theta_{2,m}\right)} & S_{22,m}^{\mathrm{mode}}e^{j2\theta_{2,\mathit{m}}}
\end{array}\right].\label{overall_scattering_matrix}
\end{equation}

Denoting $S_{11,m}^{\mathrm{mode}}=\left|S_{11,m}^{\mathrm{mode}}\right|e^{jc_{1,m}}$
and $S_{21,m}^{\mathrm{mode}}=\left|S_{21,m}^{\mathrm{mode}}\right|e^{jc_{2,m}}$,
where $c_{\mathrm{1},m}=\arg\left(S_{\mathrm{11},m}^{\mathrm{mode}}\right)$
and $c_{\mathrm{2},m}=\arg\left(S_{\mathrm{21},m}^{\mathrm{mode}}\right)$,
and comparing \eqref{overall_scattering_matrix} with \eqref{scattering_matrix_of_each_cell},
we have that
\begin{equation}
\phi_{\mathrm{r},m}=S_{11,m}^{\mathrm{total}}=\left|S_{11,m}^{\mathrm{mode}}\right|e^{j\left(2\theta_{1,m}+c_{1,m}\right),}\label{reflected signals}
\end{equation}
\begin{equation}
\phi_{\mathrm{t},m}=S_{21,m}^{\mathrm{total}}=\left|S_{21,m}^{\mathrm{mode}}\right|e^{j\left(\theta_{1,m}+\theta_{2,m}+c_{2,m}\right).}\label{transmitted signals}
\end{equation}

Therefore, for the $m$th cell of the hybrid transmitting and reflecting
BD-RIS, the phase shifts of the reflected and transmitted signals,
and the power ratio of reflected signals over the transmitted waves
can be expressed respectively as
\begin{equation}
\theta_{\mathrm{r},m}=2\theta_{1,m}+c_{1,m},\label{reflected_phase_shift}
\end{equation}
\begin{equation}
\theta_{\mathrm{t},m}=\theta_{1,m}+\theta_{2,m}+c_{2,m},\label{transmitted_phase_shift}
\end{equation}
\begin{equation}
\left|\frac{\phi_{\mathrm{r},m}}{\phi_{\mathrm{t},m}}\right|^{2}=\left|\frac{S_{11,m}^{\mathrm{mode}}}{S_{21,m}^{\mathrm{mode}}}\right|^{2}.\label{power_ratio}
\end{equation}

As specified in Section III.A and based on the reflection and transmission
phase relationship \eqref{reflected_phase_shift} and \eqref{transmitted_phase_shift},
to realize at least 1-bit independent phase reconfigurability between
$\theta_{\mathrm{r},m}$ and $\theta_{\mathrm{r},m}$, and for structural
symmetry consideration, the phase tunability of these two phase shifters
should be at least 2-bit, that is, $\theta_{1,m},\theta_{2,m}\in\left\{ 0,\frac{\pi}{2},\pi,\frac{3\pi}{2}\right\} $.
In this case, the hybrid transmitting and reflecting BD-RIS can always
achieve 1-bit independent control for the reflected waves and 2-bit
independent control for the transmitted waves, regardless of the direction
from which the incident wave arrives. Moreover, to realize dynamic
mode switching, we need to reconfigure the two-port power splitter
so that the power ratio given by \eqref{power_ratio} can be tuned
to 0, 1, and to $\gg$1, which corresponds to the three modes specified
in Section III.A.

In this way, $\mathbf{S}_{\mathrm{mode},m}$ is fixed for a specific
mode accordingly, and as expressed from \eqref{reflected_phase_shift}
to \eqref{power_ratio}, the phase reconfigurability of $\theta_{\mathrm{r},m}$
and $\theta_{\mathrm{t},m}$ is only related to those two phase shifters
via $\theta_{1,m}$ and $\theta_{2,m}$. Furthermore, the phase reconfiguration
will not affect the power ratio. It should also be noted that by cascading
this two-port network with the same polarization port of the reflecting
and transmitting antenna, our design also avoids introducing polarization
conversion.

Also note that implementing two extra phase shifters with 2-bit phase
reconfigurability will make the reconfigurable impedance network too
bulky to be integrated with two antennas, and cause extra amplitude
variation for different phase states, thereby affecting the power
ratio controlling. Therefore, for the consideration of reducing energy
loss and making the design compact, we incorporate those 2-bit phase
shifters into the antennas. Namely to design two 2-bit phase reconfigurable
antennas for each cell. Our proposed scheme for one cell of the hybrid
transmitting and reflecting BD-RIS can then be depicted by Fig. \ref{proposed_scheme}(b),
which is a symmetrical and integrated scheme involving no polarization
conversion.

\section{BD-RIS Cell Implementation}

In this section, we provide details of the BD-RIS cell implementation
that meets the design criteria developed in Section III. Prototypes
and measurements are also provided.

\subsection{Overall Cell Architecture}

\begin{figure}[t]
\begin{centering}
\includegraphics[width=0.9\columnwidth]{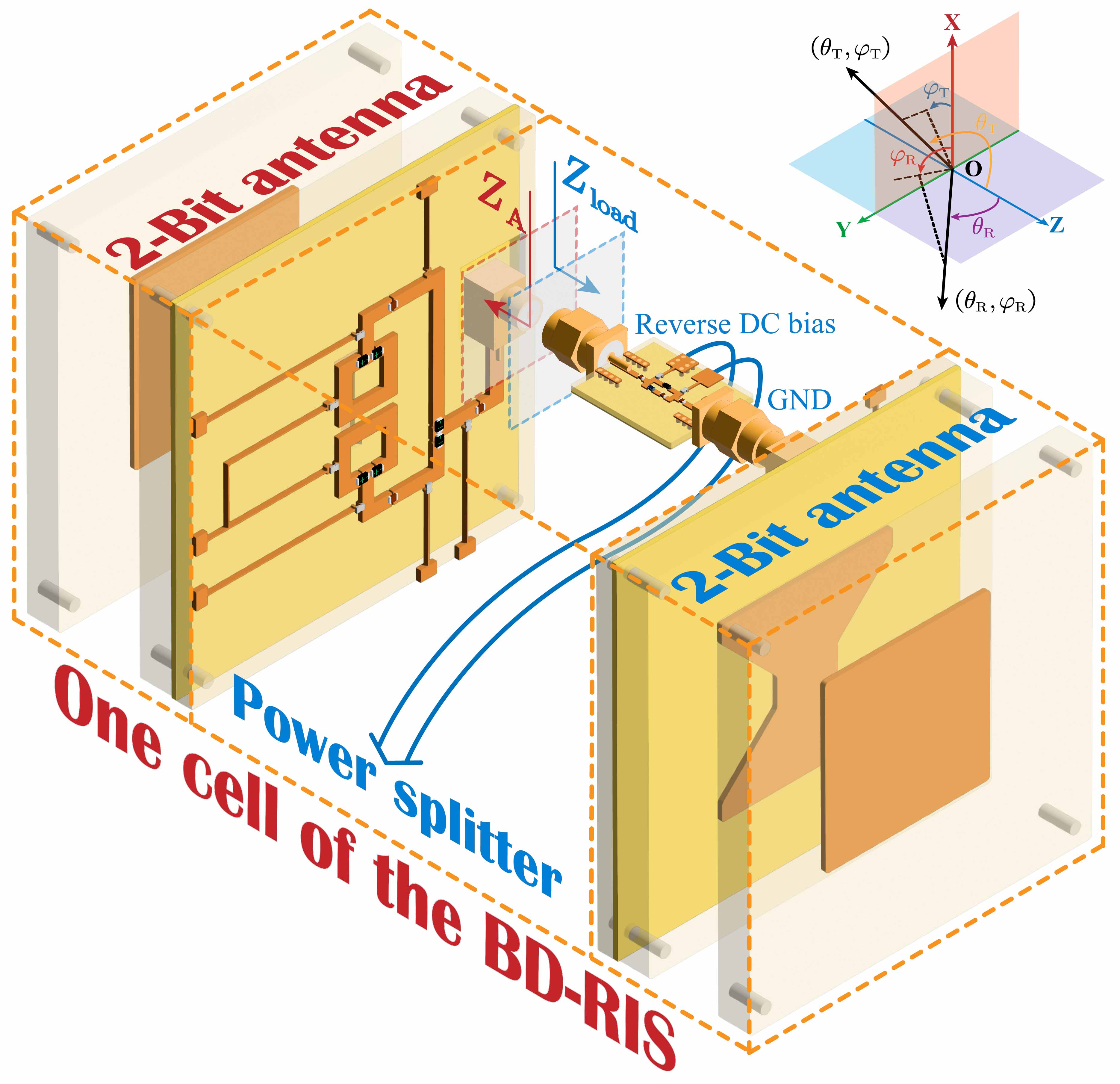}
\par\end{centering}
\centering{}\caption{\label{one cell of the hybrid transmitting and reflecting BD-RIS}
One cell of the hybrid transmitting and reflecting BD-RIS including
two 2-bit phase reconfigurable antennas and a two-port tunable power
splitter.}
\end{figure}

The specific structure based on the proposed scheme in Fig. \ref{proposed_scheme}(b)
is shown in Fig. \ref{one cell of the hybrid transmitting and reflecting BD-RIS}
for one cell of the hybrid transmitting and reflecting BD-RIS. As
illustrated, each cell is composed of three cascaded modules, including
one 2-bit phase reconfigurable antenna serving as the reflecting antenna,
a tunable two-port power splitter and another 2-bit phase reconfigurable
antenna serving as the transmitting antenna. As mentioned, implementing
the same 2-bit antennas on both sides makes the architecture symmetrical.
In addition, two SMA connectors with 90$^{\circ}$ right-angle bends
are used to achieve the vertical transitions from the microstrip line
circuit of the antenna to that of the power splitter at two sides.
Details of how we design these modules are elaborated in the following
subsections.

\subsection{Tunable Two-port Power Splitter}

\begin{figure}[t]
\begin{centering}
\textsf{\includegraphics[width=1\columnwidth]{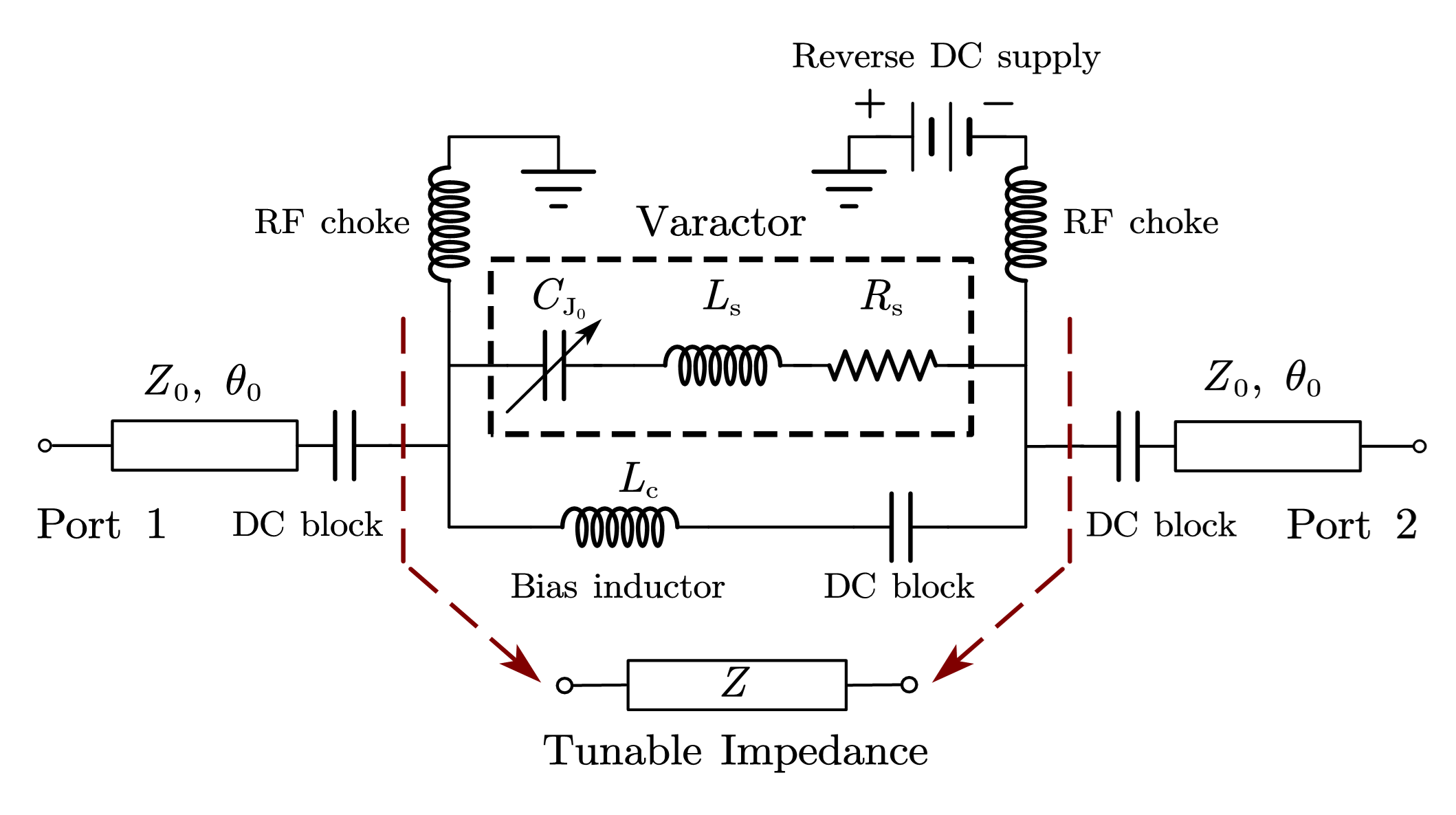}}
\par\end{centering}
\begin{raggedright}
\hspace*{0.5\columnwidth} (a)
\par\end{raggedright}
\begin{centering}
\textsf{\includegraphics[width=0.9\columnwidth]{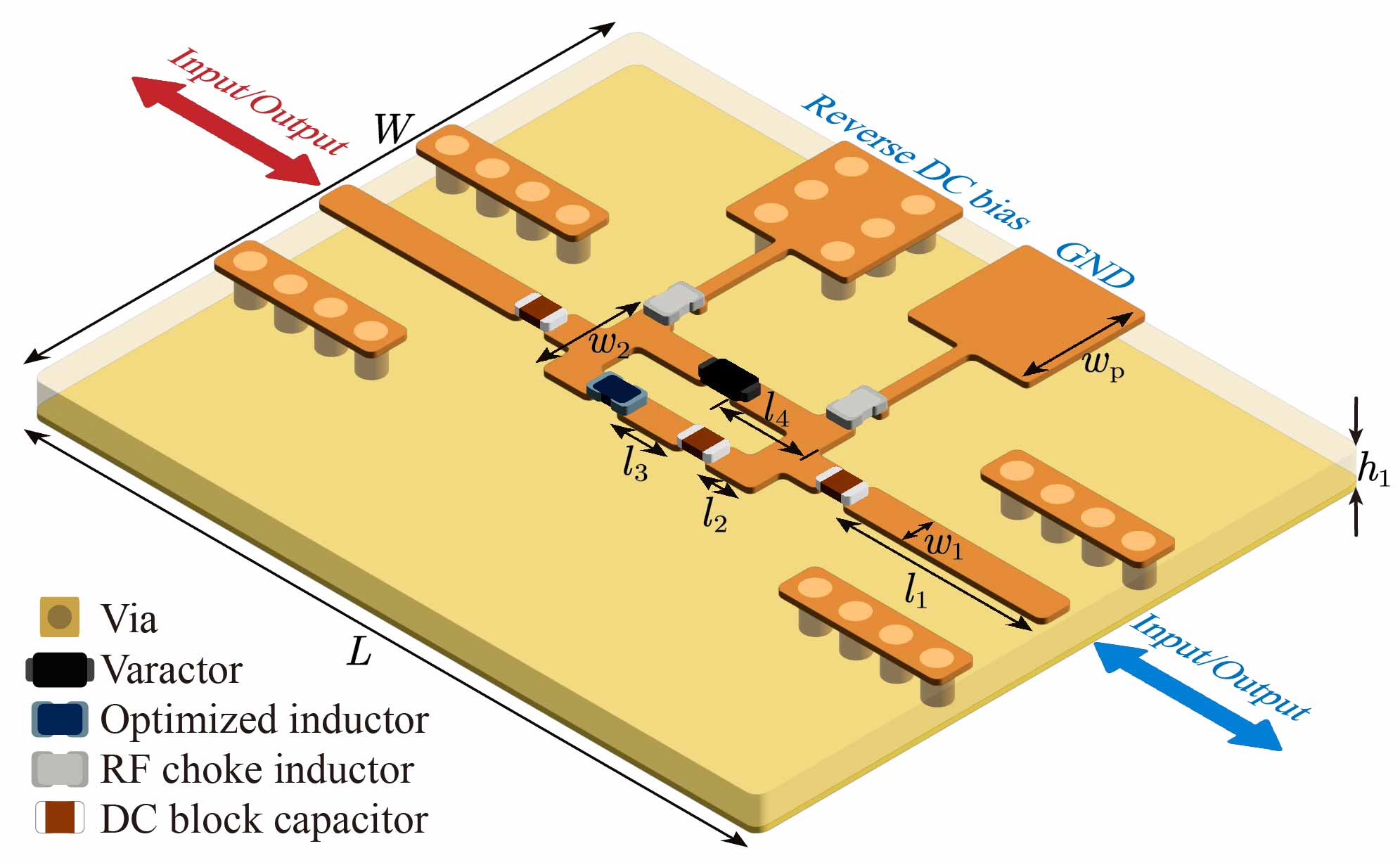}}
\par\end{centering}
\begin{raggedright}
\hspace*{0.5\columnwidth} (b)
\par\end{raggedright}
\begin{centering}
\textsf{\includegraphics[width=0.95\columnwidth]{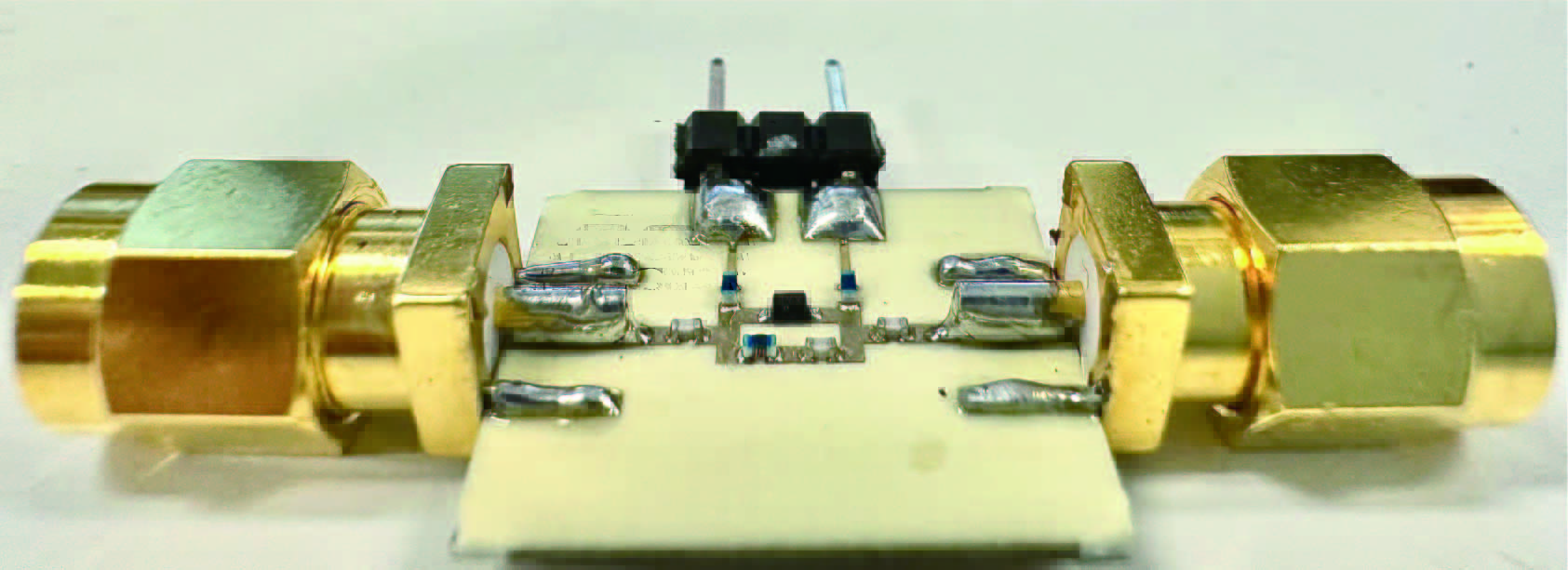}}
\par\end{centering}
\begin{raggedright}
\hspace*{0.5\columnwidth} (c)
\par\end{raggedright}
\caption{Tunable two-port power splitter design for a BD-RIS cell. (a) Schematic
of the two-port tunable power splitter. (b) Overall structure. The
dimensions are $W=13.5$, $L=18$, $h_{1}=0.813$, $l_{1}=5$, $l_{2}=1.2$,
$l_{3}=1.6$, $l_{4}=2.3$, $w_{1}=0.7$, $w_{2}=2.4$, $w_{\mathrm{p}}=3$
(unit: mm). Light yellow color: dielectric layer. Dark orange color:
copper. Yellow color: ground. (c) Photograph of the fabricated two-port
power splitter.}
\label{power_splitter_design}
\end{figure}

As shown in Fig. \ref{proposed_scheme}(b), our objective is to design
a tunable two-port power splitter to realize mode switching. Based
on the idea of utilizing reconfigurable components to exert impedance
variations on the transmission lines, we propose a straightforward
two-port network as shown in Fig. \ref{power_splitter_design}(a),
where a tunable component of impedance $Z$ is embedded in series
with two transmission lines with characteristic impedance $Z_{0}$.
The corresponding scattering matrix $\mathbf{S}_{\mathrm{mode}}$
of this two-port network is given by
\begin{equation}
\mathbf{S}_{\mathrm{mode}}=\frac{1}{Z+2Z_{0}}\left[\begin{array}{cc}
Z & 2Z_{0}\\
2Z_{0} & Z
\end{array}\right].\label{tunable 2-port network}
\end{equation}

Given $Z_{0}$ as 50 $\Omega$, we can simplify the power ratio of
the reflected signal over the transmitted signal as
\begin{equation}
P=20\textrm{log}\left|\frac{S_{11}^{\mathrm{mode}}}{S_{21}^{\mathrm{mode}}}\right|=20\textrm{log}\left|\frac{Z}{100}\right|\textrm{dB}.\label{power_ratio in dB}
\end{equation}

\begin{figure*}[t]
\begin{centering}
\textsf{\includegraphics[width=0.66\columnwidth]{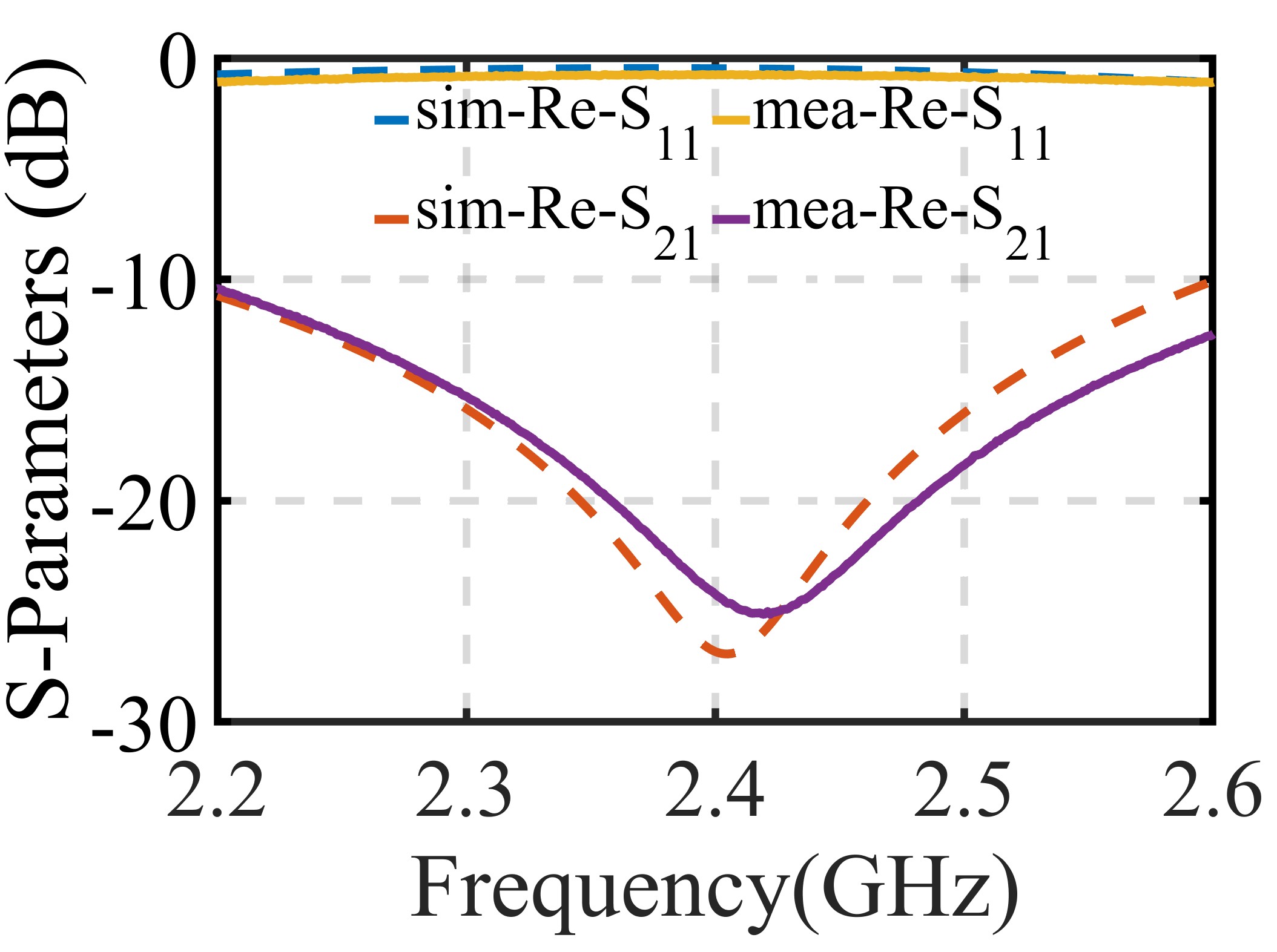}\hspace*{0\columnwidth}\includegraphics[width=0.66\columnwidth]{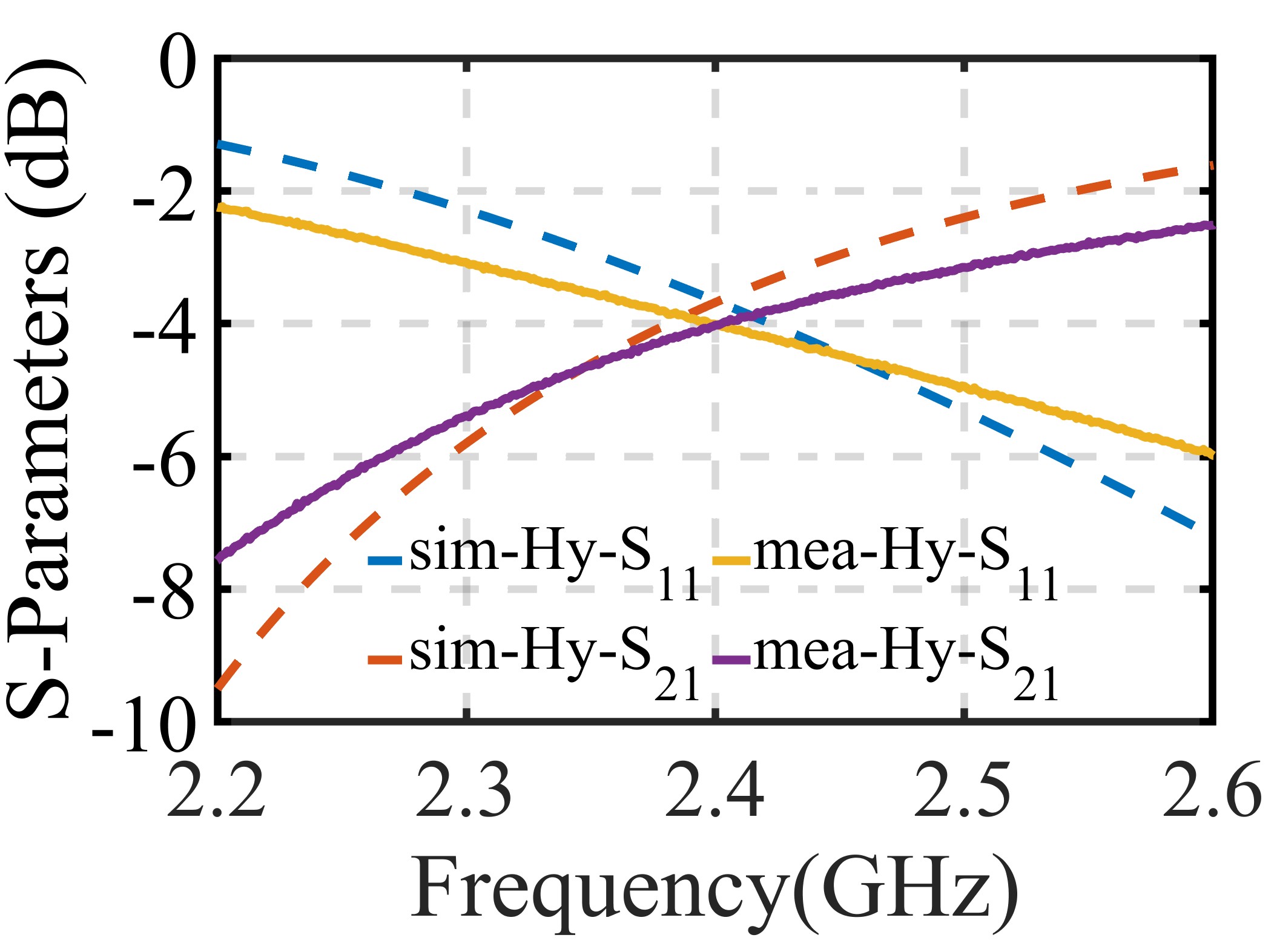}\hspace*{0\columnwidth}\includegraphics[width=0.66\columnwidth]{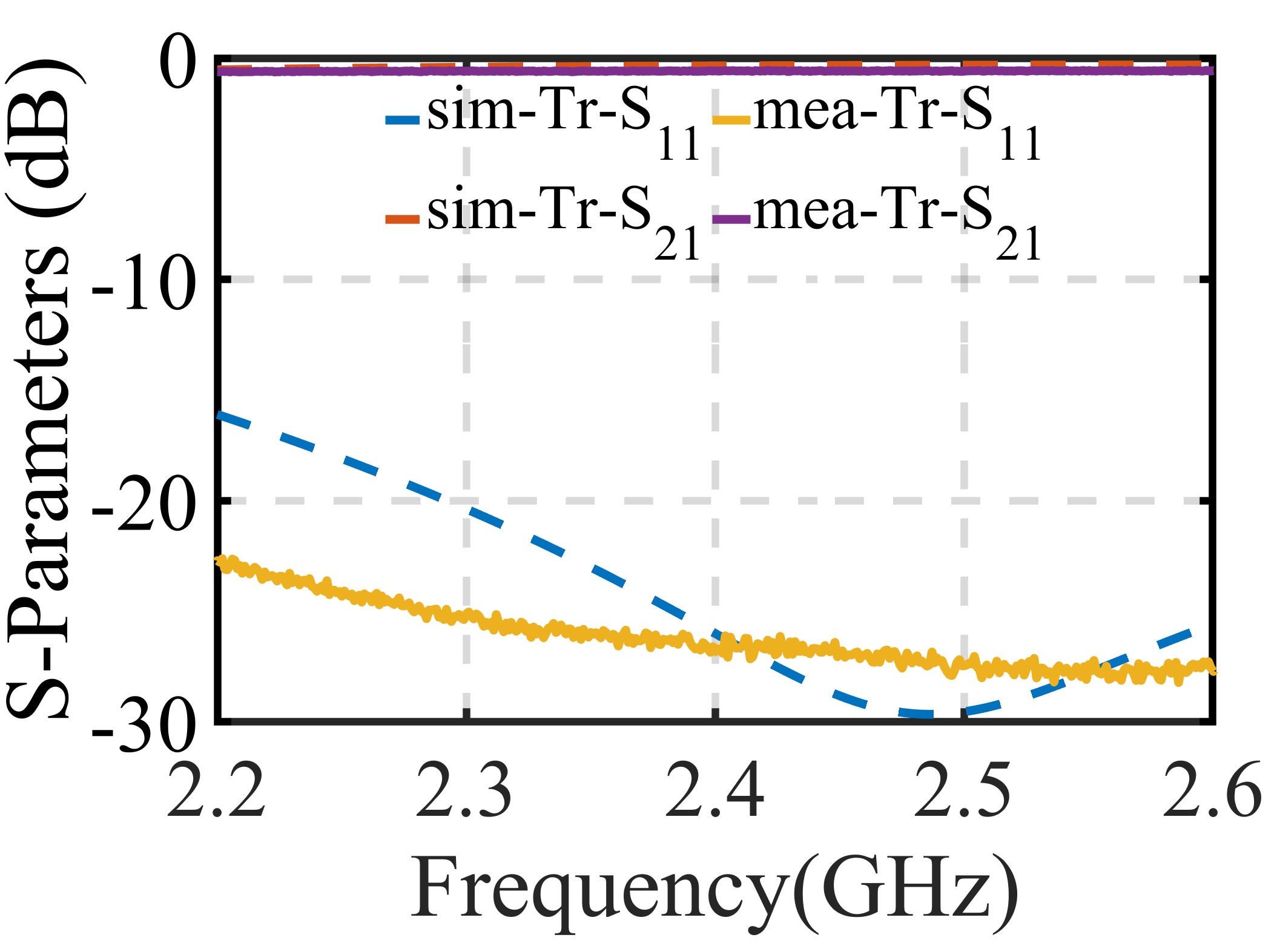}}
\par\end{centering}
\begin{raggedright}
\hspace*{0.45\columnwidth} (a)\hspace*{0.5\columnwidth} (b)\hspace*{0.55\columnwidth}
(c)
\par\end{raggedright}
\caption{Performance of the fabricated two-port power splitter when the reverse
DC voltage is set as $-$10.8 V, $-$7.4 V, and 0 V, which respectively
corresponds to: (a) Reflection mode; (b) Hybrid mode; (c) Transmission
mode.}
\label{power splitter performance}
\end{figure*}

To achieve switching between transmission, hybrid, and reflection
modes, as proposed in Section III.A, we can set the tuning range for
the power ratio $P$, to $-$20 dB, 0 dB, and to 20 dB. Correspondingly,
the tunable range of $Z$ should at least satisfy $\mathit{|Z|}\in\left[10,10^{3}\right]$.
Setting the center frequency as 2.4 GHz, and considering that the
adjustable capacitance range of commonly used commercial varactors
is from a few tenths of a picofarad (pF) to tens of pF, we propose
the circuit structure shown in Fig. \ref{power_splitter_design}(a)
to achieve the required equivalent impedance tuning capability for
dynamic mode switching. Specifically, a commercial varactor SMV2020-079LF
from Skyworks is connected in parallel with a carefully chosen bias
inductor 0402DC-3N9XJRW from Coilcraft to form the core part of the
two-port power splitter. The tunable impedance $Z$ can then be given
by
\begin{equation}
Z=\frac{\left(\frac{1}{jwC_{\mathrm{J}_{0}}}+jwL_{\mathrm{s}}+R_{\mathrm{s}}\right)jwL_{\mathrm{c}}}{\frac{1}{jwC_{\mathrm{J}_{0}}}+jwL_{\mathrm{s}}+R_{\mathrm{s}}+jwL_{\mathrm{c}}},\label{tunable impedance-1}
\end{equation}
where $0.35\:\textrm{pF}\leq C_{\mathrm{J}_{0}}\leq3.2\:\textrm{pF}$
is the tunable capacitance of varactor SMV2020-079LF when provided
with reverse voltage from $-$20 V to 0 V. The series parasitic inductance
$L_{\mathrm{s}}$ and resistance $R_{\mathrm{s}}$ introduced by the
package and transistor is 0.7 nH and 2.5 $\Omega$, respectively.
The inductance of the parallel inductor $L_{\mathrm{c}}$ is optimized
as 3.9 nH. This provides crucial admittance bias for the tunable admittance
range of the varactor so that the corresponding $\mathit{|Z|}$ can
fulfill the requirements for dynamic mode switching. Specifically,
the designed tunable two-port power splitter is implemented on a Rogers
4003C substrate ($\epsilon_{\mathrm{r}}=3.55$ and $\tan\delta=0.0027$)
with a thickness of 0.813 mm as shown in Fig. \ref{power_splitter_design}(b).
Except for the above mentioned essential components, other components
include two inductors 0402DC-R10XGRW from Coilcraft serving as the
RF chokes and three capacitors GJM1555C1H130GB01D from Murata serving
as the DC chokes.

The fabricated circuit is shown in Fig. \ref{power_splitter_design}(c),
and it was measured by a calibrated Rohde \& Schwarz ZVA40 4-Port
vector network analyzer. As demonstrated in Fig. \ref{power splitter performance}(a)-(c),
by setting reverse DC voltage $V_{\mathrm{p}}$ as $-$10.8 V, $-$7.4
V, and 0 V, using a 50-watt E3620A dual-output power supply, we can
realize mode switching between reflection, hybrid and transmission
modes, respectively. The measured insertion loss for the reflection
and transmission modes at 2.4 GHz is less than 0.8 dB, which is around
0.3 dB larger than simulated results due to extra loss introduced
by SMA connectors and cables. The simulated $S_{11}$ and $S_{21}$
for the hybrid mode at 2.4 GHz are $-$3.68 dB and $-$3.66 dB, while
the measured $S_{11}$ and $S_{21}$ are $-$4.01 dB and $-$4.02
dB, which shows that equal power splitting can be approximately obtained
. Besides, it can be seen from Fig. \ref{power splitter performance}(b)
that the 1-dB variation bandwidth for hybrid mode is from 2.35GHz
to 2.45GHz.

\subsection{2-Bit Phase Reconfigurable Antenna}

\begin{figure*}[tp]
\begin{centering}
\textsf{\includegraphics[width=0.66\columnwidth]{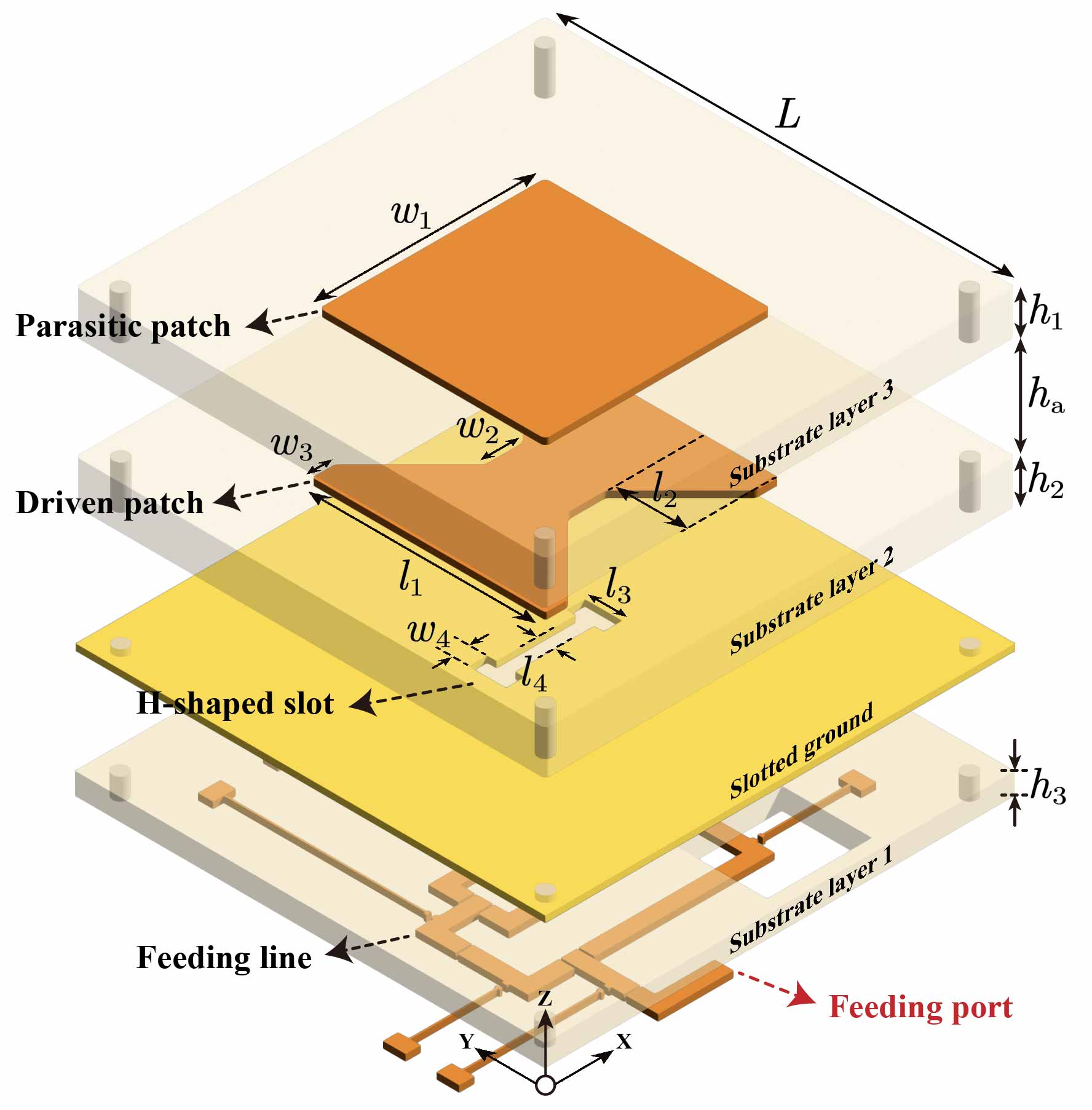}\includegraphics[width=0.66\columnwidth]{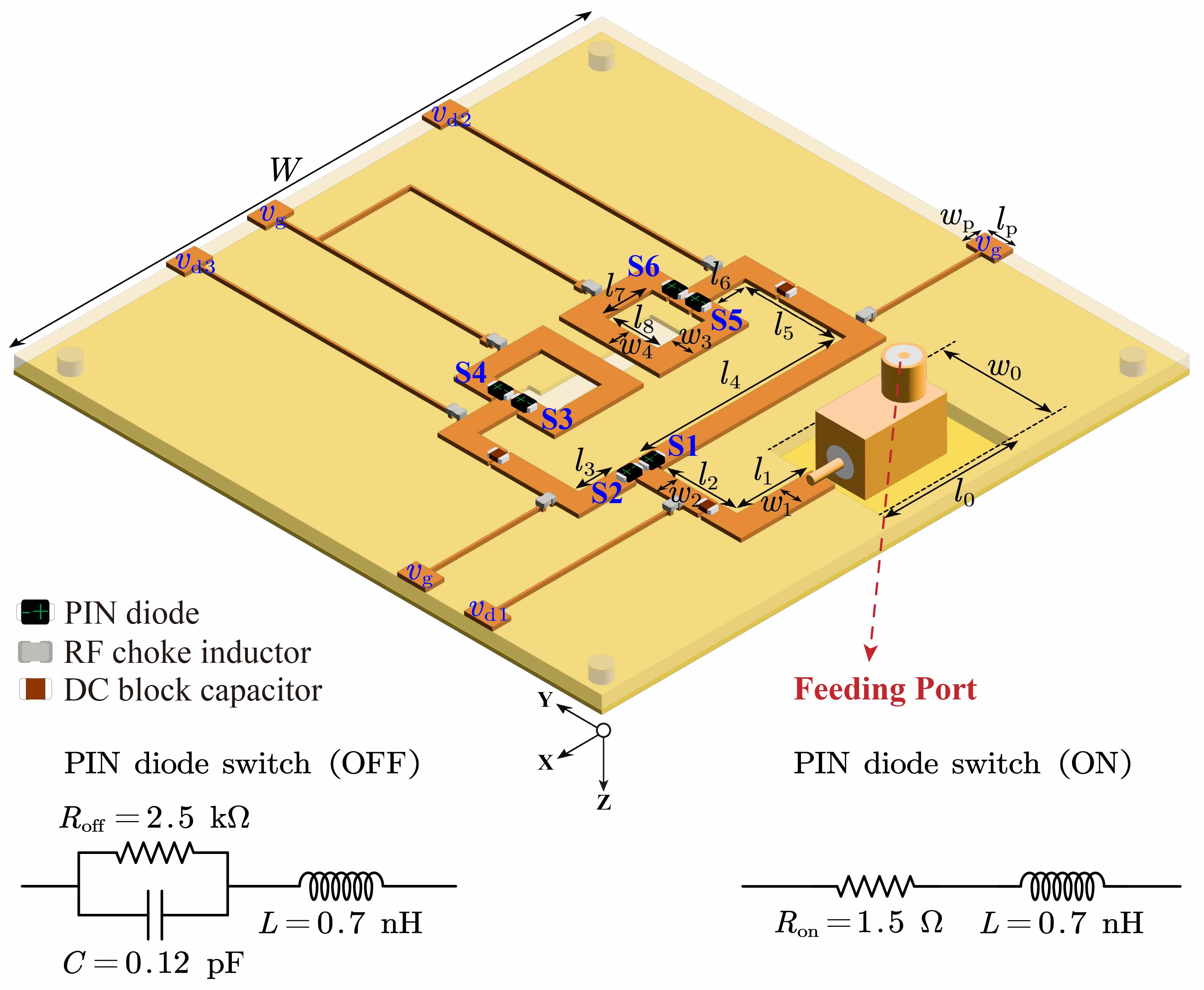}\includegraphics[width=0.66\columnwidth]{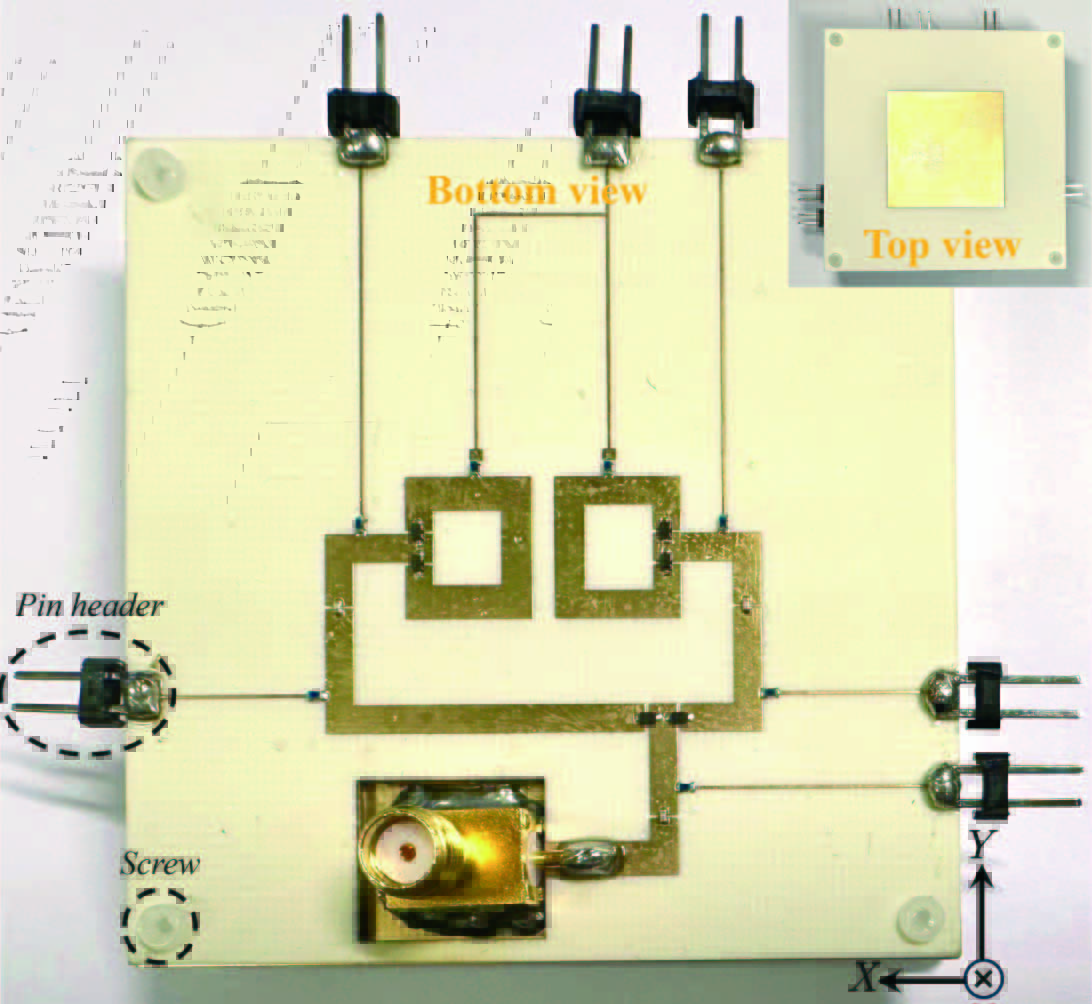}}
\par\end{centering}
\begin{raggedright}
\hspace*{0.33\columnwidth} (a)\hspace*{0.6\columnwidth} (b)\hspace*{0.6\columnwidth}
(c)
\par\end{raggedright}
\caption{2-Bit phase reconfigurable antenna design for the proposed BD-RIS.
(a) Overall structure of the designed aperture coupled phase reconfigurable
antenna. The dimensions are $L=62.5$, $h_{1}=3.048$, $h_{2}=3.048$,
$h_{3}=1.524$, $h_{\mathrm{a}}=8$, $w_{1}=30$, $w_{2}=5$, $w_{3}=3$,
$w_{4}=2.5$, $l_{1}=31$, $l_{2}=10$, $l_{3}=5$, $l_{4}=2.5$ (unit:
mm). (b) Schematic of the integrated 2-bit phase shifter design and
equivalent circuit models for the on and off states of diode SMP1345-079LF.
The dimensions are $W=62.5$, $w_{0}=12$, $w_{1}=2.5$, $w_{2}=2.1$,
$w_{3}=2.5$, $w_{4}=2.3$, $w_{\mathrm{p}}=2$, $l_{0}=14$, $l_{1}=7.9$,
$l_{2}=8.3$, $l_{3}=4$, $l_{4}=7.9$, $l_{5}=8.3$, $l_{6}=4$,
$l_{7}=5.1$, $l_{8}=5.7$, $l_{\mathrm{p}}=3$ (unit: mm). Light
yellow color: dielectric layer. Dark orange color: copper. Yellow
color: ground. (c) Photograph of the fabricated 2-bit antenna.}
\label{antenna}
\end{figure*}

It is useful here, and also later in the experimental section, to
define the structural and antenna scattering of antennas. When electromagnetic
waves impinge on the surface of an antenna, the scattered pattern
can be decomposed into structural and antenna scattering parts. The
structural scattering part is defined here as the scattered pattern
when the antenna is perfectly conjugate matched. The antenna scattering
counterpart is then defined as that part of the radiation pattern
that arises when the load deviates from the conjugate match. As illustrated
in \cite{balanis2016antenna,rao2023active}, the scattered pattern
$E^{\mathrm{s}}(Z_{\mathrm{load}})$ when the antenna is terminated
with a load $Z_{\mathrm{load}}$ can be expressed as

\begin{equation}
E^{\mathrm{s}}(Z_{\mathrm{load}})=E^{\mathrm{s}}(Z_{\mathrm{A}}^{\ast})-\frac{\varGamma^{\ast}Z_{\mathrm{A}}I_{\mathrm{s}}E^{\mathrm{unit}}}{2R_{\mathrm{A}}},\label{simplified scattered pattern}
\end{equation}
where $Z_{\mathrm{A}}^{\ast}$ is the perfect conjugate match of the
antenna input impedance $Z_{\mathrm{A}}$, $R_{\mathrm{A}}$ is the
real part of $Z_{\mathrm{A}}$, and $I_{\mathrm{s}}$ is the current
induced by the incident wave at the antenna's port when the port is
shorted. $E^{\mathrm{unit}}$ is the radiation pattern when the antenna
is excited by a unit current at the input port. $\varGamma^{\ast}$
is related to $Z_{\mathrm{load}}$ by

\begin{equation}
\varGamma^{\ast}=\frac{Z_{\mathrm{load}}-Z_{\mathrm{A}}^{\ast}}{Z_{\mathrm{load}}+Z_{\mathrm{A}}^{\ast}},\label{Z_L and Tps relationship-1-1}
\end{equation}
This leads to the typical approach to achieve reconfiguration in which
changing the load of the antenna will alter the antenna scattering.

The novelty in our design is that we alter the antenna scattering
part by changing the current $I_{\mathrm{s}}$ and $E^{\mathrm{unit}}$.
This is performed by integrating a 2-bit phase shifter into the antenna
structure. The advantage of this approach is that there is less loss,
because there is no need for a load and its transmission line. In
addition it enables a much more compact design.

Based on the above observations, the overall geometry of the proposed
2-bit phase reconfigurable antenna is shown in Fig. \ref{antenna}(a),
which consists of three dielectric layers and four metallic layers.
The bottom metallic layer is the microstrip feeding network integrated
with three pairs of switches to realize 2-bit phase reconfigurability.
The specific circuit topology is shown in Fig. \ref{antenna}(b).
By selecting one of the first pair of switches, S1 and S2, that are
embedded on two transmission line paths with a quarter-wavelength
electrical length difference, a 90$^{\circ}$ phase difference can
then be achieved. Furthermore, at the end of each path, there is a
split ring mounted with another pair of switches, S3 and S4 or S5
and S6. Similarly, by selecting one of the pair of switches, the current
flowing through the split ring will be reversed, thus generating a
180$^{\circ}$ phase difference. By properly controlling three pairs
of switches as shown in Table \ref{table 1}, we can realize 2-bit
phase reconfigurability by controlling $I_{\mathrm{s}}$ and $E^{\mathrm{unit}}$
in \eqref{simplified scattered pattern}.

The RF switch we use is the diode SMP1345-079LF from Skyworks and
its equivalent circuit models for the on and off states are shown
in Fig. \ref{antenna}(b). Each pair of switches are mounted in series
on three branches of transmission lines and the left and right branches
are always connected to the reference ground plane whose voltage $V_{\mathrm{g}}$
is kept as 0 V. The DC bias voltage $V_{\mathrm{d}i}$, for $i=1,2,3$,
is then added on the center branch and when $V_{\mathrm{d}i}=$1.6
V, the corresponding diode S1 or S3 or S5 will be turned on, and the
counterpart S2 or S4 or S6 will be turned off. When $V_{\mathrm{d}i}=$$-$1.6
V, the situation is the opposite. When $V_{\mathrm{d}i}=$0 V, then
no diode will be turned on. By properly controlling three pairs of
switches as shown in Table \ref{table 1}, we can then realize 2-bit
phase reconfigurability. It should be noted that, $V_{\mathrm{d}1}$
can be provided by a single I/O pin (3.3-/0-V output) of a commercial
FPGA whose GND is biased to be $-$1.6 V. While to provide three states
of bias voltages for $V_{\mathrm{d}2}$ or $V_{\mathrm{d}3}$, we
connect two I/O pins in parallel, so that by assigning one of the
output as 1.6 V, another one as $-$1.6 V, or both of the two outputs
as 1.6 V, or $-$1.6V, we can control the parallel output as 0 V,
1.6 V, and $-$1.6 V, respectively. Hence, for a single 2-bit antenna,
it should be controlled by 5 I/O pins.

To improve the bandwidth of the patch antenna, we implement the aperture
coupled technique, reshape the slot coupled square patch into a bowtie-like
and introduce another layer of parasitic patch separated by an air
gap of height $h_{\mathrm{a}}$. All three dielectric layers are Rogers
4003C ($\varepsilon_{\mathrm{r}}=3.55$, $\tan\delta=0.0027$), where
the upper two layers are of a thickness of 3.048 mm and the bottom
layer is 1.524 mm. The upper two metallic layers are the parasitic
patch and bowtie-like patch radiators. The third metallic layer is
the slotted ground plane, which couples energy from the lower microstrip
feeding network to the upper patch radiators.

\begin{table}[tp]
\caption{States of the 2-bit Phase Reconfigurable Antenna}
\label{table 1}
\centering{}%
\begin{tabular}{>{\centering}m{0.6cm}>{\centering}m{0.6cm}>{\centering}m{0.6cm}>{\centering}m{0.6cm}>{\centering}m{0.6cm}>{\centering}m{0.6cm}>{\centering}m{0.6cm}>{\centering}m{0.6cm}}
\hline 
\noalign{\vskip\doublerulesep}
{\footnotesize States} & {\footnotesize S1} & {\footnotesize S2} & {\footnotesize S3} & {\footnotesize S4} & {\footnotesize S5} & {\footnotesize S6} & {\footnotesize Phase}\tabularnewline[\doublerulesep]
\hline 
\noalign{\vskip\doublerulesep}
\noalign{\vskip\doublerulesep}
{\footnotesize 00} & {\footnotesize ON} & {\footnotesize OFF} & {\footnotesize ON} & {\footnotesize OFF} & {\footnotesize OFF} & {\footnotesize OFF} & {\footnotesize 0°}\tabularnewline[\doublerulesep]
\noalign{\vskip\doublerulesep}
\noalign{\vskip\doublerulesep}
{\footnotesize 01} & {\footnotesize OFF} & {\footnotesize ON} & {\footnotesize OFF} & {\footnotesize OFF} & {\footnotesize OFF} & {\footnotesize ON} & {\footnotesize 90°}\tabularnewline[\doublerulesep]
\noalign{\vskip\doublerulesep}
\noalign{\vskip\doublerulesep}
{\footnotesize 10} & {\footnotesize ON} & {\footnotesize OFF} & {\footnotesize OFF} & {\footnotesize ON} & {\footnotesize OFF} & {\footnotesize OFF} & {\footnotesize 180°}\tabularnewline[\doublerulesep]
\noalign{\vskip\doublerulesep}
\noalign{\vskip\doublerulesep}
{\footnotesize 11} & {\footnotesize OFF} & {\footnotesize ON} & {\footnotesize OFF} & {\footnotesize OFF} & {\footnotesize ON} & {\footnotesize OFF} & {\footnotesize 270°}\tabularnewline[\doublerulesep]
\hline 
\noalign{\vskip\doublerulesep}
\end{tabular}
\end{table}

\begin{figure}[t]
\begin{centering}
\textsf{\includegraphics[width=0.5\columnwidth]{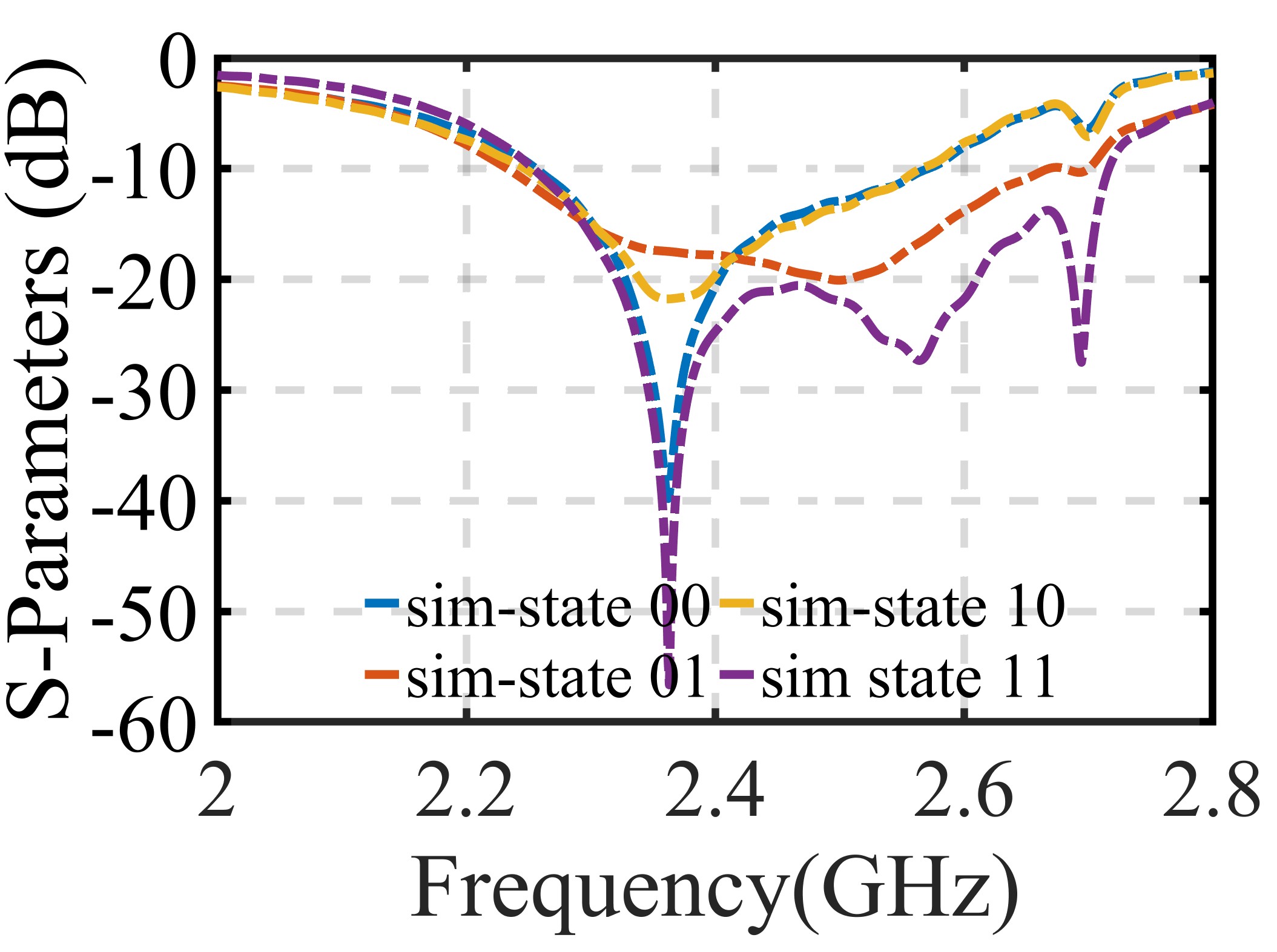}\hspace*{0.03\columnwidth}\includegraphics[width=0.5\columnwidth]{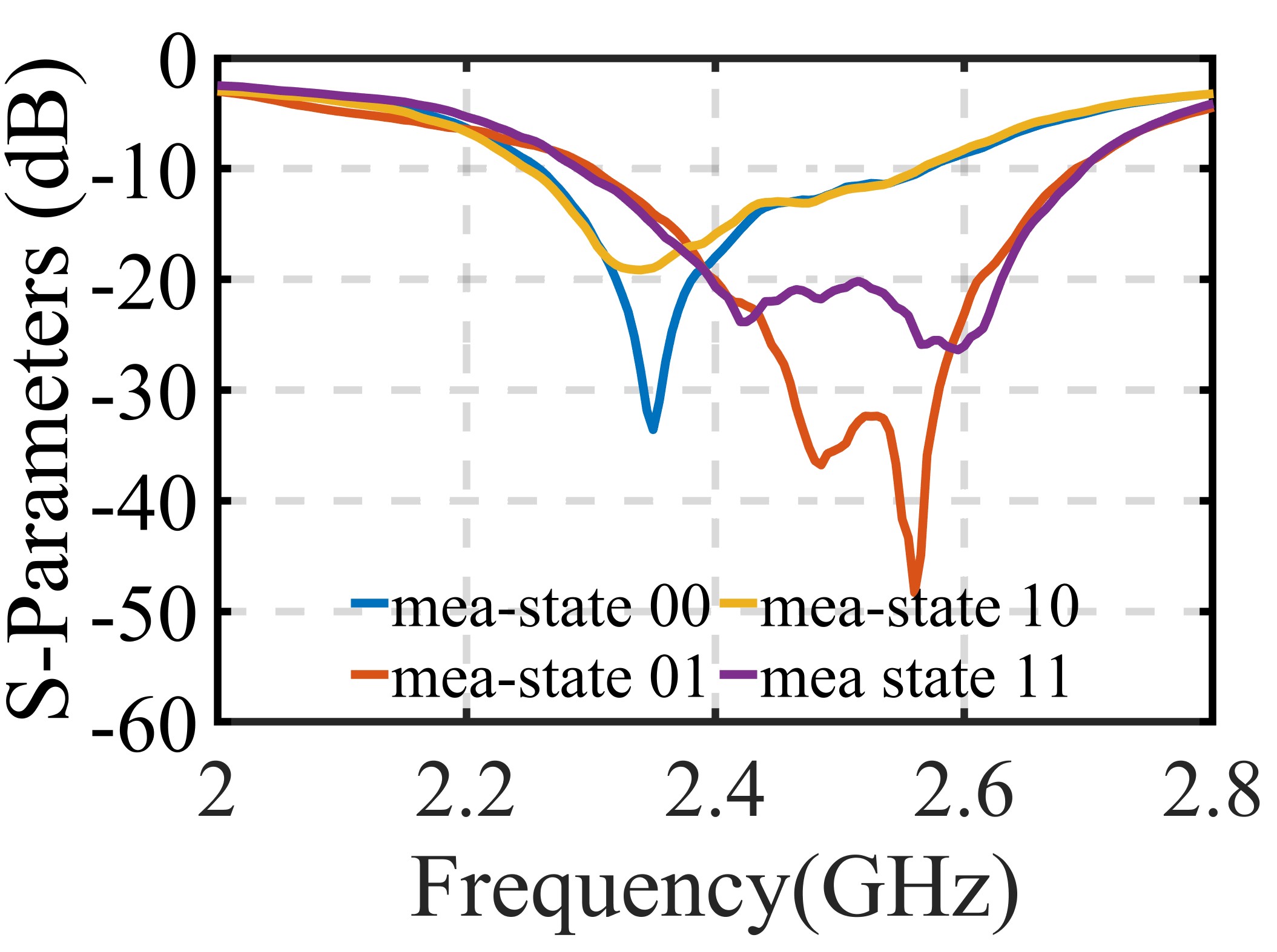}}
\par\end{centering}
\begin{raggedright}
\hspace*{0.24\columnwidth} (a)\hspace*{0.45\columnwidth} (b)
\par\end{raggedright}
\begin{centering}
\textsf{\includegraphics[width=0.5\columnwidth]{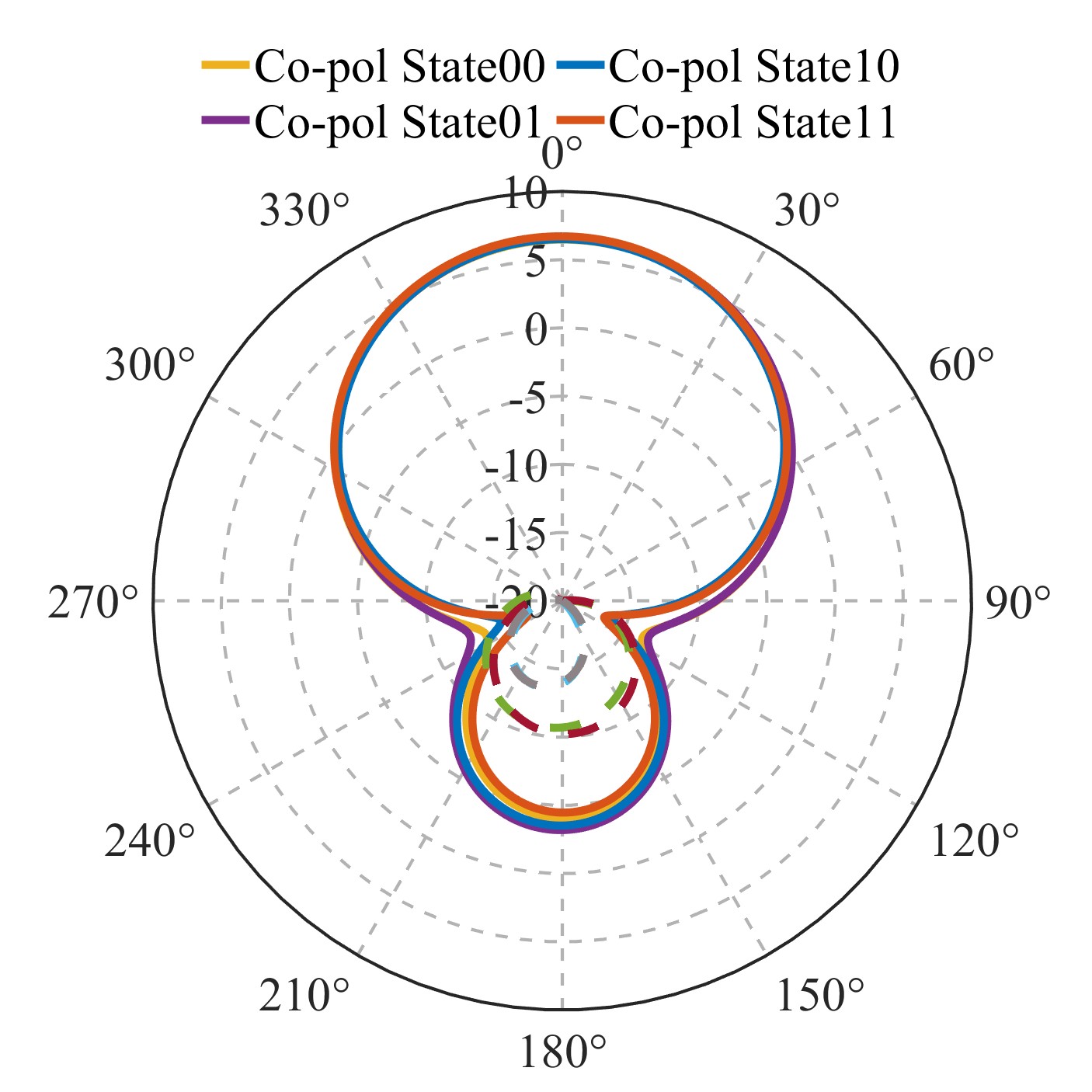}\hspace*{0.03\columnwidth}\includegraphics[width=0.5\columnwidth]{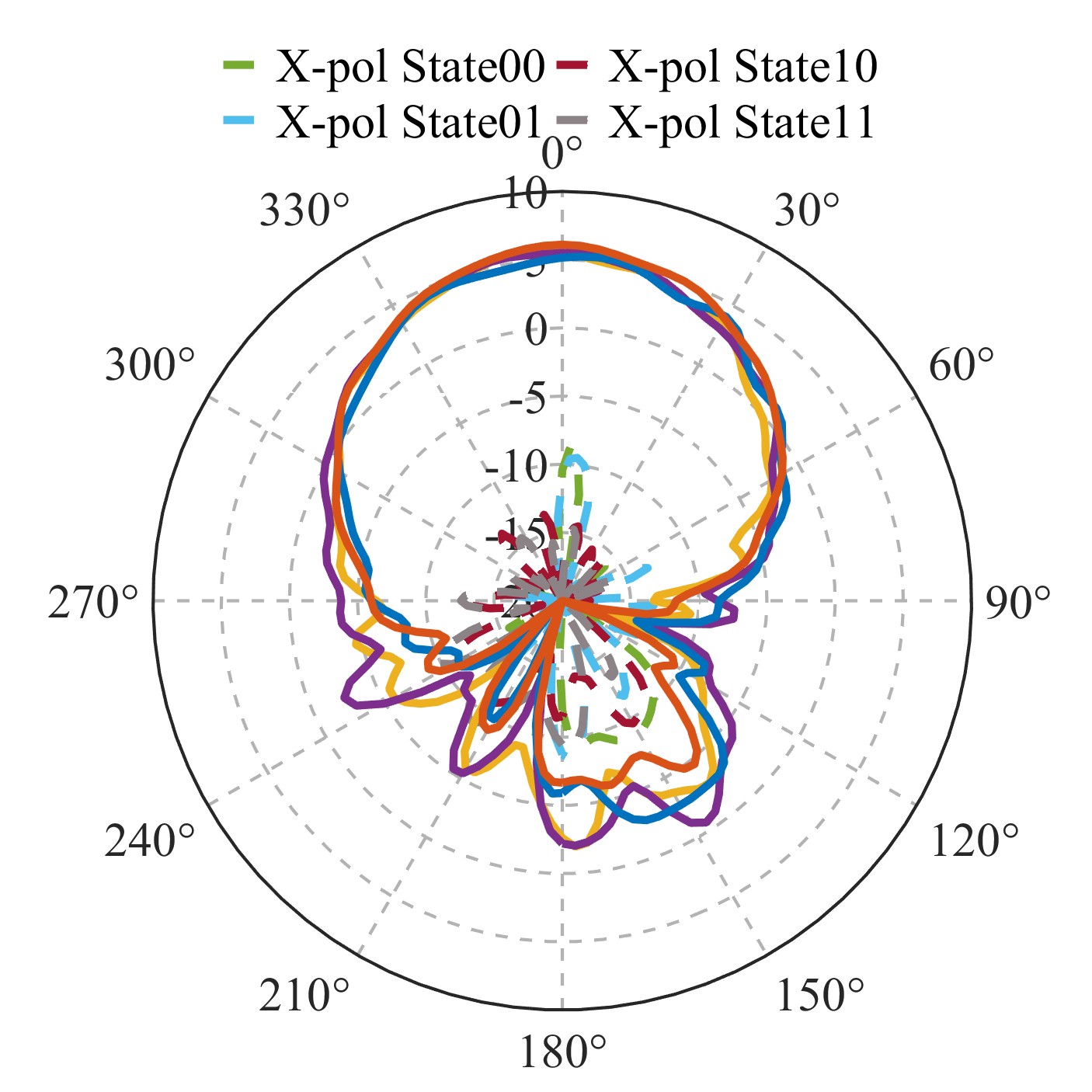}}
\par\end{centering}
\begin{raggedright}
\hspace*{0.24\columnwidth} (c)\hspace*{0.4\columnwidth} (d)
\par\end{raggedright}
\begin{centering}
\textsf{\includegraphics[width=0.5\columnwidth]{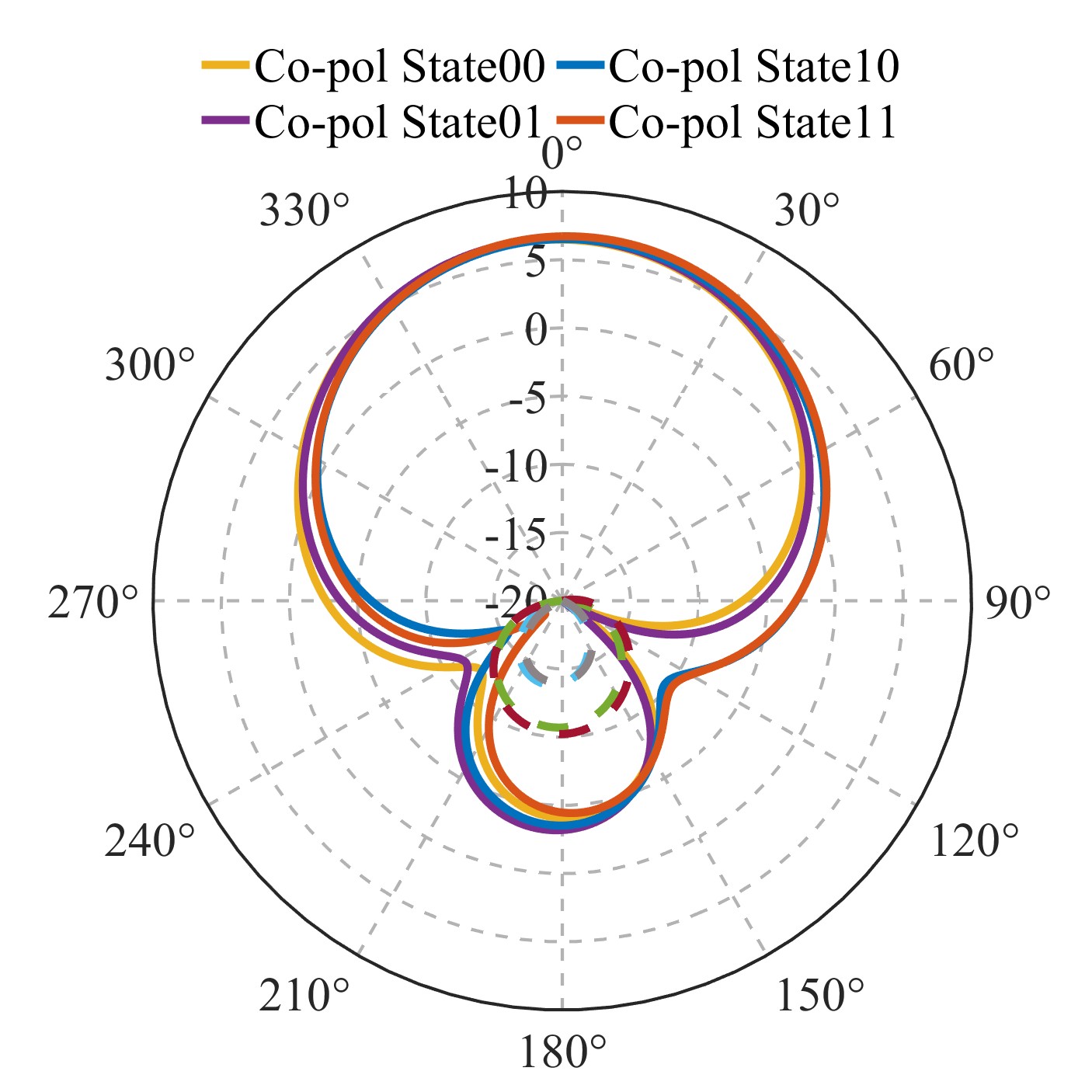}\hspace*{0.03\columnwidth}\includegraphics[width=0.5\columnwidth]{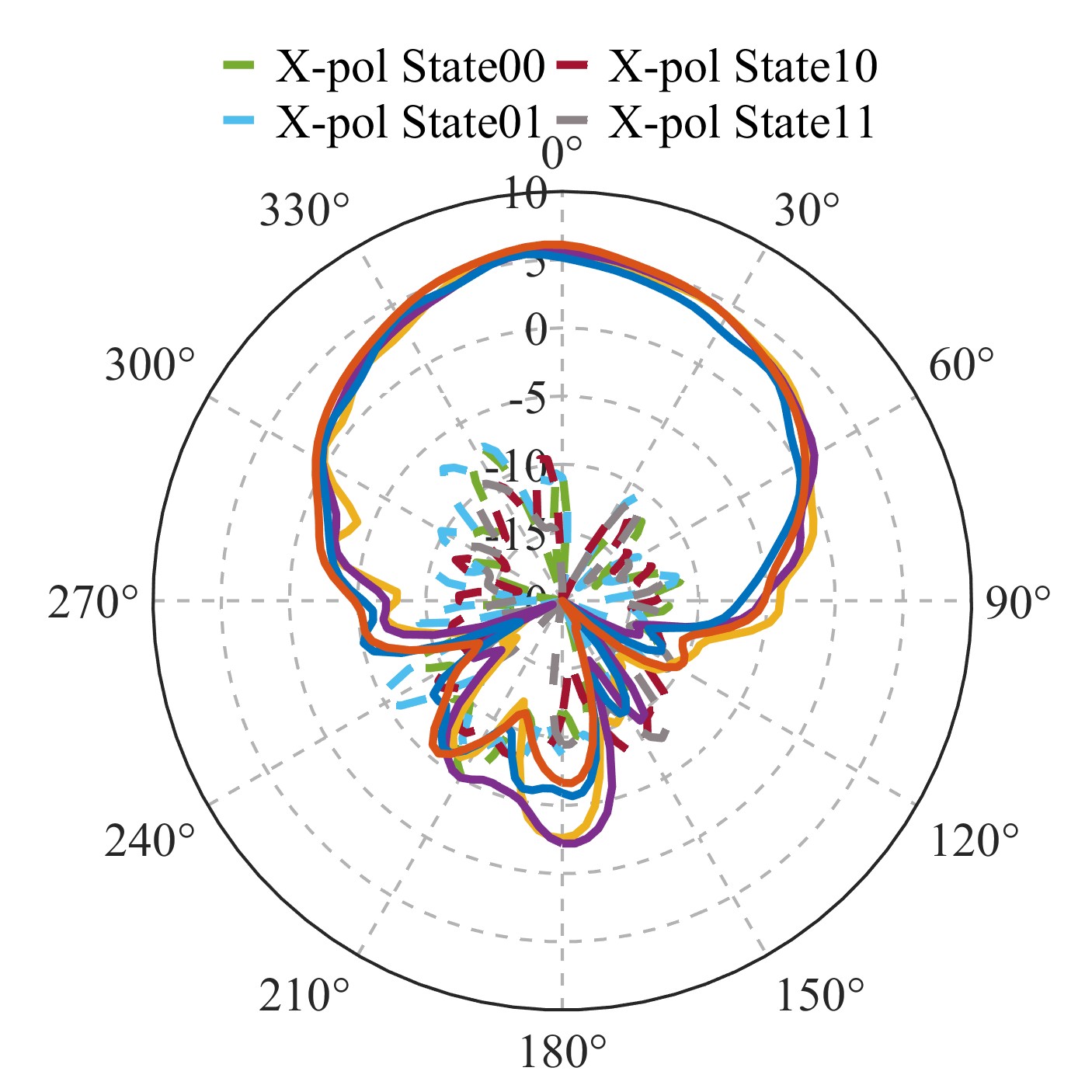}}
\par\end{centering}
\begin{raggedright}
\hspace*{0.24\columnwidth} (e)\hspace*{0.4\columnwidth} (f)
\par\end{raggedright}
\begin{centering}
\textsf{\includegraphics[width=0.5\columnwidth]{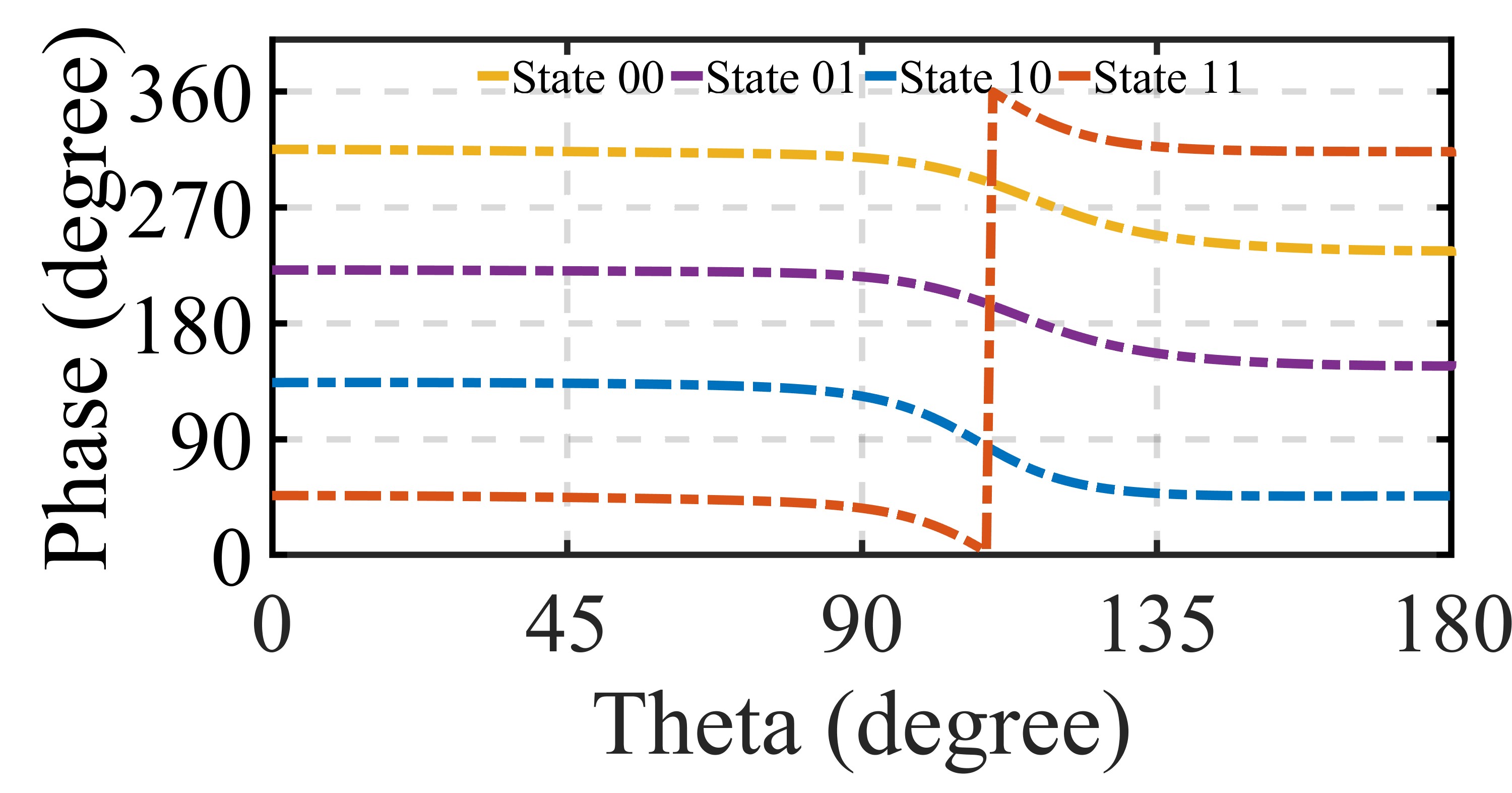}\hspace*{0.03\columnwidth}\includegraphics[width=0.5\columnwidth]{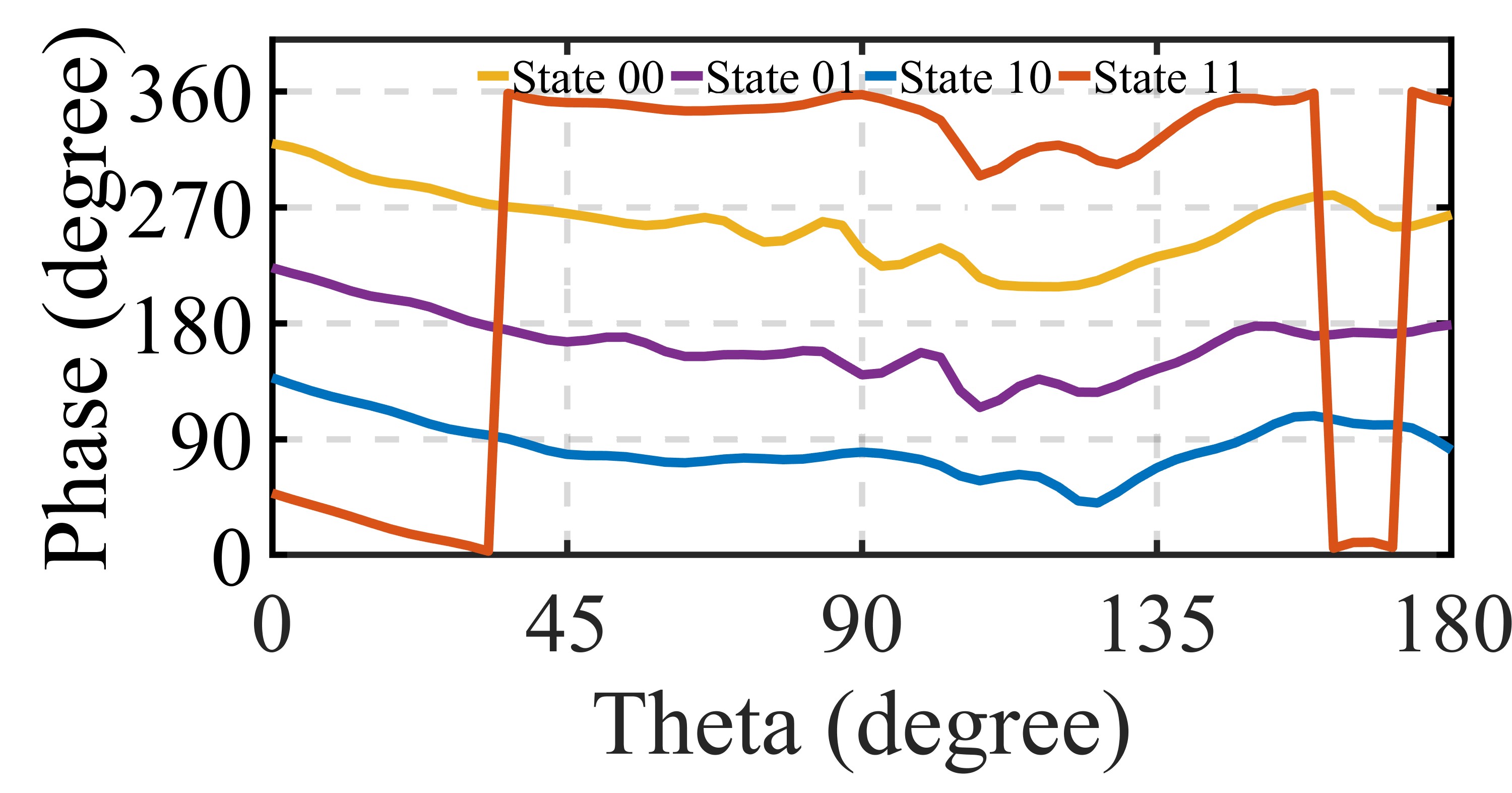}}
\par\end{centering}
\begin{raggedright}
\hspace*{0.24\columnwidth} (g)\hspace*{0.45\columnwidth} (h)
\par\end{raggedright}
\caption{Performance of the fabricated 2-bit phase reconfigurable antenna.
(a) Simulated input return loss. (b) Measured input return loss. (c)
Simulated radiation pattern in the E plane (unit: dBi). (d) Measured
radiation pattern in the E plane (unit: dBi). (e) Simulated radiation
pattern in the H plane (unit: dBi). (f) Measured radiation pattern
in the H plane (unit: dBi). (g) Simulated E-field phase responses.
(h) Measured E-field phase responses.}
\label{antenna performance}
\end{figure}

The simulated and measured $S_{11}$ parameters of the fabricated
2-bit antenna is shown in Fig. \ref{antenna performance}(a) and (b),
respectively. Despite a slight frequency shift for state 01 and state
11 due to fabrication error, the measured S-parameters of the four
states are generally in good agreement with the simulated results
and the bandwidth around 2.4 GHz is over 200 MHz (2.3-2.5 GHz). Fig.
\ref{antenna performance}(c)-(f) demonstrate the simulated and measured
E-plane (XOZ plane) and H-plane (YOZ plane) radiation patterns of
the four states at 2.4 GHz, respectively. The radiation patterns and
broadside co-polarization gains among the four states are very similar
and stable. Fig. \ref{antenna performance}(g) and (h) depicts the
simulated and measured E-field phase responses of four states at 2.4
GHz, where the four phases are evenly distributed over 360$^{\circ}$
with a 90$^{\circ}$ phase difference between adjacent states. Therefore,
a 2-bit phase reconfigurable antenna design showing favorable performance
for the proposed BD-RIS architecture is achieved.

\section{Multiple Cells Performance Analysis}

\begin{figure*}[t]
\begin{centering}
\textsf{\includegraphics[width=2\columnwidth]{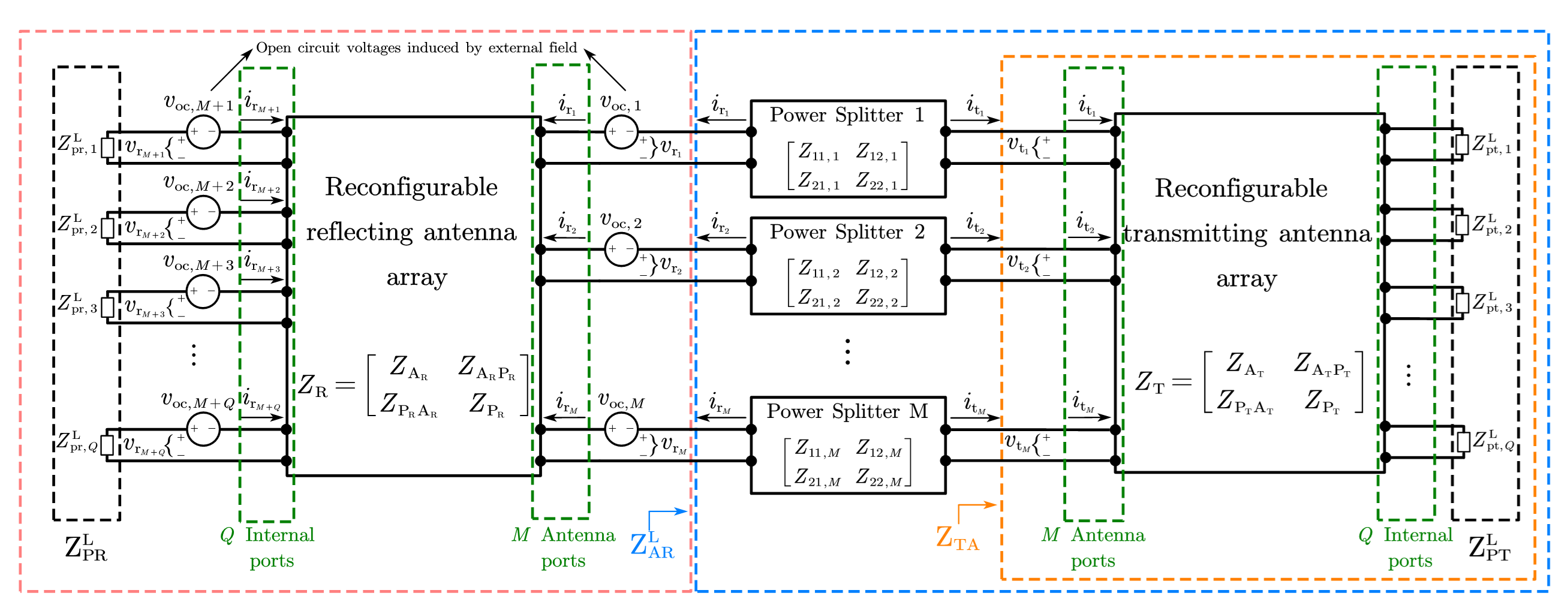}}
\par\end{centering}
\caption{\label{fig:6}Thévenin equivalent model of the hybrid transmitting
and reflecting BD-RIS illuminated by an external electromagnetic field.}
\label{thevenin equivalent}
\end{figure*}

In this section we establish models and perform analysis to characterize
the performance of the reflection and transmission electromagnetic
responses of our BD-RIS with multiple cells. Subsequently, we employ
the genetic algorithm (GA) to conduct beamforming optimization for
our proposed BD-RIS under different operation modes.

\subsection{Physical Model and Analytical Method}

Physical models have been proposed to model the electromagnetic responses
of a hybrid transmitting and reflecting BD-RIS, such as the phase
shift model, the load impedance model and the generalized sheet transition
conditions model \cite{xu2022simultaneously}. However, most of these
models have not been verified by hardware implementations. Here we
propose a physical model for the hybrid transmitting and reflecting
BD-RIS that is based upon the hardware scheme shown in Fig. \ref{proposed_scheme}(b).
Its electromagnetic performance can be characterized by using Thévenin
equivalent circuits \cite{mautz1973modal} and linear superposition
of electromagnetic waves in free space \cite{balanis2012advanced}.

The Thévenin equivalent model of the hybrid transmitting and reflecting
BD-RIS is shown in Fig. \ref{fig:6}. Given a BD-RIS consisting of
$M$ cells, it can be viewed as a phase reconfigurable reflecting
antenna array, as enclosed by the pink dashed line box in Fig. \ref{fig:6},
loaded with an $M$-port reconfigurable network, as enclosed by the
blue dashed line box. The $M$-port reconfigurable network is further
constructed by $M$ two-port tunable power splitters cascaded with
another $M$-element phase reconfigurable transmitting antenna array,
as enclosed by the orange dashed line box.

For each phase reconfigurable antenna, there are $q$ switches ($q=6$
in our proposed design), so that the antenna array will have in total
$q\times M=Q$ switches ($q=6$ and $M=16$ and therefore, $Q=96$).
To model the effect of these $Q$ switches we can replace them by
$Q$ internal ports, as highlighted by the green dashed line. The
internal ports are then terminated with equivalent loads to denote
the on and off states of the switches. Thus, the entire phase reconfigurable
antenna array, e.g., reflecting or transmitting antenna array, can
be characterized by an $\left(M+Q\right)\times\left(M+Q\right)$ impedance
matrix $\mathbf{Z}_{\chi}$ (where we use $\mathit{\chi}\in\left\{ \mathrm{R},\mathrm{T}\right\} $
with $\mathrm{R}$ and $\mathrm{T}$ representing the reflecting and
transmitting antenna array, respectively) as
\begin{equation}
\mathbf{Z}_{\chi}=\left[\begin{array}{cc}
\mathbf{Z}_{\mathrm{A}_{\chi}} & \mathbf{Z}_{\mathrm{A}_{\chi}\mathrm{P}_{\chi}}\\
\mathbf{Z}_{\mathrm{P}_{\chi}\mathrm{A}_{\chi}} & \mathbf{Z}_{\mathrm{P}_{\chi}}
\end{array}\right],\label{receiving or transmitting multiport impedance matrix}
\end{equation}
where matrix $\mathbf{Z}_{\mathrm{A}_{\chi}}\in\mathbb{C}^{M\times M}$
represents the impedance sub-matrix for the $M$ antenna ports, $\mathbf{Z}_{\mathrm{P}_{\chi}}\in\mathbb{C}^{Q\times Q}$
represents the impedance sub-matrix for the $Q$ internal ports, $\mathbf{Z}_{\mathrm{A}_{\chi}\mathrm{P}_{\chi}}\in\mathbb{C}^{M\times Q}$
represents the trans-impedance between the voltages of the $M$ antenna
ports and the currents of the $Q$ internal ports, and $\mathbf{Z}_{\mathrm{P}_{\chi}\mathrm{A}_{\chi}}$
is the transpose of $\mathbf{Z}_{\mathrm{A}_{\chi}\mathrm{P}_{\chi}}$.

For the $Q$ internal ports of the reflecting antenna, we express
the load impedance at the $q$th internal port as $Z_{\mathrm{pr},q}^{\mathrm{L}}$.
This is based on the equivalent circuit models given in Fig. \ref{antenna}(b),
with $Z_{\mathrm{pr},q}^{L}\in\left\{ Z_{\mathrm{on}},Z_{\mathrm{off}}\right\} $,
where $Z_{\mathrm{on}}=R_{\mathrm{on}}+j\omega L$ and $Z_{\mathrm{off}}=\frac{R_{\mathrm{off}}/jwC}{R_{\mathrm{off}}+1/jwC}+j\omega L$,
with $R_{\mathrm{on}}$ = 1.5 $\Omega$, $L$ = 0.7 nH and $R_{\mathrm{off}}$
= 2.5 $\mathrm{k}\Omega$, $C$ = 0.12 pF. By collecting $Z_{\mathrm{pr},q}^{L}\forall\,q$,
a diagonal matrix $\mathbf{Z}_{\mathrm{PR}}^{\mathrm{L}}=\textrm{diag}(Z_{\mathrm{pr},1}^{\mathrm{L}},\cdots,Z_{\mathrm{pr},Q}^{\mathrm{L}})$
can be formed as marked by the black dashed line on the left side.

We also denote the load impedance of $M$ antenna ports as $\mathbf{Z}_{\mathrm{AR}}^{\mathrm{L}}\mathbf{\in\mathbb{C}^{\mathit{M\times M}}}$,
which can also be viewed as the input impedance of the $M$-port reconfigurable
network as marked by the blue dashed line in Fig. \ref{fig:6}. Thus,
we can build a block diagonal matrix $\mathbf{Z}_{\mathrm{L}}=\textrm{blkdiag}\left(\mathbf{Z}_{\mathrm{AR}}^{\mathrm{L}},\mathbf{Z}_{\mathrm{PR}}^{\mathrm{L}}\right)\in\mathbb{C}^{\left(M+Q\right)\times\left(M+Q\right)}$
to denote the overall load impedance matrix of this $\left(M+Q\right)$-port
network. Here $\textrm{blkdiag}\left(\cdot\right)$ denotes a block
matrix such that the main-diagonal blocks are matrices and all off-diagonal
blocks are zero matrices.

To analyze the entire system we denote the voltage and current vectors
of the $\left(M+Q\right)$ ports embedded in the reflecting antenna
array as $\mathbf{v}_{\mathrm{R}}=\left[v_{\mathrm{r},1},\ldots,v_{\mathrm{r},M},\ldots,v_{\mathrm{r},M+Q}\right]^{T}$
and $\mathbf{i}_{\mathrm{R}}=\left[i_{\mathrm{r},1},\ldots,i_{\mathrm{r},M},\ldots,i_{\mathrm{r},M+Q}\right]^{T}$,
where $^{T}$ is the transpose operation. For $m=1,\ldots,M$, $v_{\mathrm{r},m}$
and $i_{\mathrm{r},m}$ refers to the voltage and current at the $m$th
reflecting antenna, respectively, and for $q=1,\ldots,Q$, $v_{\mathrm{r},M+q}$
and $i_{\mathrm{r},M+q}$ refers to the voltage and current at the
$q$th internal port of the reflecting antenna array, respectively.
When the phase reconfigurable reflecting antenna array is illuminated
by the external electromagnetic waves, $\mathbf{Z}_{\mathrm{R}}$,
$\mathbf{v}_{\mathrm{R}}$, and $\mathbf{i}_{\mathrm{R}}$ can be
related as
\begin{equation}
\mathbf{v}_{\mathrm{oc}}=\mathbf{v}_{\mathrm{R}}-\mathbf{Z}_{\mathrm{R}}\mathbf{i}_{\mathrm{R}},\label{entire array impedance matrix}
\end{equation}
where $\mathbf{v}_{\mathrm{oc}}=\left[v_{\mathrm{oc},1},\ldots,v_{\mathrm{oc},M},\ldots,v_{\mathrm{oc},M+Q}\right]^{T}$
with $v_{\mathrm{oc},m}$ and $v_{\mathrm{oc},M+q}$ denoting the
open-circuit voltage excited at the $m$th antenna port and the $q$th
internal port for $m=1,\ldots,M$, and $q=1,\ldots,Q$, respectively.
For this $\left(M+Q\right)$-port network, we also have the following
relationship between $\mathbf{Z}_{\mathrm{L}}$, $\mathbf{v}_{\mathrm{R}}$,
and $\mathbf{i}_{\mathrm{R}}$,
\begin{equation}
\mathbf{v}_{\mathrm{R}}=-\mathbf{Z}_{\mathrm{L}}\mathbf{i}_{\mathrm{R}}.\label{load impedance}
\end{equation}
By substituting \eqref{load impedance} into \eqref{entire array impedance matrix},
the currents at all the $M$ antenna ports and $Q$ internal ports
of the reflecting antenna array excited by external electromagnetic
waves is given by
\begin{equation}
\mathbf{i}_{\mathrm{R}}=-\left(\mathbf{Z}_{\mathrm{R}}+\mathbf{Z}_{\mathrm{L}}\right)^{-1}\mathbf{v}_{\mathrm{oc}}.\label{port current}
\end{equation}
The total controllable scattered pattern, which corresponds to the
total controllable reflected fields, can then be calculated using
the expression
\begin{equation}
\begin{array}{c}
\mathbf{E}_{\mathrm{r}}\left(\Omega\right)=\sum_{m=1}^{M}i_{\mathrm{r},m}\mathbf{E}_{m}\left(\Omega\right)+\sum_{q=1}^{Q}i_{\mathrm{r},M+q}\mathbf{E}_{M+q}\left(\Omega\right)\\
+\mathbf{E}_{\mathrm{oc}}\left(\Omega\right)-\mathbf{E}_{\mathrm{r}}^{\mathrm{str}}\left(\Omega\right)
\end{array},\label{reflected fields}
\end{equation}
where $\Omega=\left(\theta,\varphi\right)$ with $\theta\in\left[0,180^{\circ}\right]$
and $\varphi\in\left[0,360^{\circ}\right]$. $\mathbf{E}_{m}\left(\Omega\right)$
and $\mathbf{E}_{M+q}\left(\Omega\right)$ are the electric fields
radiated by the $m$th reflecting antenna port and $q$th internal
port excited by a unit current source when all the other antenna and
internal ports are open-circuited, respectively. $\mathbf{E}_{\mathrm{oc}}\left(\Omega\right)$
is the scattered pattern when all the $M$ antenna ports and $Q$
internal ports are open-circuited. $\mathbf{E}_{\mathrm{r}}^{\mathrm{str}}\left(\Omega\right)$
is the structural mode scattering field when all the ports of the
antenna array are conjugated-matched. Therefore, $\mathbf{E}_{\mathrm{r}}\left(\Omega\right)$
also corresponds to the antenna scattering field as defined in Section
IV.C.

To calculate the reflected fields, the key lies in the construction
of load impedance $\mathbf{Z}_{\mathrm{L}}$. This is because in \eqref{port current}
and \eqref{reflected fields}, $\mathbf{Z}_{\mathrm{R}}$, $\mathbf{v}_{\mathrm{oc}}$,
$\mathbf{E}_{m}\left(\Omega\right)$$\forall\,m$, $\mathbf{E}_{M+q}\left(\Omega\right)$$\forall\,q$,
$\mathbf{E}_{\mathrm{oc}}\left(\Omega\right)$, $\mathbf{E}_{\mathrm{r}}^{\mathrm{str}}\left(\Omega\right)$
depict the fundamental electromagnetic characteristics of the designed
structure and can be acquired using electromagnetic solvers once,
such as CST Microwave Studio. As part of $\mathbf{Z}_{\mathrm{L}}$,
the load impedance matrix of internal ports, $\mathbf{Z}_{\mathrm{PR}}^{\mathrm{L}}$
can be fully constructed once the states of all the switches in the
reflecting antenna array are known (see later Section), while the
load impedance $\mathbf{Z}_{\mathrm{AR}}^{\mathrm{L}}$ of $M$ antenna
ports still needs to be found.

The input impedance matrix of the loaded $M$-port network $\mathbf{Z}_{\mathrm{AR}}^{\mathrm{L}}$,
consists of $M$ two-port power splitters cascaded with the $M$-element
reconfigurable transmitting antenna array. Denote the antenna impedance
matrix of the cascaded transmitting antenna array as $\mathbf{Z}_{\mathrm{TA}}\in\mathbb{C}^{M\times M}$,
marked by the orange dashed line in Fig. \ref{fig:6}. Define $\mathbf{v}_{\mathrm{T}}=\left[v_{\mathrm{t},1},\ldots,v_{\mathrm{t},M}\right]^{T}$
and $\mathbf{i}_{\mathrm{T}}=\left[i_{\mathrm{t},1},\ldots,i_{\mathrm{t},M}\right]^{T}$
to denote the voltage and current vectors of the $M$ antenna ports
embedded in the transmitting antenna array. For the transmitting antenna
array, $\mathbf{v}_{\mathrm{T}}$ and $\mathbf{i}_{\mathrm{T}}$ are
related by
\begin{equation}
\mathbf{v}_{\mathrm{T}}=\mathbf{Z}_{\mathrm{TA}}\mathbf{i}_{\mathrm{T}}.\label{transmit array impedance relation}
\end{equation}
Besides, considering that each reflecting antenna and transmitting
antenna are interconnected via a power splitter, for the $m$th power
splitter, we can relate $v_{\mathrm{r},m}$, $i_{\mathrm{r},m}$ and
$v_{\mathrm{t},m}$, $i_{\mathrm{t},m}$ by
\begin{equation}
\left[\begin{array}{c}
v_{\mathrm{r},m}\\
v_{\mathrm{t},m}
\end{array}\right]=\left[\begin{array}{cc}
z_{11,m} & z_{12,m}\\
z_{21,m} & z_{22,m}
\end{array}\right]\left[\begin{array}{c}
-i_{\mathrm{r},m}\\
-i_{\mathrm{t},m}
\end{array}\right],\label{power splitter impedance relation}
\end{equation}
where each entry of the impedance matrix of power splitter $z_{ij,m}(i,j=1,2)\forall\,m$
is determined by the operation mode of the power splitters and can
be obtained based on our measured S-parameters for different modes
in Section IV.B. Therefore, by combining \eqref{transmit array impedance relation}
and \eqref{power splitter impedance relation}, the load impedance
$\mathbf{Z}_{\mathrm{AR}}^{\mathrm{L}}$ can be interpreted as

\begin{equation}
\mathbf{Z}_{\mathrm{AR}}^{\mathrm{L}}=-(\mathbf{C}-\mathbf{Z}_{\mathrm{TA}}\mathbf{A})^{-1}(\mathbf{Z}_{\mathrm{TA}}\mathbf{B-D}),\label{input impedance of multi-port network}
\end{equation}
where $\mathbf{A}=\textrm{diag}(a_{1},a_{2},\cdots,a_{M})$ with $a_{m}=-1/z_{12,m}$,
$\mathbf{B}=\textrm{diag}(b_{1},b_{2},\cdots,b_{M})$ with $b_{m}=-z_{11,m}/z_{12,m}$,
$\mathbf{C}=\textrm{diag}(c_{1},c_{2},\cdots,c_{M})$ with $c_{m}=-z_{22,m}/z_{12,m}$,
and $\mathbf{D}=\textrm{diag}(d_{1},d_{2},\cdots,d_{M})$ with $d_{m}=z_{22,m}z_{11,m}/z_{12,m}-z_{21,m}$,
for $m=1,\ldots,M$. While $\mathbf{Z}_{\mathrm{TA}}$ refers to the
antenna impedance matrix of the transmitting antenna array, it should
be expressed by the impedance sub-matrix $\mathbf{Z}_{\mathrm{A}_{\mathrm{T}}}$
for the $M$ transmitting antenna ports as defined from \eqref{receiving or transmitting multiport impedance matrix}
and a perturbation term characterizing the effects of different phase
states of transmitting antennas \cite{zhang2021compact} as follows
\begin{equation}
\mathbf{Z}_{\mathrm{TA}}=\mathbf{Z}_{\mathrm{A}_{\mathrm{T}}}-\mathbf{Z}_{\mathrm{A}_{\mathrm{T}}\mathrm{P}_{\mathrm{T}}}(\mathbf{Z}_{\mathrm{P}_{\mathrm{T}}}+\mathbf{Z}_{\mathrm{PT}}^{\mathrm{L}})^{-1}\mathbf{Z}_{\mathrm{P}_{\mathrm{T}}\mathrm{A}_{\mathrm{T}}},\label{transmit array  antenna impedance matrix}
\end{equation}
where $\mathbf{Z}_{\mathrm{P}_{\mathrm{T}}}$, $\mathbf{Z}_{\mathrm{A}_{\mathrm{T}}\mathrm{P}_{\mathrm{T}}}$,
$\mathbf{Z}_{\mathrm{P}_{\mathrm{T}}\mathrm{A}_{\mathrm{T}}}$ are
also given from \eqref{receiving or transmitting multiport impedance matrix},
and similar to $\mathbf{Z}_{\mathrm{PR}}^{\mathrm{L}}$, we define
$\mathbf{Z}_{\mathrm{PT}}^{\mathrm{L}}=\textrm{diag}(Z_{\mathrm{pt},1}^{\mathrm{L}},\cdots,Z_{\mathrm{pt},Q}^{\mathrm{L}})$
with $Z_{\mathrm{pt},q}^{\mathrm{L}}\in\left\{ Z_{\mathrm{on}},Z_{\mathrm{off}}\right\} $,
$q=1,\ldots,Q$, to denote the load impedance matrix of all the $Q$
internal ports on the transmitting antenna array, as marked by the
black dashed line on the right side in Fig. \ref{fig:6}. Hence, by
calculating $\mathbf{Z}_{\mathrm{TA}}$, we can include the states
information of all the antenna elements on the transmitting antenna
array.

Till now, $\mathbf{Z}_{\mathrm{L}}=\textrm{blkdiag}\left(\mathbf{Z}_{\mathrm{AR}}^{\mathrm{L}},\mathbf{Z}_{\mathrm{PR}}^{\mathrm{L}}\right)\in\mathbb{C}^{\left(M+Q\right)\times\left(M+Q\right)}$
can be fully expressed by collecting $\mathbf{Z}_{\mathrm{PR}}^{\mathrm{L}}$
and $\mathbf{Z}_{\mathrm{AR}}^{\mathrm{L}}$, which contains the states
information of all the reflecting and transmitting antennas and the
operation mode information of all the power splitters. Therefore,
based on \eqref{port current} and \eqref{reflected fields}, for
any combination of the states and any operation mode of the power
splitters, the corresponding $\mathbf{i}_{\mathrm{R}}$, $\mathbf{v}_{\mathrm{R}}$
and controllable reflected fields $\mathbf{E}_{\mathrm{r}}\left(\Omega\right)$
can be obtained.

To calculate the transmitted fields, we can derive $i_{\mathrm{t},m}$$\forall\,m$
at each transmitting antenna port from \eqref{port current} and \eqref{power splitter impedance relation}
by
\begin{equation}
i_{\mathrm{t},m}=c_{m}v_{\mathrm{r},m}+d_{m}i_{\mathrm{r},m}.\label{transmitted current}
\end{equation}
Then by linear superposition of the excited electromagnetic waves
at each transmitting antenna port, the overall transmitted field can
be expressed as
\begin{equation}
\mathbf{E}_{\mathrm{t}}\left(\Omega\right)=\sum_{m=1}^{M}i_{\mathrm{t},m}\mathbf{E}_{\mathrm{t},m}\left(\Omega,\mathbf{Z}_{\mathrm{PT}}^{\mathrm{L}}\right),\label{transmitted fields}
\end{equation}
where $\mathbf{E}_{\mathrm{t},m}(\Omega\mathrm{,}\mathbf{Z}_{\mathrm{PT}}^{\mathrm{L}})$
denotes the electric field radiated by the $m$th transmitting antenna
port excited by a unit current source when all the other transmitting
antenna ports are open-circuited and the $Q$ internal ports are terminated
with load impedance $\mathbf{Z}_{\mathrm{PT}}^{\mathrm{L}}$ according
to the specific combination of phase states of transmitting antennas.
Defining $\mathbf{E}_{\mathrm{t}}(\Omega,\mathbf{Z}_{\mathrm{PT}}^{\mathrm{L}})=\left[\mathbf{E}_{\mathrm{t},1}(\Omega,\mathbf{Z}_{\mathrm{PT}}^{\mathrm{L}}),\ldots,\mathbf{E}_{\mathrm{t},M}(\Omega,\mathbf{Z}_{\mathrm{PT}}^{\mathrm{L}})\right]$,
$\mathbf{E}_{\mathrm{t}}(\Omega,\mathbf{Z}_{\mathrm{PT}}^{\mathrm{L}})$
can be found by \cite{zhang2021compact}
\begin{equation}
\mathbf{E}_{\mathrm{t}}\left(\Omega,\mathbf{Z}_{\mathrm{PT}}^{\mathrm{L}}\right)=\mathbf{E}_{\mathrm{A}}\left(\Omega\right)-\mathbf{E}_{\mathrm{P}}\left(\Omega\right)(\mathbf{Z}_{\mathrm{P}_{\mathrm{T}}}+\mathbf{Z}_{\mathrm{PT}}^{\mathrm{L}})^{-1}\mathbf{Z}_{\mathrm{P}_{\mathrm{T}}\mathrm{A}_{\mathrm{T}}},\label{antenna port radiation pattern}
\end{equation}
where $\mathbf{E}_{\mathrm{A}}(\Omega)=\left[\mathbf{E}_{\mathrm{t},1}(\Omega),\ldots,\mathbf{E}_{\mathrm{t},M}(\Omega)\right]$
and $\mathbf{E}_{\mathrm{P}}(\Omega)=\left[\mathbf{E}_{\mathrm{t},M+1}(\Omega),\ldots,\mathbf{E}_{\mathrm{t},M+Q}(\Omega)\right]$
with $\mathbf{E}_{\mathrm{t},m}(\Omega)$ and $\mathbf{E}_{\mathrm{t},M+q}(\Omega)$
being the electric fields radiated by the $m$th transmitting antenna
port and by the $q$th internal port on the transmitting antenna array
excited by a unit current source when all the other antenna and internal
ports are open-circuited, respectively. $\mathbf{Z}_{\mathrm{P}_{\mathrm{T}}}$
and $\mathbf{Z}_{\mathrm{A}_{\mathrm{T}}\mathrm{P}_{\mathrm{T}}}$
can be obtained from $\mathbf{Z}_{\mathrm{T}}$ by \eqref{receiving or transmitting multiport impedance matrix}
and similarly, $\mathbf{Z}_{\mathrm{T}}$, $\mathbf{E}_{\mathrm{t},m}\left(\Omega\right)$$\forall\,m$
, $\mathbf{E}_{\mathrm{t},M+q}\left(\Omega\right)$$\forall\,q$ can
also be acquired using CST Microwave Studio once.

Therefore, when a vertically incident plane wave impinges on one side
of the hybrid transmitting and reflecting BD-RIS, as long as the operation
mode of power splitters and the phase configurations of the reflecting
and transmitting antenna arrays are settled, we can efficiently calculate
the corresponding reflected and transmitted fields based on \eqref{port current},
\eqref{reflected fields}, \eqref{transmitted current}, and \eqref{transmitted fields},
respectively.

\subsection{Beamforming Optimization}

For the proposed BD-RIS design, for one cell with two antennas interconnected
with each other through a power splitter as shown in Fig. \ref{one cell of the hybrid transmitting and reflecting BD-RIS},
there are 16 combinations (since each antenna has four states). Therefore
for a BD-RIS with 4$\times$4 cells, a total of $16^{16}$ configurations
are available, which is impractical for direct calculation for beam
steering applications. To efficiently optimize the radiation patterns
for BD-RIS operating in different modes, we adopt GA to optimize the
phase states of the reflecting and transmitting antennas to achieve
desired beamforming targets.

To illustrate, denote the state of the $m$th reflecting and transmitting
antenna as $x_{\mathrm{r},m}$ and $x_{\mathrm{t},m}$, $m=1,\ldots,16$,
respectively. Each of them should be depicted by two binary bits to
distinguish four states, i.e., $x_{\mathrm{r},m}$, $\mathit{x}_{\mathrm{t},m}\in\left\{ 00,01,10,11\right\} $.
Once given the specific states of $x_{\mathrm{r},m}$ and $x_{\mathrm{t},m}$,
the corresponding load impedances, from $Z_{\mathrm{pr},6\left(m-1\right)+1}^{\mathrm{L}}$
to $Z_{\mathrm{pr},6m}^{\mathrm{L}}$, and from $Z_{\mathrm{pt},6\left(m-1\right)+1}^{\mathrm{L}}$
to $Z_{\mathrm{pt},6m}^{\mathrm{L}}$, for 6 internal ports on the
$m$th reflecting and transmitting antenna can be found by loading
different combinations of $Z_{\mathrm{on}}$ and $Z_{\mathrm{off}}$
based on the reconfigurable control states in Table \ref{table 1}
in Section IV.C, respectively. Define the phase state configuration
vector as $\mathbf{x}=\left[x_{\mathrm{r},1},\ldots,x_{\mathrm{r},M},x_{\mathrm{t},1},\ldots,x_{\mathrm{t},M}\right]^{T}$,
which collects all the phase states information of the reflecting
and transmitting antennas. Thus, a general optimization process can
be formulated as
\begin{equation}
\begin{array}{cc}
\underset{\mathbf{x}}{\min} & -\left|E_{\mathrm{r}}(f_{\mathrm{c}},\Omega_{\mathrm{R}},\mathbf{x})\right|\left|E_{\mathrm{t}}(f_{\mathrm{c}},\Omega_{\mathrm{T}},\mathbf{x})\right|\\
\mathrm{s.t.} & \mathbf{x}\in\left\{ 00,01,10,11\right\} ^{2M}
\end{array},\label{beamforming optimization}
\end{equation}
where $f_{\mathrm{c}}$ is the center frequency set as 2.4 GHz for
the designed BD-RIS. $\Omega_{\mathrm{R}}=\left(\theta_{\mathrm{R}},\varphi_{\mathrm{R}}\right)$
and $\Omega_{\mathrm{T}}=\left(\theta_{\mathrm{T}},\varphi_{\mathrm{T}}\right)$
are the two target beam angles for the reflected and transmitted radiation
patterns, respectively. Taking the coordinate system shown in Fig.
\ref{4=0000D74_BD-RIS} as reference, then we have $\varphi_{\mathrm{R}},\varphi_{\mathrm{T}}\in\left[0,360^{\circ}\right)$,
$\theta_{\mathrm{R}}\in\left[0,90^{\circ}\right)$, and $\theta_{\mathrm{T}}\in\left(90^{\circ},180^{\circ}\right]$.
When we design the reflection (transmission) mode, and considering
that the transmitted (reflected) signals are almost negligible, we
only need to conduct half-space beamforming optimization by assigning
$\left|E_{\mathrm{t}}\right|$ ($\left|E_{\mathrm{r}}\right|$) as
unity in the cost function in \eqref{beamforming optimization}.

\begin{figure}[t]
\begin{centering}
\textsf{\includegraphics[width=0.5\columnwidth]{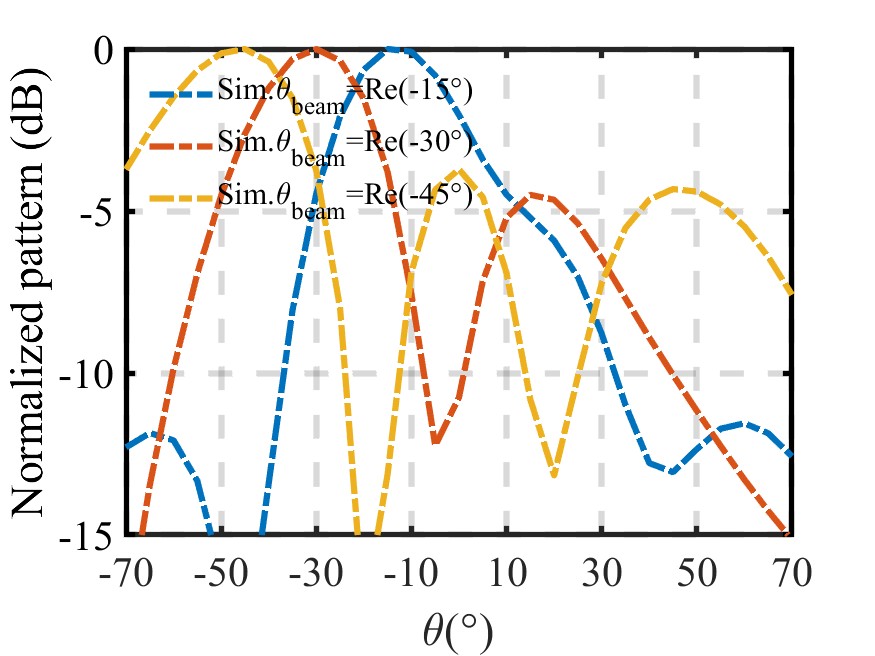}\hspace*{0\columnwidth}\includegraphics[width=0.5\columnwidth]{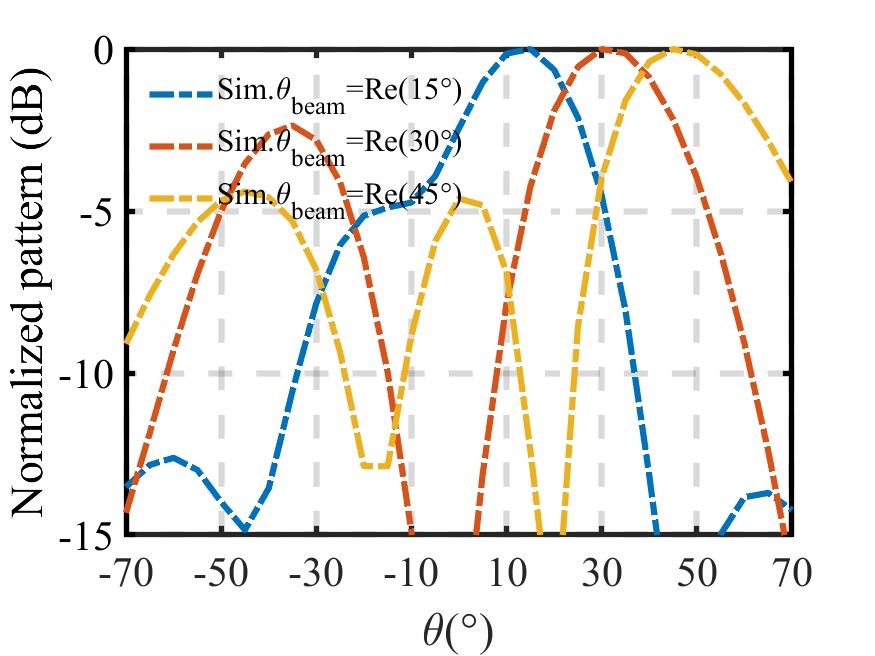}}
\par\end{centering}
\begin{raggedright}
\hspace*{0.2\columnwidth} (a)\hspace*{0.5\columnwidth} (b)
\par\end{raggedright}
\begin{centering}
\textsf{\includegraphics[width=0.5\columnwidth]{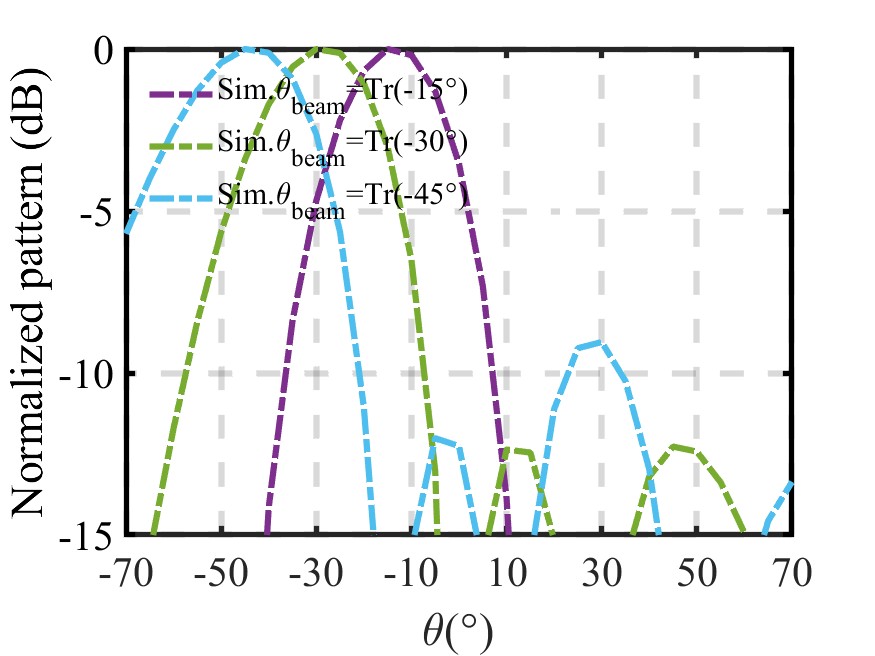}\hspace*{0\columnwidth}\includegraphics[width=0.5\columnwidth]{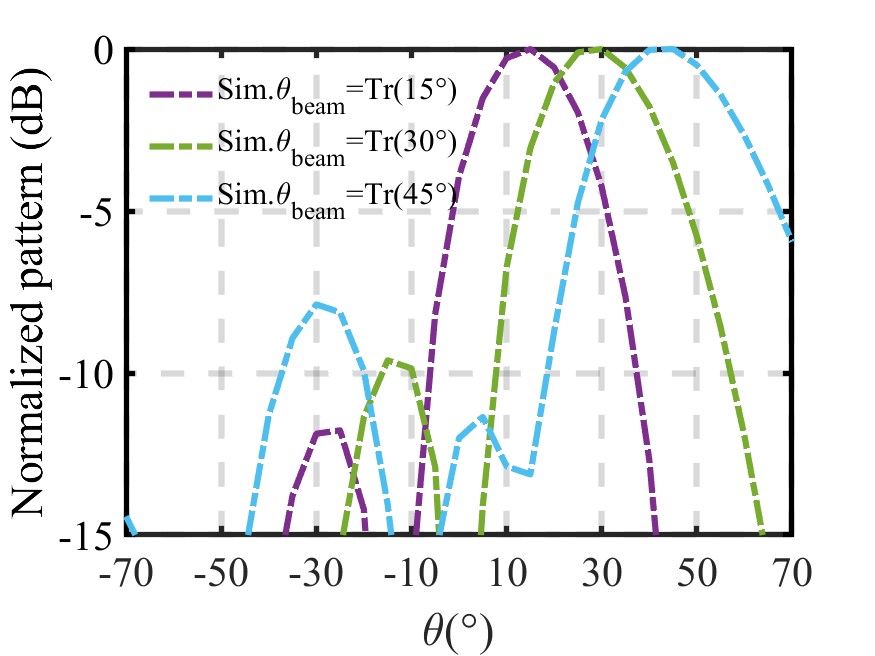}}
\par\end{centering}
\begin{raggedright}
\hspace*{0.2\columnwidth} (c)\hspace*{0.5\columnwidth} (d)
\par\end{raggedright}
\caption{Simulated normalized scattered wave pattern in the YOZ plane at 2.4
GHz with the incident wave perpendicular to the BD-RIS when the BD-RIS
operates in (a) Reflection mode with scanning beams $\left(\theta,\phi\right)_{\mathrm{Re,beam}}$$=\left(-45^{\circ}\sim-15^{\circ},90^{\circ}\right)$,
(b) Reflection mode with scanning beams $\left(\theta,\phi\right)_{\mathrm{Re,beam}}$$=\left(15^{\circ}\sim45^{\circ},90^{\circ}\right)$,
(c) Transmission mode with scanning beams $\left(\theta,\phi\right)_{\mathrm{Tr,beam}}$$=\left(-165^{\circ}\sim-135^{\circ},90^{\circ}\right)$,
(d) Transmission mode with scanning beams $\left(\theta,\phi\right)_{\mathrm{Tr,beam}}$$=\left(135^{\circ}\sim165^{\circ},90^{\circ}\right)$.}
\label{optimization results-1}
\end{figure}

\begin{figure}[t]
\begin{centering}
\textsf{\includegraphics[width=0.5\columnwidth]{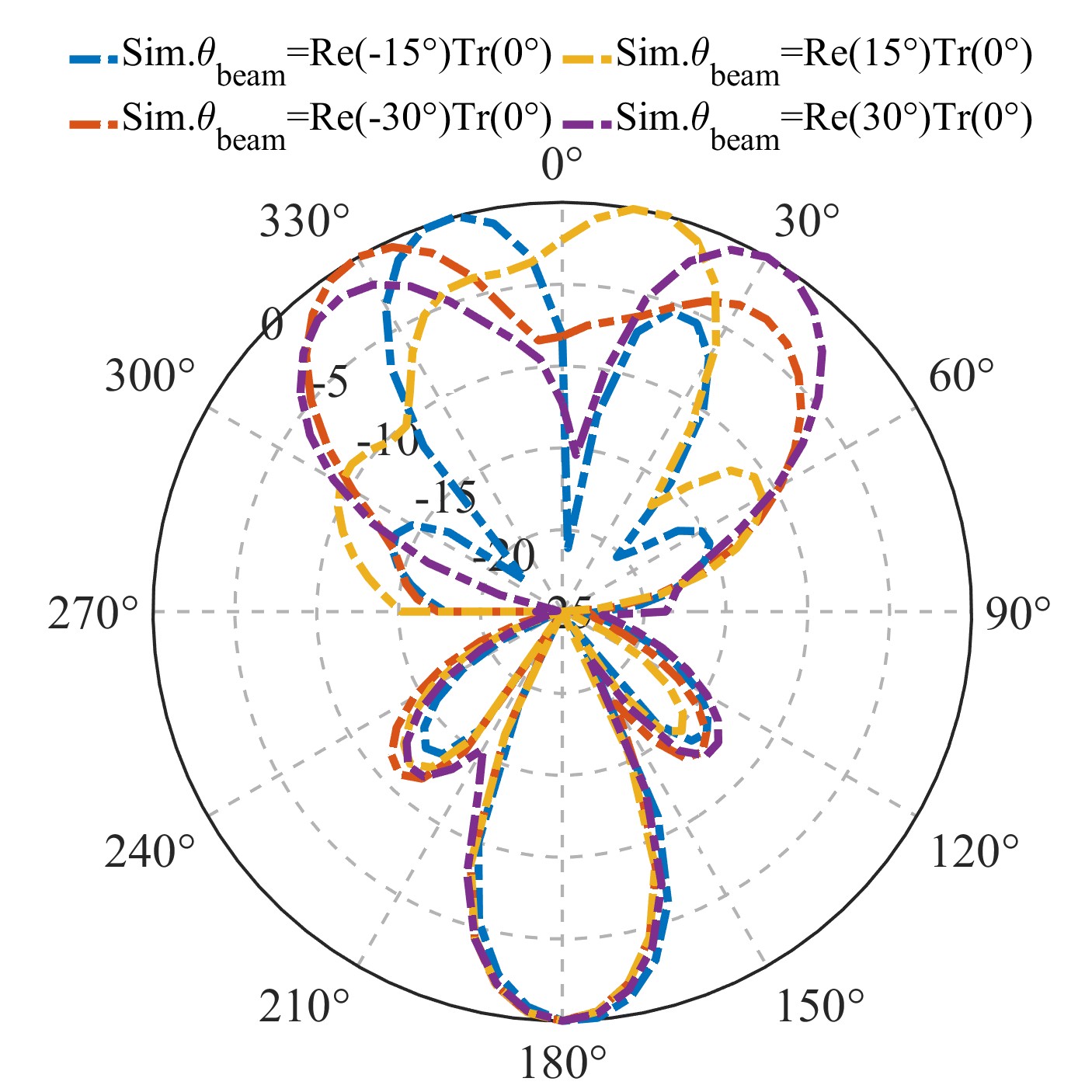}\hspace*{0.03\columnwidth}\includegraphics[width=0.5\columnwidth]{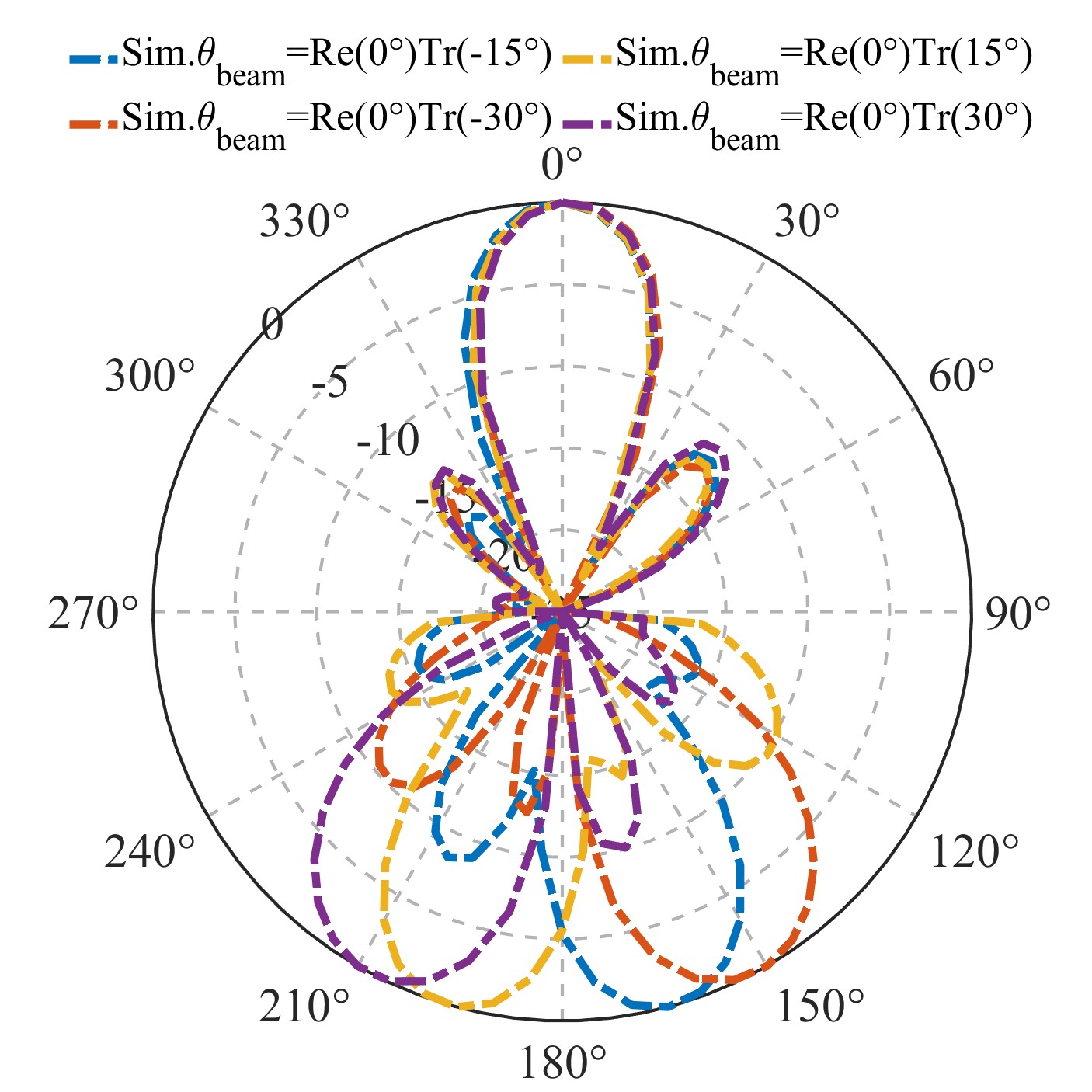}}
\par\end{centering}
\begin{raggedright}
\hspace*{0.2\columnwidth} (a)\hspace*{0.5\columnwidth} (c)
\par\end{raggedright}
\begin{centering}
\textsf{\includegraphics[width=0.5\columnwidth]{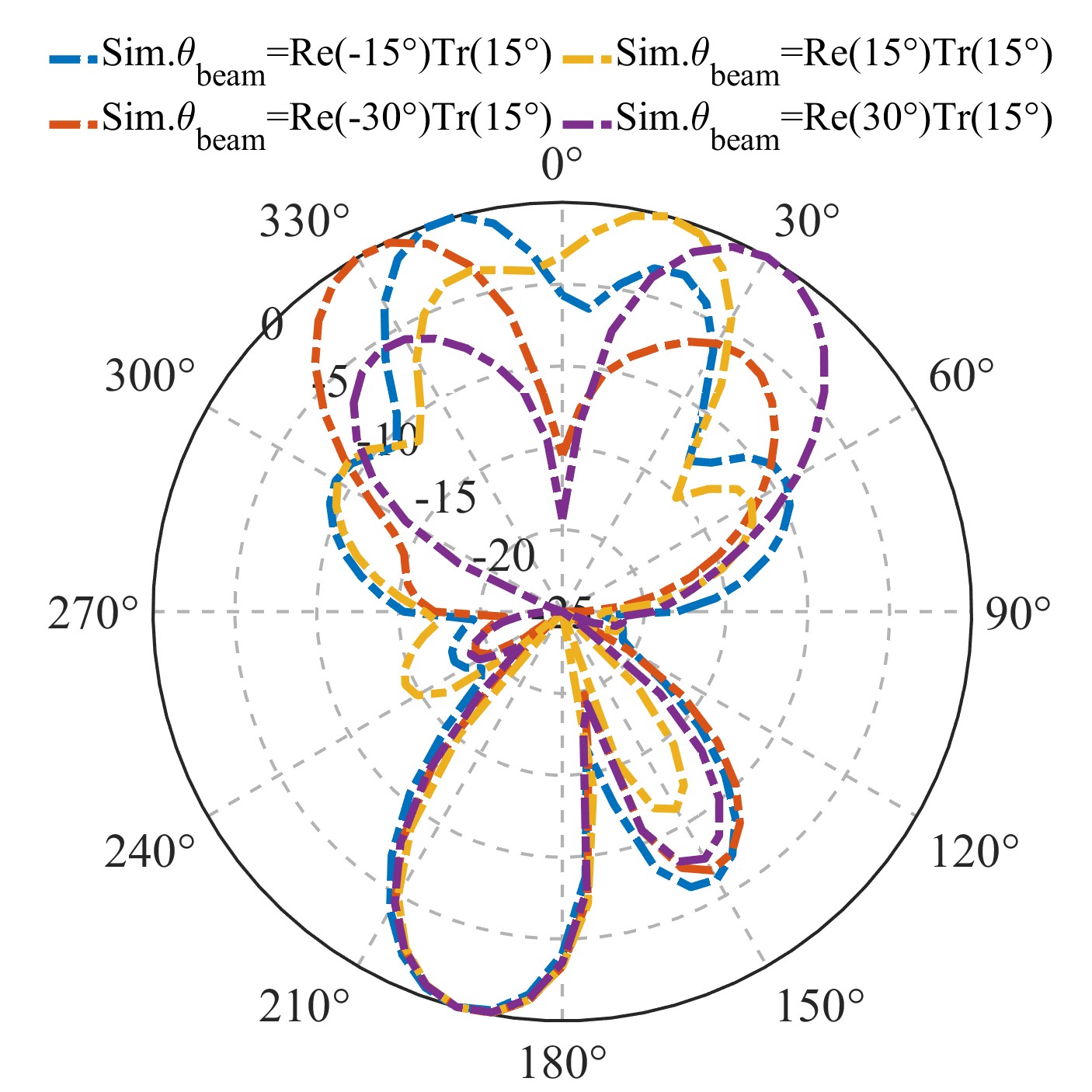}\hspace*{0.03\columnwidth}\includegraphics[width=0.5\columnwidth]{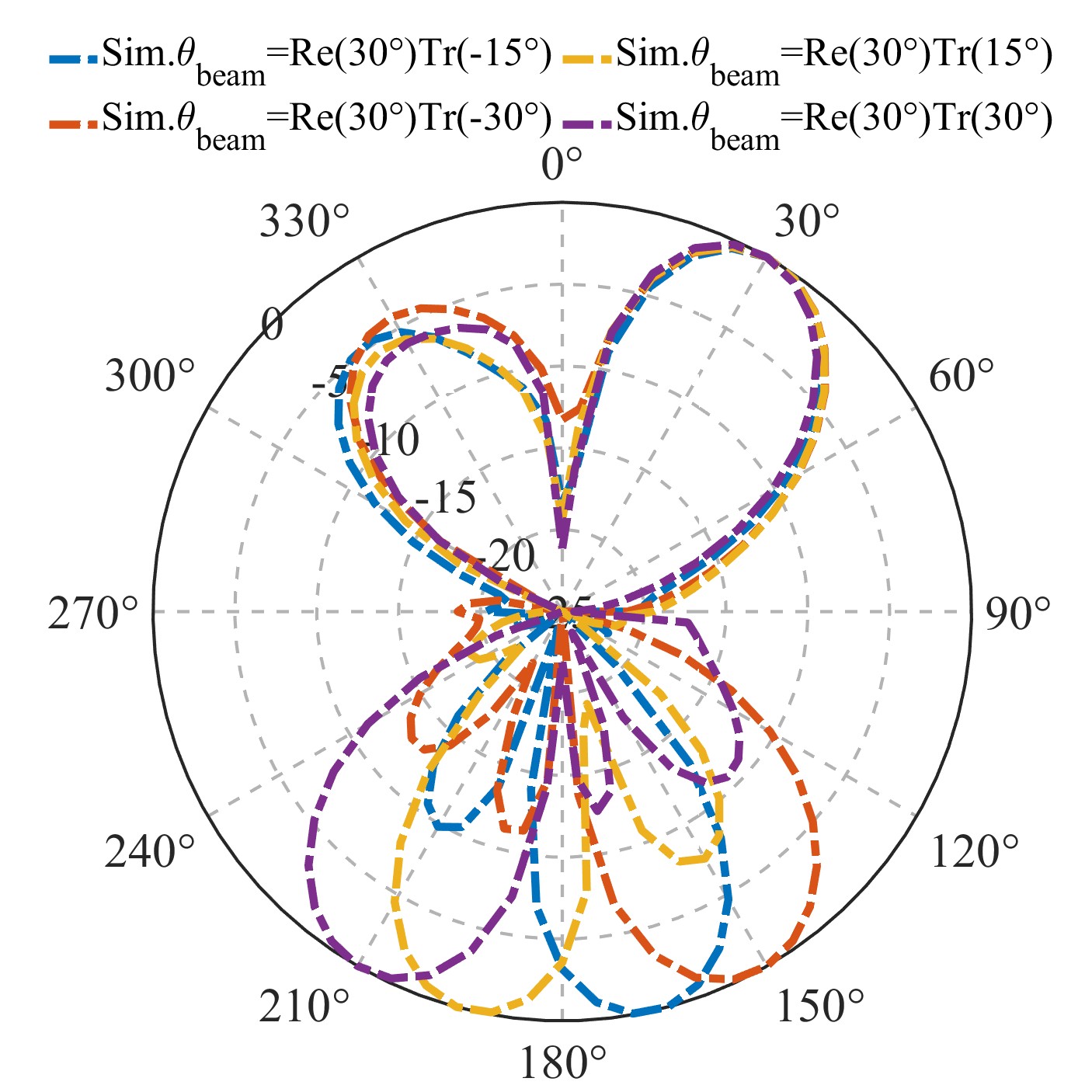}}
\par\end{centering}
\begin{raggedright}
\hspace*{0.2\columnwidth} (b)\hspace*{0.5\columnwidth} (d)
\par\end{raggedright}
\caption{Simulated normalized scattered wave pattern in the YOZ plane at 2.4
GHz with the incident wave perpendicular to the BD-RIS when the BD-RIS
operates in hybrid mode. (a) Scanning reflected scattered patterns
$\left(\theta,\phi\right)_{\mathrm{Re,beam}}=\left(-30^{\circ}\sim30^{\circ},90^{\circ}\right)$
with fixed transmitted scattered patterns $\left(\theta,\phi\right)_{\mathrm{Tr,beam}}=\left(180^{\circ},90^{\circ}\right)$.
(b) Scanning reflected scattered patterns $\left(\theta,\phi\right)_{\mathrm{Re,beam}}=\left(-30^{\circ}\sim30^{\circ},90^{\circ}\right)$
with fixed transmitted scattered patterns $\left(\theta,\phi\right)_{\mathrm{Tr,beam}}=\left(-165^{\circ},90^{\circ}\right)$.
(c) Scanning transmitted scattered patterns $\left(\theta,\phi\right)_{\mathrm{Tr,beam}}=\left(-150^{\circ}\sim150^{\circ},90^{\circ}\right)$
with fixed reflected scattered patterns $\left(\theta,\phi\right)_{\mathrm{Re,beam}}=\left(0^{\circ},90^{\circ}\right)$.
(d) Scanning transmitted scattered patterns $\left(\theta,\phi\right)_{\mathrm{Tr,beam}}=\left(-150^{\circ}\sim150^{\circ},90^{\circ}\right)$
with fixed reflected scattered patterns $\left(\theta,\phi\right)_{\mathrm{Re,beam}}=\left(30^{\circ},90^{\circ}\right)$.}
\label{optimization results-2}
\end{figure}

The optimized patterns for reflection mode and transmission mode with
$-45^{\circ}$ to $45^{\circ}$ scanning range in the YOZ plane are
presented in Fig. \ref{optimization results-1}(a)-(b) and (c)-(d),
respectively. To verify the independent control over reflected and
transmitted beams when operating in hybrid mode, the transmitted beams
are fixed toward $180^{\circ}$ or $-165^{\circ}$ when scanning the
reflected beams from $-30^{\circ}$ to $30^{\circ}$ as shown in Fig.
\ref{optimization results-2}(a)-(b); the reflected beams are fixed
toward $0^{\circ}$ and $30^{\circ}$ when scanning the transmitted
beams from $-150^{\circ}$ to $150^{\circ}$ as shown in Fig. \ref{optimization results-2}(c)-(d).
As illustrated in Section III, the effective reflection phase control
of one cell is 1-bit, less than the 2-bit phase reconfigurability
of transmission phase. This makes the reflected beams all feature
wider beamwidth and higher sidelobes than the transmitted ones. These
discrepancies can be improved by employing more powerful phase reconfigurable
antennas with higher resolution and scaling up the array scale of
the BD-RIS. Besides, It should be noted that, due to the symmetric
characteristic of the BD-RIS with 4$\times$4 cells, its optimized
scattering performance in the XOZ plane is very similar to that of
the YOZ plane. Therefore, the above results demonstrate that the 4$\times$4
BD-RIS can realize dynamic mode switching with independent beam control.

\section{Experimental Results}

\begin{figure}[t]
\begin{centering}
\textsf{\includegraphics[width=0.9\columnwidth]{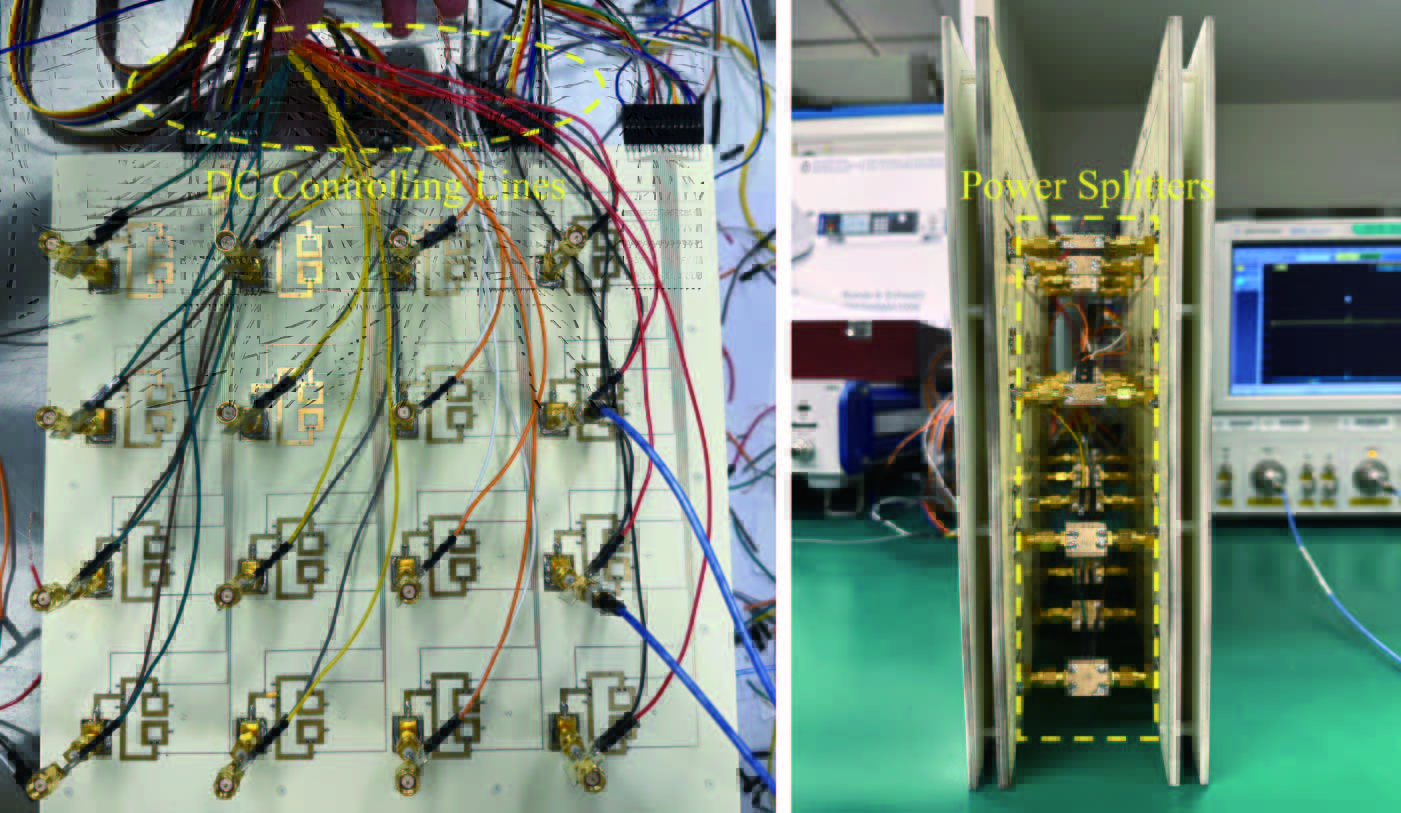}\hspace*{0\columnwidth}}
\par\end{centering}
\begin{raggedright}
\hspace*{0.25\columnwidth} (a)\hspace*{0.45\columnwidth} (b)
\par\end{raggedright}
\begin{centering}
\textsf{\includegraphics[width=1\columnwidth]{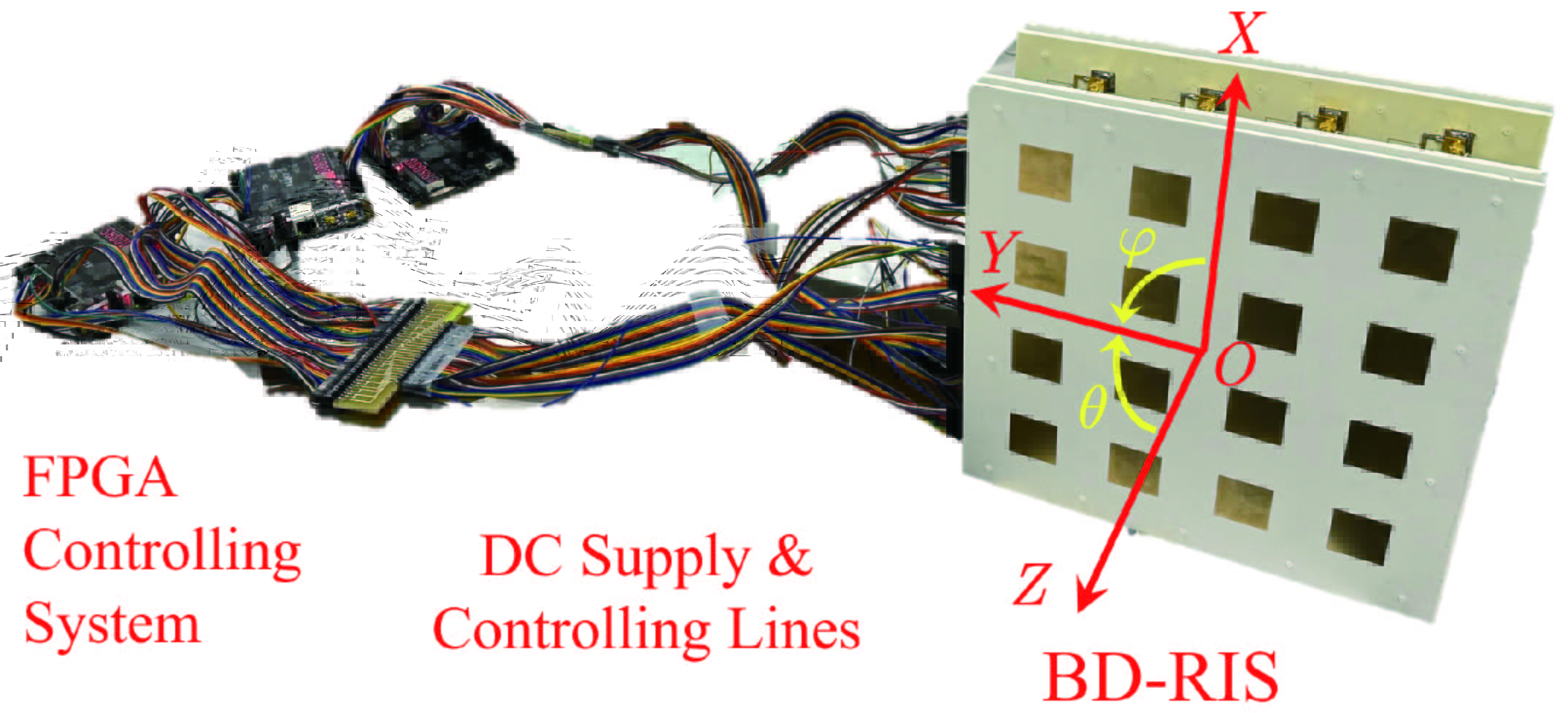}}
\par\end{centering}
\begin{raggedright}
\hspace*{0.5\columnwidth} (c)
\par\end{raggedright}
\caption{Photograph of the prototype of the hybrid transmitting and reflecting
BD-RIS with 4$\mathbf{\times}$4 cells and the controlling systems.
(a) Inner view. (b) Side view. (c) Overall view.}
\label{STAR-RIS control systems}
\end{figure}

To experimentally verify the proposed BD-RIS design, we fabricated
a prototype of the BD-RIS with 4$\mathbf{\times}$4 cells as shown
in Fig. \ref{STAR-RIS control systems}, where the DC control circuits
are also integrated with the reflecting and transmitting antenna arrays,
respectively. As illustrated in Section IV.C, we need 5 I/O pins to
control one 2-bit antenna. Thus, for the BD-RIS with 4$\mathbf{\times}$4$\mathbf{\times}$2
antennas, we need in total 160 pins and then three FPGAs AX7035, each
of which has 64 I/O pins, are used to configure the states of all
the diodes to realize the desired beamforming designs. Detailed experimental
setups and results are provided as follows.

\begin{figure}[t]
\begin{centering}
\textsf{\includegraphics[width=0.95\columnwidth]{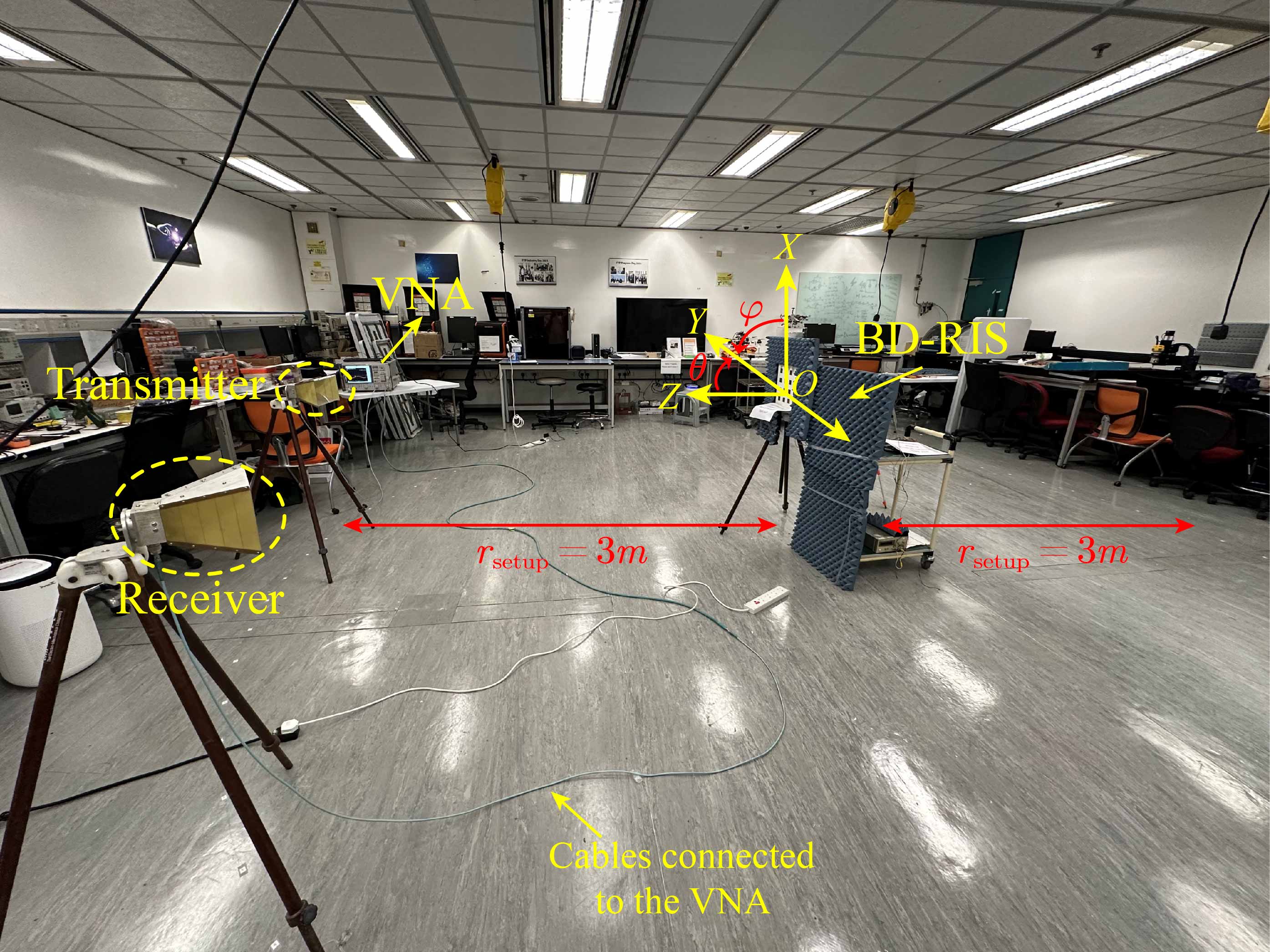}}
\par\end{centering}
\caption{Photograph of the experiment setup for measuring the reflected and
transmitted radiation patterns of the hybrid transmitting and reflecting
BD-RIS. The configuration of the coordinate system and its origin
O are also shown.}
\label{testbed}
\end{figure}

\subsection{Experimental Setup}

To measure the reflected and transmitted fields of the BD-RIS, we
set up a testbed with a circular space of 6 m diameter as shown in
Fig. \ref{testbed}. Two horn antennas were located 3 m away from
the fabricated BD-RIS, and situated within its far-field region. One
horn antenna was used as the transmitter offering a vertically incident
plane wave, while the other acted as a receiver to test the scattered
signals by the BD-RIS. A vector network analyzer, Rohde \& Schwarz
ZVA40, was utilized to measure the $S_{21}$ parameter between the
transmitter and receiver. By fixing the transmitter and rotating the
receiver at 5$^{\circ}$ intervals, we can measure the scattered fields
of the BD-RIS over all angles.

To avoid repeatedly moving the fabricated BD-RIS and its controlling
devices, while still considering the elimination of environmental
effects, we propose an efficient testing method based on structural
and antenna scattering as described in Section IV.C. That is we obtain

\begin{equation}
S_{21,\mathrm{scat}}^{\mathrm{ant}}\left(\Omega\right)=S_{21,\mathrm{total}}\left(\Omega\right)-S_{21,\mathrm{total}}^{\mathrm{str}}\left(\Omega\right),\label{scattered pattern without structural mode}
\end{equation}
instead of measuring an environmental $S_{21}\left(\Omega\right)$
as in the experimental sections of \cite{rao2022novel,rao2023active},
we can keep the BD-RIS stationary and measure and subtract the corresponding
structural scattering part for the reflected or transmitted waves.
This is because the subtracted total structural part already includes
the effects of the environment. In this way, the left term $S_{21,\mathrm{scat}}^{\mathrm{ant}}\left(\Omega\right)$
corresponds to the pure antenna scattering part. It is also consistent
with our theoretical calculation by \eqref{reflected fields} and
\eqref{transmitted fields}. As defined, the structural scattering
part refers to the scattered pattern when the antenna is conjugate-matched.
In the experiments, for the reflected signals, the situation where
all the reflecting antennas are conjugate-matched corresponds to the
situation where all the power splitters operate in transmission mode.
For the transmitted waves, the situation where all the transmitting
antennas are conjugate-matched corresponds to the situation where
all the power splitters operate in reflection mode.

\subsection{Measurement Results}

\begin{figure}[t]
\begin{centering}
\textsf{\includegraphics[width=0.5\columnwidth]{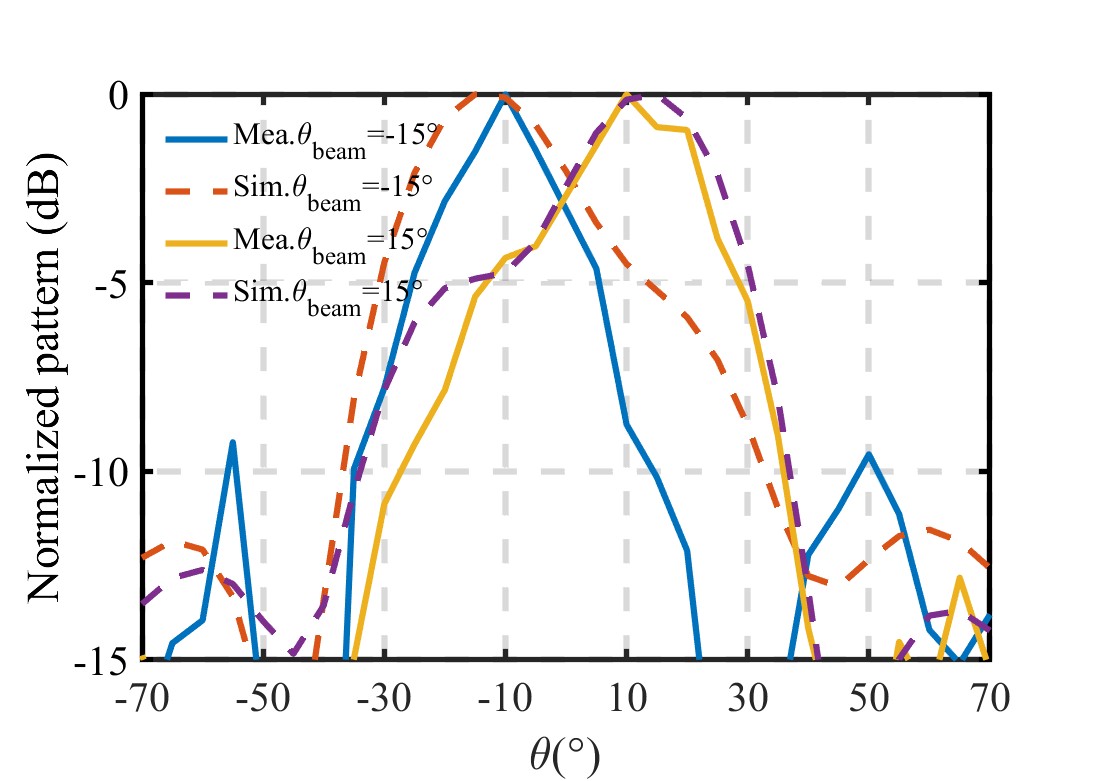}\hspace*{0\columnwidth}\includegraphics[width=0.5\columnwidth]{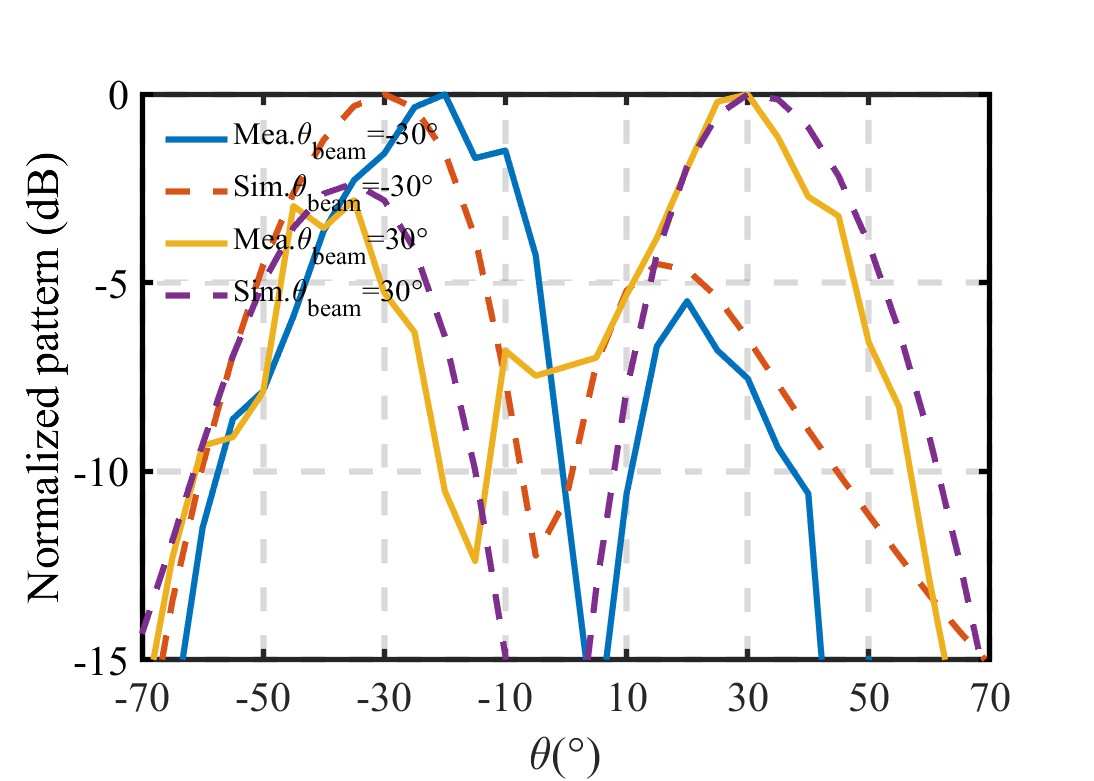}}
\par\end{centering}
\begin{raggedright}
\hspace*{0.2\columnwidth} (a)\hspace*{0.5\columnwidth} (b)
\par\end{raggedright}
\begin{centering}
\textsf{\includegraphics[width=0.5\columnwidth]{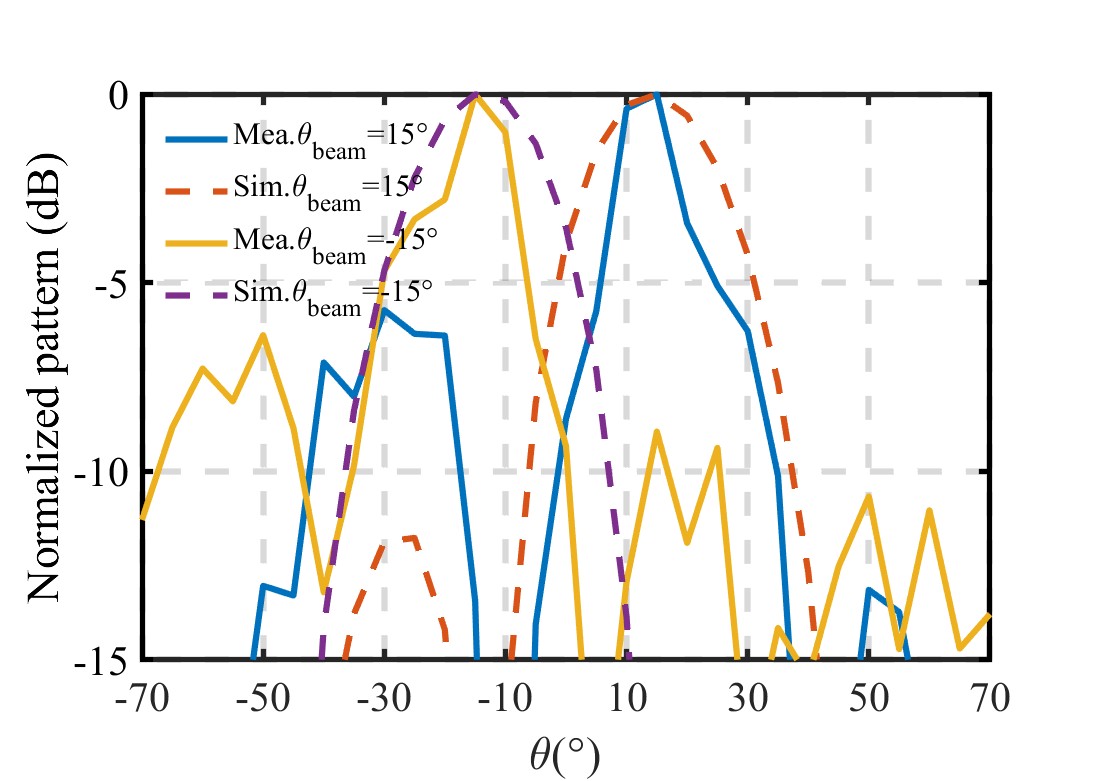}\hspace*{0\columnwidth}\includegraphics[width=0.5\columnwidth]{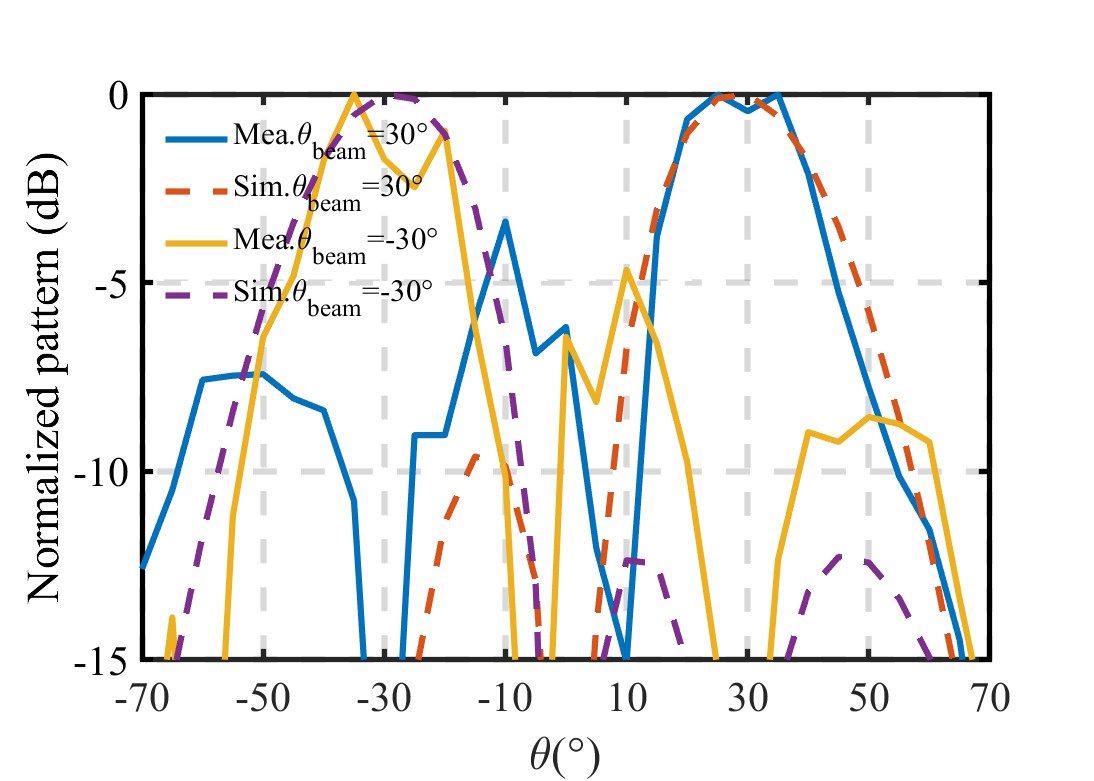}}
\par\end{centering}
\begin{raggedright}
\hspace*{0.2\columnwidth} (c)\hspace*{0.5\columnwidth} (d)
\par\end{raggedright}
\caption{Measured and simulated normalized scattered wave pattern in the YOZ
plane at 2.4 GHz with the incident wave perpendicular to the BD-RIS
when the BD-RIS operates in (a) Reflection mode with $\left(\theta,\phi\right)_{\mathrm{Re,beam}}=\left(\pm15^{\circ},90^{\circ}\right)$,
(b) Reflection mode with $\left(\theta,\phi\right)_{\mathrm{Re,beam}}=\left(\pm30^{\circ},90^{\circ}\right)$,
(c) Transmission mode with $\left(\theta,\phi\right)_{\mathrm{Tr,beam}}=\left(\pm165^{\circ},90^{\circ}\right)$,
(d) Transmission mode with $\left(\theta,\phi\right)_{\mathrm{Tr,beam}}=\left(\pm150^{\circ},90^{\circ}\right)$.}
\label{pure modes of STAR-RIS}
\end{figure}

\begin{figure}[t]
\begin{centering}
\textsf{\includegraphics[width=0.5\columnwidth]{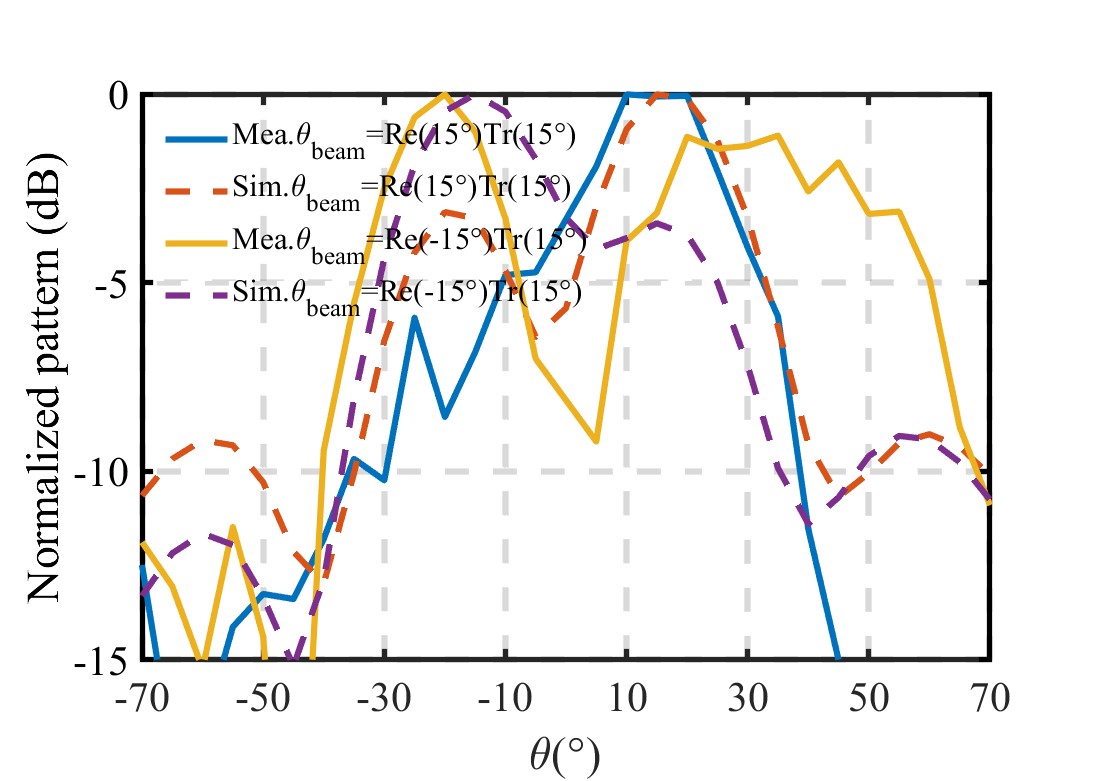}\hspace*{0.03\columnwidth}\includegraphics[width=0.5\columnwidth]{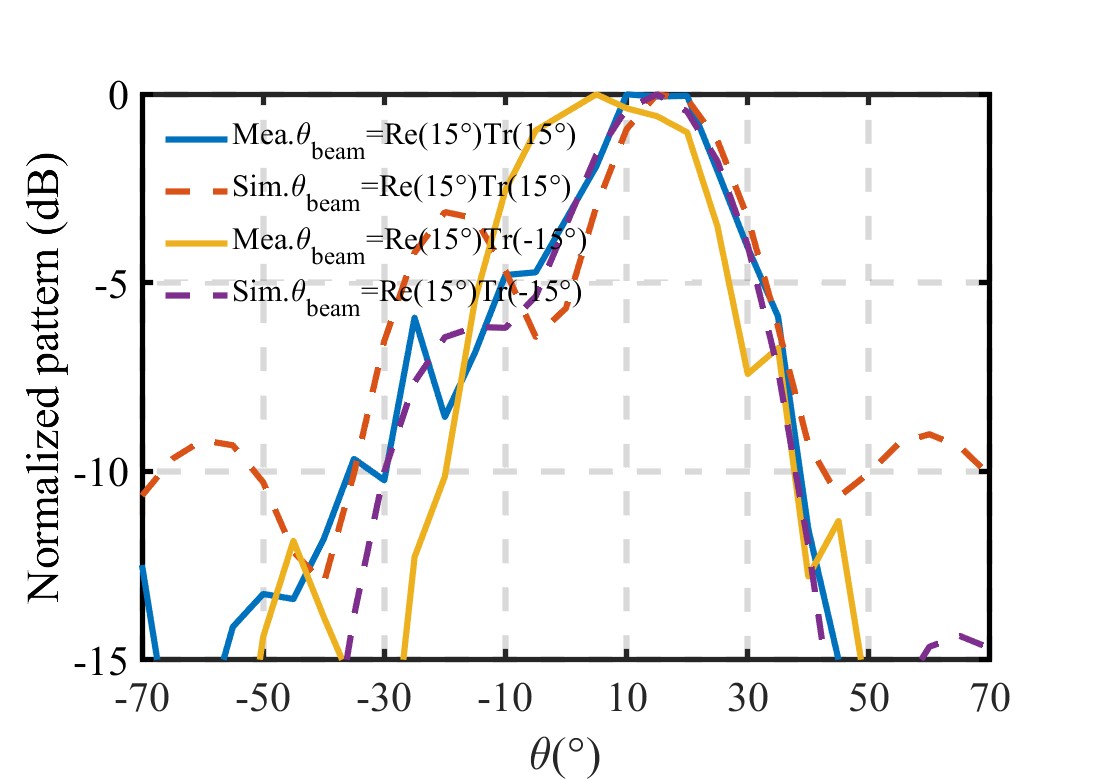}}
\par\end{centering}
\begin{raggedright}
\hspace*{0.2\columnwidth} (a)\hspace*{0.5\columnwidth} (c)
\par\end{raggedright}
\begin{centering}
\textsf{\includegraphics[width=0.5\columnwidth]{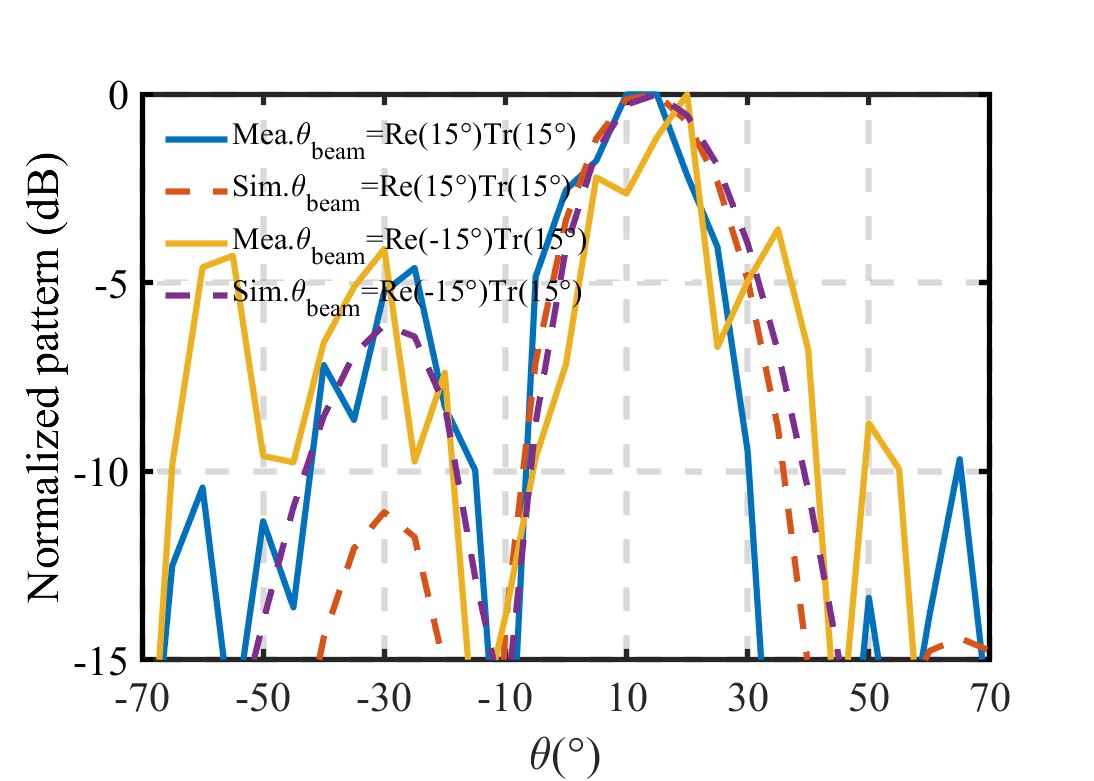}\hspace*{0.03\columnwidth}\includegraphics[width=0.5\columnwidth]{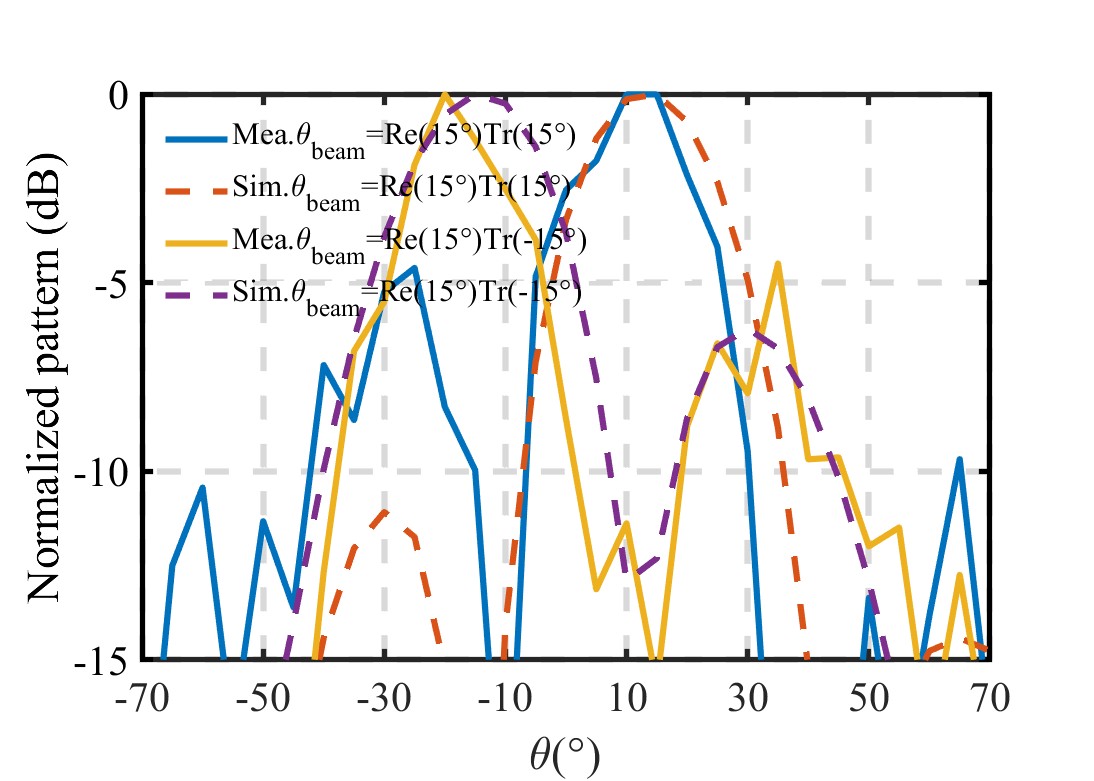}}
\par\end{centering}
\begin{raggedright}
\hspace*{0.2\columnwidth} (b)\hspace*{0.5\columnwidth} (d)
\par\end{raggedright}
\caption{Measured and simulated normalized scattered wave pattern in the YOZ
plane at 2.4 GHz with the incident wave perpendicular to the BD-RIS
when the BD-RIS operates in hybrid mode (a) Reflected scattered patterns
when $\left(\theta,\phi\right)_{\mathrm{Re,beam}}=\left(\pm15^{\circ},90^{\circ}\right)$
and $\left(\theta,\phi\right)_{\mathrm{Tr,beam}}=\left(165^{\circ},90^{\circ}\right)$,
(b) Transmitted scattered patterns when $\left(\theta,\phi\right)_{\mathrm{Re,beam}}=\left(\pm15^{\circ},90^{\circ}\right)$
and $\left(\theta,\phi\right)_{\mathrm{Tr,beam}}=\left(165^{\circ},90^{\circ}\right)$,
(c) Reflected scattered patterns when $\left(\theta,\phi\right)_{\mathrm{Re,beam}}=\left(15^{\circ},90^{\circ}\right)$
and $\left(\theta,\phi\right)_{\mathrm{Tr,beam}}=\left(\pm165^{\circ},90^{\circ}\right)$,
(d) Transmitted scattered patterns when $\left(\theta,\phi\right)_{\mathrm{Re,beam}}=\left(15^{\circ},90^{\circ}\right)$
and $\left(\theta,\phi\right)_{\mathrm{Tr,beam}}=\left(\pm165^{\circ},90^{\circ}\right)$.}
\label{hybrid mode of STAR-RIS}
\end{figure}
\begin{table*}[tp]
\centering{}\caption{Comparison with related work}
\label{table 2}%
\begin{tabular}{ccccccccc}
\toprule 
\addlinespace
{\footnotesize Ref.} & \begin{cellvarwidth}[m]
\centering
{\footnotesize Frequency}{\footnotesize\par}

{\footnotesize (GHz)}
\end{cellvarwidth} & \begin{cellvarwidth}[m]
\centering
{\footnotesize Reconfigurable}{\footnotesize\par}

{\footnotesize design}
\end{cellvarwidth} & \begin{cellvarwidth}[m]
\centering
{\footnotesize Reflection}{\footnotesize\par}

{\footnotesize mode}
\end{cellvarwidth} & \begin{cellvarwidth}[m]
\centering
{\footnotesize Hybrid}{\footnotesize\par}

{\footnotesize mode}
\end{cellvarwidth} & \begin{cellvarwidth}[m]
\centering
{\footnotesize Transmission}{\footnotesize\par}

{\footnotesize mode}
\end{cellvarwidth} & \begin{cellvarwidth}[m]
\centering
{\footnotesize Polarization}{\footnotesize\par}

{\footnotesize conversion}
\end{cellvarwidth} & \begin{cellvarwidth}[m]
\centering
{\footnotesize Incident wave}{\footnotesize\par}

{\footnotesize range}
\end{cellvarwidth} & \begin{cellvarwidth}[m]
\centering
{\footnotesize Excitation}{\footnotesize\par}

{\footnotesize method}
\end{cellvarwidth}\tabularnewline\addlinespace
\midrule
\addlinespace
{\footnotesize\cite{duan2024non}} & {\footnotesize 11.5} & {\footnotesize No} & \multirow{1}{*}{{\footnotesize Fixed}} & {\footnotesize Fixed} & {\footnotesize Fixed} & {\footnotesize Required} & {\footnotesize Single side} & {\footnotesize Near-field}\tabularnewline
{\footnotesize\cite{li2022transmission}} & {\footnotesize 4.75} & {\footnotesize No} & \multirow{1}{*}{{\footnotesize Fixed}} & {\footnotesize Fixed} & {\footnotesize Fixed} & {\footnotesize Required} & {\footnotesize Single side} & {\footnotesize Near-field}\tabularnewline
{\footnotesize\cite{yin2024highly}} & {\footnotesize 5} & {\footnotesize Yes} & {\footnotesize 1-bit} & / & {\footnotesize 1-bit} & {\footnotesize Not required} & {\footnotesize Single side} & {\footnotesize Near-field}\tabularnewline
{\footnotesize\cite{yu2022reconfigurable}} & {\footnotesize 9.5} & {\footnotesize Yes} & {\footnotesize 1-bit} & / & {\footnotesize 1-bit} & {\footnotesize Required} & {\footnotesize Single side} & {\footnotesize Near-field}\tabularnewline
{\footnotesize\cite{xiang2023simultaneous}} & {\footnotesize 5.8} & {\footnotesize Yes} & / & {\footnotesize 1-bit (Re)\&1-bit (Tr)} & / & {\footnotesize Required} & {\footnotesize Single side} & {\footnotesize Near-field}\tabularnewline
{\footnotesize This work} & {\footnotesize 2.4} & {\footnotesize Yes} & {\footnotesize 1-bit} & {\footnotesize 1-bit (Re)\&2-bit (Tr)} & {\footnotesize 2-bit} & {\footnotesize Not required} & {\footnotesize Dual sides} & {\footnotesize Far-field}\tabularnewline
\midrule 
\multicolumn{9}{l}{{\scriptsize Re: Reflection, Tr: Transmission.}}\tabularnewline
\end{tabular}
\end{table*}

Fig. \ref{pure modes of STAR-RIS}(a)-(b) show the simulated and measured
scattered pattern in the YOZ plane with four representative directions
of reflected beams when the BD-RIS is operating in reflection mode
with a vertically incident plane wave excitation. Fig. \ref{pure modes of STAR-RIS}(c)-(d)
refers to the situation when the BD-RIS operates in transmission mode
with four transmitted beams. It can be observed that the measured
reflected and transmitted patterns generally agree with the simulated
results utilizing \eqref{reflected fields} and \eqref{transmitted fields},
respectively. The small disagreement is due to the non-ideal wave
excitation, background scattering, and fabrication and positioning
errors. Besides, it can be observed that, the measured reflected beams
also feature wider beamwidth and higher sidelobes than the transmitted
ones, which align with the fact that the effective reflection phase
reconfigurability is less than that of transmission phase. Anyhow,
the fabricated BD-RIS can realize reflected and transmitted beam steering
from $\mathbf{-}$30$^{\circ}$ to 30$^{\circ}$ in the reflection
and transmission space of the YOZ plane in reflection mode and transmission
mode, respectively. While in this experiment our purpose is to demonstrate
the viability of beam steering, wider beam steering angles will be
able to be achieved using BD-RIS with more cells.

The key result for our design is the demonstration of the hybrid mode.
Fig. \ref{hybrid mode of STAR-RIS} showcases the independent beam
steering capabilities of the reflected and transmitted fields when
the BD-RIS operates in hybrid mode. To illustrate, the left column
depicts the case when the reflected beam is directed toward $\mathbf{-}$15$^{\circ}$
and 15$^{\circ}$ in the reflection space shown in Fig. \ref{pure modes of STAR-RIS}(a),
while the transmitted beam is fixed toward 15$^{\circ}$ in the transmission
space shown in Fig. \ref{pure modes of STAR-RIS}(b). Conversely,
the right column depicts the case when the reflected beam is fixed
toward 15$^{\circ}$ as shown in Fig. \ref{pure modes of STAR-RIS}(c),
while the transmitted beam is directed toward $\mathbf{-}$15$^{\circ}$
and 15$^{\circ}$ as shown in Fig. \ref{pure modes of STAR-RIS}(d).
Similarly, the beam steering performance of the transmitted pattern
is still better than that of the reflected pattern in terms of thinner
beamwidth and lower sidelobes. One reason for the discrepancies in
the sidelobes of reflected beams in Fig. \ref{pure modes of STAR-RIS}(a)
is that the loaded impedance of the reflecting antenna array is not
exactly 50 $\Omega$ and therefore the structural scattering pattern
of the reflected waves is not accurately modeled. Overall however
these results provide solid validation for the mode switching and
independent beam steering capabilities of the proposed BD-RIS.

\subsection{Comparison with Related Work}

Table \ref{table 2} compares the performance of the proposed hybrid
transmitting and reflecting BD-RIS with related work. Although all
the listed previous works \cite{li2022transmission,xiang2023simultaneous,yu2022reconfigurable,yin2024highly,duan2024non}
have avoided using the frequency multiplexing technique to realize
bidirectional functionalities, some of them still suffer from the
limitations of being non-reconfigurable \cite{li2022transmission,duan2024non},
or involving polarization conversion \cite{li2022transmission,xiang2023simultaneous,yu2022reconfigurable}.
More importantly, for the reconfigurable designs, \cite{xiang2023simultaneous,yu2022reconfigurable,yin2024highly},
they can only realize the hybrid mode, or reflection or transmission
mode, but cannot directly switch between the three modes and ensure
independent control of the reflected and transmitted beams. In addition,
none of the abovementioned works feature a symmetrical structure,
which restricts their applications in real-life wireless communication
environments where waves are incident from both sides. In our work,
to overcome this limitation, we separate the phase and power splitting
reconfigurability control by designing a 2-bit phase reconfigurable
antenna and a tunable two-port power splitter. The interconnection
of the identical polarization ports of the two 2-bit antennas with
the two-port power splitter not only enables independent beam control
and power splitting, but also keeps the polarization of reflected
and transmitted waves unchanged. Furthermore, this configuration ensures
a symmetrical structure. Based on our proposed architecture, the proposed
BD-RIS cell can also be extended into a dual-polarized version, by
using a dual-polarized phase reconfigurable antenna with two sets
of two-port power splitter. This still secures independent control
of each polarization, thus increasing polarization diversity.

\section{Conclusions}

A reconfigurable and symmetric hybrid transmitting and reflecting
BD-RIS design has been proposed, which boasts independent beam steering
and mode switching capabilities among reflection, transmission and
hybrid modes in the same aperture, frequency band and polarization.
Based on scattering matrix analysis and guided by a novel integrated
scheme, the hybrid transmitting and reflecting BD-RIS comprises two
reconfigurable antenna arrays interconnected by an array of tunable
two-port power splitters. The two-port power splitter is built upon
a varactor parallel with a bias inductor, able to realize mode switching
with less than 1 dB insertion loss. To integrate the phase shifter
with an antenna, a 2-bit phase reconfigurable antenna with 200 MHz
bandwidth around 2.4 GHz, stable radiation patterns among four states
and evenly distributed phases over 360$^{\circ}$ has been promoted
to serve as the building block for those two reconfigurable antenna
arrays facing toward the reflection and transmission spaces. To characterize
and optimize the electromagnetic response of the proposed BD-RIS design,
a Thévenin equivalent physical model and corresponding analytical
method have been described. A prototype of the proposed BD-RIS with
4$\mathbf{\times}$4 cells was fabricated and measured. Simulation
and experimental results have successfully verified the multifunctionalities
of the hybrid transmitting and reflecting BD-RIS design. This work
helps fill the gap between realizing practical hardware design and
establishing accurate physical model for the hybrid transmitting and
reflecting BD-RIS, enabling hybrid transmitting and reflecting BD-RIS
assisted wireless communications.

\appendices{}
\begin{lyxlist}{00.00.0000}
\item [{}]~
\end{lyxlist}
\bibliographystyle{IEEEtran}
\bibliography{STAR_RIS}

\begin{thebibliography}{10}
\providecommand{\url}[1]{#1}
\csname url@samestyle\endcsname
\providecommand{\newblock}{\relax}
\providecommand{\bibinfo}[2]{#2}
\providecommand{\BIBentrySTDinterwordspacing}{\spaceskip=0pt\relax}
\providecommand{\BIBentryALTinterwordstretchfactor}{4}
\providecommand{\BIBentryALTinterwordspacing}{\spaceskip=\fontdimen2\font plus
\BIBentryALTinterwordstretchfactor\fontdimen3\font minus
  \fontdimen4\font\relax}
\providecommand{\BIBforeignlanguage}[2]{{%
\expandafter\ifx\csname l@#1\endcsname\relax
\typeout{** WARNING: IEEEtran.bst: No hyphenation pattern has been}%
\typeout{** loaded for the language `#1'. Using the pattern for}%
\typeout{** the default language instead.}%
\else
\language=\csname l@#1\endcsname
\fi
#2}}
\providecommand{\BIBdecl}{\relax}
\BIBdecl

\bibitem{saad2019vision}
W.~Saad, M.~Bennis, and M.~Chen, ``A vision of 6g wireless systems:
  Applications, trends, technologies, and open research problems,'' \emph{IEEE
  Network}, vol.~34, no.~3, pp. 134--142, 2019.

\bibitem{wang2023road}
C.-X. Wang, X.~You, X.~Gao, X.~Zhu, Z.~Li, C.~Zhang, H.~Wang, Y.~Huang,
  Y.~Chen, H.~Haas \emph{et~al.}, ``On the road to 6g: Visions, requirements,
  key technologies, and testbeds,'' \emph{IEEE Commun. Surv. Tutorials},
  vol.~25, no.~2, pp. 905--974, 2023.

\bibitem{DiRenzo2020}
M.~Di~Renzo, A.~Zappone, M.~Debbah, M.-S. Alouini, C.~Yuen, J.~de~Rosny, and
  S.~Tretyakov, ``Smart radio environments empowered by reconfigurable
  intelligent surfaces: How it works, state of research, and the road ahead,''
  \emph{IEEE J. Sel. Areas Commun.}, vol.~38, no.~11, pp. 2450--2525, 2020.

\bibitem{wu2019towards}
Q.~Wu and R.~Zhang, ``Towards smart and reconfigurable environment: Intelligent
  reflecting surface aided wireless network,'' \emph{IEEE Commun. Mag.},
  vol.~58, no.~1, pp. 106--112, 2019.

\bibitem{wu2021intelligent}
Q.~Wu, S.~Zhang, B.~Zheng, C.~You, and R.~Zhang, ``Intelligent reflecting
  surface-aided wireless communications: A tutorial,'' \emph{IEEE Trans.
  Commun.}, vol.~69, no.~5, pp. 3313--3351, 2021.

\bibitem{li2022beyond}
H.~Li, S.~Shen, and B.~Clerckx, ``Beyond diagonal reconfigurable intelligent
  surfaces: From transmitting and reflecting modes to single-, group-, and
  fully-connected architectures,'' \emph{IEEE Trans. Wireless Commun.},
  vol.~22, no.~4, pp. 2311--2324, 2022.

\bibitem{li2023reconfigurable}
H.~Li, S.~Shen, M.~Nerini, and B.~Clerckx, ``Reconfigurable intelligent
  surfaces 2.0: Beyond diagonal phase shift matrices,'' \emph{IEEE Commun.
  Mag.}, 2023.

\bibitem{zhou2023optimizing}
Y.~Zhou, Y.~Liu, H.~Li, Q.~Wu, S.~Shen, and B.~Clerckx, ``Optimizing power
  consumption, energy efficiency, and sum-rate using beyond diagonal ris a
  unified approach,'' \emph{IEEE Trans. Wireless Commun.}, vol.~23, no.~7, pp.
  7423--7438, 2023.

\bibitem{li2024beyond}
H.~Li, S.~Shen, M.~Nerini, M.~Di~Renzo, and B.~Clerckx, ``Beyond diagonal
  reconfigurable intelligent surfaces with mutual coupling: Modeling and
  optimization,'' \emph{IEEE Commun. Lett.}, vol.~28, no.~4, pp. 937--941,
  2024.

\bibitem{xu2021star}
J.~Xu, Y.~Liu, X.~Mu, and O.~A. Dobre, ``Star-riss: Simultaneous transmitting
  and reflecting reconfigurable intelligent surfaces,'' \emph{IEEE Commun.
  Lett.}, vol.~25, no.~9, pp. 3134--3138, 2021.

\bibitem{Zhang2020}
S.~Zhang, H.~Zhang, B.~Di, Y.~Tan, Z.~Han, and L.~Song, ``Beyond intelligent
  reflecting surfaces: Reflective-transmissive metasurface aided communications
  for full-dimensional coverage extension,'' \emph{IEEE Trans. Veh. Technol.},
  vol.~69, no.~11, pp. 13\,905--13\,909, 2020.

\bibitem{mu2021simultaneously}
X.~Mu, Y.~Liu, L.~Guo, J.~Lin, and R.~Schober, ``Simultaneously transmitting
  and reflecting (star) ris aided wireless communications,'' \emph{IEEE Trans.
  Wireless Commun.}, vol.~21, no.~5, pp. 3083--3098, 2021.

\bibitem{zhang2022intelligent}
H.~Zhang and B.~Di, ``Intelligent omni-surfaces: Simultaneous refraction and
  reflection for full-dimensional wireless communications,'' \emph{IEEE Commun.
  Surv. Tutorials}, vol.~24, no.~4, pp. 1997--2028, 2022.

\bibitem{ahmed2023survey}
M.~Ahmed, A.~Wahid, S.~S. Laique, W.~U. Khan, A.~Ihsan, F.~Xu, S.~Chatzinotas,
  and Z.~Han, ``A survey on star-ris: Use cases, recent advances, and future
  research challenges,'' \emph{IEEE Internet Things J.}, vol.~10, no.~16, pp.
  14\,689--14\,711, 2023.

\bibitem{niu2021weighted}
H.~Niu, Z.~Chu, F.~Zhou, P.~Xiao, and N.~Al-Dhahir, ``Weighted sum rate
  optimization for star-ris-assisted mimo system,'' \emph{IEEE Trans. Veh.
  Technol.}, vol.~71, no.~2, pp. 2122--2127, 2021.

\bibitem{katwe2023improved}
M.~Katwe, K.~Singh, B.~Clerckx, and C.-P. Li, ``Improved spectral efficiency in
  star-ris aided uplink communication using rate splitting multiple access,''
  \emph{IEEE Trans. Wireless Commun.}, vol.~22, no.~8, pp. 5365--5382, 2023.

\bibitem{wang2021reconfigurable}
H.~L. Wang, H.~F. Ma, M.~Chen, S.~Sun, and T.~J. Cui, ``A reconfigurable
  multifunctional metasurface for full-space control of electromagnetic
  waves,'' \emph{Adv. Funct. Mater.}, vol.~31, no.~25, p. 2100275, 2021.

\bibitem{bao2021programmable}
L.~Bao, Q.~Ma, R.~Y. Wu, X.~Fu, J.~Wu, and T.~J. Cui, ``Programmable
  reflection--transmission shared-aperture metasurface for real-time control of
  electromagnetic waves in full space,'' \emph{Adv. Sci.}, vol.~8, no.~15, p.
  2100149, 2021.

\bibitem{hu2022intelligent}
Q.~Hu, J.~Zhao, K.~Chen, K.~Qu, W.~Yang, J.~Zhao, T.~Jiang, and Y.~Feng, ``An
  intelligent programmable omni-metasurface,'' \emph{Laser Photonics Rev.},
  vol.~16, no.~6, p. 2100718, 2022.

\bibitem{yin2024reconfigurable}
T.~Yin, J.~Ren, B.~Zhang, P.~Li, Y.~Luan, and Y.~Yin, ``Reconfigurable
  transmission-reflection-integrated coding metasurface for full-space
  electromagnetic wavefront manipulation,'' \emph{Adv. Opt. Mater.}, vol.~12,
  no.~2, p. 2301326, 2024.

\bibitem{duan2024non}
K.~Duan, K.~Chen, T.~Jiang, J.~Zhao, and Y.~Feng, ``Non-interleaved
  bidirectional metasurface for spatial energy allocation,'' \emph{IEEE Trans.
  Antennas Propag.}, 2024.

\bibitem{yin2024highly}
T.~Yin, J.~Ren, Y.~Chen, K.~Xu, and Y.~Yin, ``Highly integrated reconfigurable
  shared-aperture em surface with multifunctionality in transmission,
  reflection and absorption,'' \emph{IEEE Trans. Antennas Propag.}, 2024.

\bibitem{bao2021full}
L.~Bao, X.~Fu, R.~Y. Wu, Q.~Ma, and T.~J. Cui, ``Full-space manipulations of
  electromagnetic wavefronts at two frequencies by encoding both amplitude and
  phase of metasurface,'' \emph{Adv. Mater. Technol.}, vol.~6, no.~4, p.
  2001032, 2021.

\bibitem{wang2018simultaneous}
X.~Wang, J.~Ding, B.~Zheng, S.~An, G.~Zhai, and H.~Zhang, ``Simultaneous
  realization of anomalous reflection and transmission at two frequencies using
  bi-functional metasurfaces,'' \emph{Sci. Rep.}, vol.~8, no.~1, p. 1876, 2018.

\bibitem{xu2016multifunctional}
H.-X. Xu, S.~Tang, G.-M. Wang, T.~Cai, W.~Huang, Q.~He, S.~Sun, and L.~Zhou,
  ``Multifunctional microstrip array combining a linear polarizer and focusing
  metasurface,'' \emph{IEEE Trans. Antennas Propag.}, vol.~64, no.~8, pp.
  3676--3682, 2016.

\bibitem{cai2017high}
T.~Cai, G.~Wang, S.~Tang, H.~Xu, J.~Duan, H.~Guo, F.~Guan, S.~Sun, Q.~He, and
  L.~Zhou, ``High-efficiency and full-space manipulation of electromagnetic
  wave fronts with metasurfaces,'' \emph{Phys. Rev. Appl.}, vol.~8, no.~3, p.
  034033, 2017.

\bibitem{hou2020helicity}
H.~Hou, G.~Wang, H.~Li, W.~Guo, and T.~Cai, ``Helicity-dependent metasurfaces
  employing receiver-transmitter meta-atoms for full-space wavefront
  manipulation,'' \emph{Opt. Express}, vol.~28, no.~19, pp. 27\,575--27\,587,
  2020.

\bibitem{wu2022circular}
L.-X. Wu, K.~Chen, T.~Jiang, J.~Zhao, and Y.~Feng,
  ``Circular-polarization-selective metasurface and its applications to
  transmit-reflect-array antenna and bidirectional antenna,'' \emph{IEEE Trans.
  Antennas Propag.}, vol.~70, no.~11, pp. 10\,207--10\,217, 2022.

\bibitem{zhang2018transmission}
L.~Zhang, R.~Y. Wu, G.~D. Bai, H.~T. Wu, Q.~Ma, X.~Q. Chen, and T.~J. Cui,
  ``Transmission-reflection-integrated multifunctional coding metasurface for
  full-space controls of electromagnetic waves,'' \emph{Adv. Funct. Mater.},
  vol.~28, no.~33, p. 1802205, 2018.

\bibitem{zhang2019multifunction}
C.~Zhang, J.~Gao, X.~Cao, S.-J. Li, H.~Yang, and T.~Li, ``Multifunction tunable
  metasurface for entire-space electromagnetic wave manipulation,'' \emph{IEEE
  Trans. Antennas Propag.}, vol.~68, no.~4, pp. 3301--3306, 2019.

\bibitem{guo2019transmission}
W.-L. Guo, K.~Chen, G.-M. Wang, X.-Y. Luo, Y.-J. Feng, and C.-W. Qiu,
  ``Transmission--reflection-selective metasurface and its application to rcs
  reduction of high-gain reflector antenna,'' \emph{IEEE Trans. Antennas
  Propag.}, vol.~68, no.~3, pp. 1426--1435, 2019.

\bibitem{li2022transmission}
G.~Li, H.~Shi, J.~Yi, B.~Li, A.~Zhang, and Z.~Xu,
  ``Transmission--reflection-integrated metasurfaces design for simultaneous
  manipulation of phase and amplitude,'' \emph{IEEE Trans. Antennas Propag.},
  vol.~70, no.~7, pp. 6072--6077, 2022.

\bibitem{lau2010planar}
J.~Y. Lau and S.~V. Hum, ``A planar reconfigurable aperture with lens and
  reflectarray modes of operation,'' \emph{IEEE Trans. Microwave Theory Tech.},
  vol.~58, no.~12, pp. 3547--3555, 2010.

\bibitem{wang20191}
M.~Wang, S.~Xu, F.~Yang, and M.~Li, ``A 1-bit bidirectional reconfigurable
  transmit-reflect-array using a single-layer slot element with pin diodes,''
  \emph{IEEE Trans. Antennas Propag.}, vol.~67, no.~9, pp. 6205--6210, 2019.

\bibitem{yu2022reconfigurable}
H.~Yu, P.~Li, J.~Su, Z.~Li, S.~Xu, and F.~Yang, ``Reconfigurable bidirectional
  beam-steering aperture with transmitarray, reflectarray, and
  transmit-reflect-array modes switching,'' \emph{IEEE Trans. Antennas
  Propag.}, vol.~71, no.~1, pp. 581--595, 2022.

\bibitem{xiang2023simultaneous}
M.~Xiang, Y.~Xiao, J.~Deng, S.~Xu, and F.~Yang, ``Simultaneous transmitting and
  reflecting reconfigurable array (star-ra) with independent beams,''
  \emph{IEEE Trans. Antennas Propag.}, 2023.

\bibitem{zhao2024design}
Y.~Zhao, D.~Wang, H.~Gu, P.~Chen, Y.~Liang, M.~Du, W.~Liu, and L.~Ge, ``Design
  of a novel wideband reconfigurable transmit-reflect-array antenna utilizing
  twin lines,'' \emph{IEEE Trans. Antennas Propag.}, 2024.

\bibitem{cao20231}
X.~Cao, C.~Deng, Y.~Yin, Y.~Hao, and K.~Sarabandi, ``1-bit reconfigurable
  transmit-and reflect-array antenna using patch-ground-patch structure,''
  \emph{IEEE Antennas Wirel. Propag. Lett.}, vol.~23, no.~1, pp. 434--438,
  2023.

\bibitem{li2021dual}
X.~Li, Q.~Huang, L.~Yang, M.~Cai, S.~Yang, S.~Yang, and Y.~Li, ``Dual-band
  wideband reflect-transmit-array with different polarizations using
  three-layer elements,'' \emph{IEEE Antennas Wirel. Propag. Lett.}, vol.~20,
  no.~7, pp. 1317--1321, 2021.

\bibitem{yang2021multifunctional}
S.~Yang, Z.~Yan, T.~Zhang, M.~Cai, F.~Fan, and X.~Li, ``Multifunctional
  tri-band dual-polarized antenna combining transmitarray and reflectarray,''
  \emph{IEEE Trans. Antennas Propag.}, vol.~69, no.~9, pp. 6016--6021, 2021.

\bibitem{liu2020wideband}
S.~Liu and Q.~Chen, ``A wideband, multifunctional reflect-transmit-array
  antenna with polarization-dependent operation,'' \emph{IEEE Trans. Antennas
  Propag.}, vol.~69, no.~3, pp. 1383--1392, 2020.

\bibitem{feng2022reflect}
J.~Feng, Z.~Yan, S.~Yang, F.~Fan, T.~Zhang, X.~Liu, X.~Zhao, and Q.~Chen,
  ``Reflect--transmit-array antenna with independent dual circularly polarized
  beam control,'' \emph{IEEE Antennas Wirel. Propag. Lett.}, vol.~22, no.~1,
  pp. 89--93, 2022.

\bibitem{li2020fss}
W.~Li, Y.~Wang, S.~Sun, and X.~Shi, ``An fss-backed reflection/transmission
  reconfigurable array antenna,'' \emph{IEEE Access}, vol.~8, pp.
  23\,904--23\,911, 2020.

\bibitem{zhong2019fss}
X.~Zhong, H.-X. Xu, L.~Chen, W.~Li, H.~Wang, and X.~Shi, ``An fss-backed
  broadband phase-shifting surface array with multimode operation,'' \emph{IEEE
  Trans. Antennas Propag.}, vol.~67, no.~9, pp. 5974--5981, 2019.

\bibitem{balanis2016antenna}
C.~A. Balanis, \emph{Antenna theory: analysis and design}.\hskip 1em plus 0.5em
  minus 0.4em\relax John Wiley \& Sons, 2016.

\bibitem{rao2023active}
J.~Rao, Y.~Zhang, S.~Tang, Z.~Li, C.-Y. Chiu, and R.~Murch, ``An active
  reconfigurable intelligent surface utilizing phase-reconfigurable reflection
  amplifiers,'' \emph{IEEE Trans. Microwave Theory Tech.}, vol.~71, no.~7, pp.
  3189--3202, 2023.

\bibitem{xu2022simultaneously}
J.~Xu, Y.~Liu, X.~Mu, J.~T. Zhou, L.~Song, H.~V. Poor, and L.~Hanzo,
  ``Simultaneously transmitting and reflecting intelligent omni-surfaces:
  Modeling and implementation,'' \emph{IEEE Veh. Technol. Mag.}, vol.~17,
  no.~2, pp. 46--54, 2022.

\bibitem{mautz1973modal}
J.~Mautz and R.~Harrington, ``Modal analysis of loaded n-port scatterers,''
  \emph{IEEE Trans. Antennas Propag.}, vol.~21, no.~2, pp. 188--199, 1973.

\bibitem{balanis2012advanced}
C.~A. Balanis, \emph{Advanced engineering electromagnetics}.\hskip 1em plus
  0.5em minus 0.4em\relax John Wiley \& Sons, 2012.

\bibitem{zhang2021compact}
Y.~Zhang, S.~Shen, Z.~Han, C.-Y. Chiu, and R.~Murch, ``Compact mimo systems
  utilizing a pixelated surface: Capacity maximization,'' \emph{IEEE Trans.
  Veh. Technol.}, vol.~70, no.~9, pp. 8453--8467, 2021.

\bibitem{rao2022novel}
J.~Rao, Y.~Zhang, S.~Tang, Z.~Li, S.~Shen, C.-Y. Chiu, and R.~Murch, ``A novel
  reconfigurable intelligent surface for wide-angle passive beamforming,''
  \emph{IEEE Trans. Microwave Theory Tech.}, vol.~70, no.~12, pp. 5427--5439,
  2022.

\end{thebibliography}

\end{document}